\documentclass[aps,prx,twocolumn,
 amsmath,amssymb,
superscriptaddress,longbibliography]{revtex4-2}
\usepackage[utf8]{inputenc}

\usepackage{graphicx}
\usepackage[dvipsnames]{xcolor}
\usepackage{physics}
\usepackage[breaklinks]{hyperref}
\usepackage{dsfont}
\usepackage[T1]{fontenc}
\usepackage{longtable}

\usepackage{ulem} 

\newcommand{\addt}[1]{\textcolor{black}{#1}}
\newcommand{\outt}[1]{}

\newcommand{\out}[1]{}
\newcommand{\outnew}[1]{}

\newcommand{\add}[1]{\textcolor{black}{#1}} 

\newcommand{\vs}{E_{\text{VS}}}
\newcommand{\vsz}{E_{\text{VS,0}}}
\newcommand{\lcspin}{l_c^{\delta \omega}}
\newcommand{\ldot}{L_{\text{dot}}}
\newcommand{\TP}{\tau} 
\newcommand{\LQB}{L_{\mathrm{s}}}  
\newcommand{\LQBL}{L_{\mathrm{s}}}  
\newcommand{\vd}{v}  
\newcommand{\lcorb}{l_c^{\delta V}}
\newcommand{\lcgg}{l_c^{gg}}
\newcommand{\bus}{SQuS}  

\usepackage{graphicx}
\usepackage{dcolumn}
\usepackage{bm}

\begin{document}
\title{Blueprint of a scalable spin qubit shuttle device for coherent mid-range qubit transfer in disordered Si/SiGe/SiO\texorpdfstring{$_2$}{2}}

\author{Veit Langrock}%
\email[These authors contributed equally]{}
\affiliation{JARA-FIT Institute for Quantum Information, Forschungszentrum J\"ulich GmbH and RWTH Aachen University, Aachen, Germany}
\author{Jan A. Krzywda}%
\email[These authors contributed equally]{}
\affiliation{Institute of Physics, Polish Academy of Sciences, Warsaw, Poland}
\author{Niels Focke}%
\affiliation{JARA-FIT Institute for Quantum Information, Forschungszentrum J\"ulich GmbH and RWTH Aachen University, Aachen, Germany}
\author{Inga Seidler}%
\affiliation{JARA-FIT Institute for Quantum Information, Forschungszentrum J\"ulich GmbH and RWTH Aachen University, Aachen, Germany}
\author{Lars R. Schreiber}%
\email[email to: ]{lars.schreiber@physik.rwth-aachen.de}
\affiliation{JARA-FIT Institute for Quantum Information, Forschungszentrum J\"ulich GmbH and RWTH Aachen University, Aachen, Germany}
\author{{\L}ukasz Cywi{\'n}ski}%
\email[email to: ]{lcyw@ifpan.edu.pl}
\affiliation{Institute of Physics, Polish Academy of Sciences, Warsaw, Poland}

\date{today}

\begin{abstract}
    Silicon spin qubits stand out due to their very long coherence times, compatibility with industrial fabrication, and prospect to integrate classical control electronics. To achieve a truly scalable architecture, a coherent mid-range link that moves the electrons between qubit registers has been suggested to solve the signal fan-out problem. Here, we present a blueprint of such a $\approx 10\,\mu$m long link, called a spin qubit shuttle, which is based on connecting an array of gates into a small number of sets. To control these sets, only a few voltage control lines are needed and the number of  these sets and thus the number of required control signals is independent of the length of this link.   We discuss two different operation modes for the spin qubit shuttle: A qubit conveyor, i.e. a potential minimum that smoothly moves laterally, and a bucket brigade, in which the electron is transported through a series of tunnel-coupled quantum dots by adiabatic passage. We find the former approach more promising considering a realistic Si/SiGe device including potential disorder from the charged defects at the Si/SiO$_2$ layer, as well as typical charge noise. Focusing on the qubit transfer fidelity in the conveyor shuttling mode, we discuss in detail motional narrowing, the interplay between orbital and valley excitation and relaxation in presence of $g$-factors that depend on orbital and valley state of the electron, and effects from spin-hotspots.
    We find that a transfer fidelity of 99.9 \% is feasible in Si/SiGe at a speed of $\sim$10\,m/s, if the average valley splitting and its inhomogeneity stay within realistic bounds. Operation at low global magnetic field $\approx 20$\,mT and material engineering towards high valley splitting is favourable for reaching high fidelities of transfer.
\end{abstract}

\maketitle
\section{Introduction}
Quantum computing architecture based on gated semiconductor quantum dots (QDs) promises the necessary number of qubits for the use of quantum error correction due to their very good coherence properties and direct compatibility with established semiconductor technology \cite{Zwanenburg13,Burkard21}. By using nuclear spin-free $^{28}$Si \cite{Wild12}, the silicon-based spin qubit dephasing time is significantly increased, as it is no longer limited by hyperfine interaction with remaining $^{29}$Si, but by charge noise, which couples to the spin via local magnetic field gradients \cite{Neumann15, Yoneda18, Struck20, Kranz20}. The fidelity of single-qubit \cite{Veldhorst14,Laucht15,Muhonen15,Yoneda18} and two-qubit gates \cite{Harvey-Collard22} already exceeds the quantum error correction threshold \cite{Xue22}. Simple two-qubit gates require an overlap of the electron wave-function \cite{Watson18,Zajac18,Huang19}, which would require a very dense two-dimensional qubit matrix, in which topological quantum error correction can be realized \cite{Fowler12,Horsman12}. However, this dense matrix approach is not easily scalable, if all gates forming the QDs must be individually controllable. The size scale for the outgoing signal lines and their control electronics exceeds that of the dense qubit field by far, and leads to a signal fan-out problem \cite{Vandersypen17}. One part of the solution are multi-layer crossbar architectures \cite{Li18,Hollenberg06,Borsoi22}, in which individual qubits are addressed by the combination of signals at the gates' crossing, or alternatively continuously driven qubits controlled by the global magnetic field \cite{HansenAPR22,HansenPRA21}. Such dense qubits registers can be connected by coherent links providing a two-qubit operation at a distance of approximately 10\,$\mu$m, in order to make space for vias or tiling with cryoelectronics \cite{Vandersypen17, Geck19, Boter22}. Coulomb interaction alone is too weak for long-range high-fidelity two-qubit coupling \cite{Trifunovic12}. First successes could be achieved by an indirect interaction via a mm-long electromagnetic cavity \cite{Beaudoin16,Mi17,Stockklauser17,Borjans20,Harvey-Collard22}, but tuning of the qubit-carrying double quantum dots (DQDs) to the resonance frequency of the cavity is challenging \cite{Borjans20}. Off-resonant driving theoretically circumvents this, but requires longer operation times \cite{Warren21}. In addition, the fabrication of the cavities is hardly compatible with industrial gate-fabrication.

Another method for a medium-range coupling distance of the order of  10\,$\mu$m is the controlled shuttling of the electron \cite{Taylor05,Huang13}, carrying the quantum information in its spin degree of freedom, using a series of gates \cite{Zhao19,Boter19}. Recently, this approach was integrated in the blueprint of a sparse spin qubit array compatible with industrial fabrication without providing details on control and spin coherence of the shuttling process \cite{Boter22}. Alternatively, the shuttling can be used to distribute entangled pairs of electrons to distant arrays (cores), in order to provide coherent communication between them \cite{jnane22}. The charge of a single electron has already been transferred in Si/SiGe over a distance of nine tunnel-coupled QDs \cite{Mills19} using Landau-Zener charge transitions \cite{Li17,Ginzel20,Krzywda20,Zhao19}. In GaAs, the spin-coherent transfer \cite{Baart16,Flentje17,Fujita17} that also preserves spin entanglement \cite{Jadot20} has already been shown. Some GaAs electron conveyors employ surface acoustic waves, replacing the need for a gate array \cite{Mcneil11,Takada19,Jadot20}, but velocity of shuttling with surface acoustic waves is fixed, limiting flexibility of operation, and  furthermore GaAs lacks nuclear spin-free isotopes, making spin qubit decoherence very hard to reduce \cite{Bluhm11}. The demonstrated Si-based shuttlers require individual tuning of the gate array to compensate local potential disorder. Hence, the number of signal lines is proportional to the length of the shuttler, and thus it is not solving the fan-out problem.

We want to realize a mid-range coherent link by shuttling the electron, the spin of which constitutes the qubit, across a distance of approximately $\LQB=10$\,$\mu$m and will refer to it as a spin qubit shuttle (\bus{}). The \bus{} has to fulfil the following criteria: (I) In order to solve the fan-out problem, the number of input terminals required has to be independent of the length of the \bus{}. A scalable quantum computer architecture can be implemented by such a \bus{}. (II) The electron transfer has to be spin-coherent with a sufficiently low error of $\delta C \lesssim  10^{-3}$, in order to preserve the quantum information to the degree necessary for achieving fault-tolerance using quantum error correction codes \cite{Devitt13,Terhal15}, or for executing NISQ algorithms \cite{Preskill18}.
(III) The transfer process has to be at least as fast as the typical timescales of single qubit and near-range two-qubit gates or qubit readout, in order to avoid the situation in which it is the qubit shuttling that determines the quantum algorithm runtime. Thus, a transfer velocity $\vd \sim 10$\,m/s is sufficient. If shuttling is relatively rare compared to qubit manipulation and qubit readout, an order of magnitude lower $\vd$ might be feasible as well. The ratio of occurrences of these events will depend on details of a quantum computer architecture.

We present in this paper a blueprint of such a \bus{}: Scalability is achieved by electrically connecting control lines not to individual gates, but to a few so-called ``gate sets'', where all gates within one set are electrically connected and thus on the same potential.  We discuss two distinct transport modes: The first - the ``bucket-brigade'' (BB) mode - relies on periodic modulation of voltages controlling relative detunings between adjacent QDs in a {\it pre-existing} chain of $N\approx 100$ tunnel-coupled QDs \cite{Mills19}. 
The second - the ``conveyor belt'' (CB) mode - relies on electrostatic creation of a single deep quantum dot that is moving along a one-dimensional channel \cite{Seidler22}. 
We argue that the BB mode is less robust than the CB mode, when scalability of the quantum computing architecture is seriously taken into account in presence of realistic electrostatic disorder.
We focus thus on the theory of shuttling in the  CB mode:  we carry out theoretical optimization of the design of the Si/SiGe structure with gate sets that predicts robust dynamics of an electron-containing QD moving across a disordered channel, and calculate spin qubit decoherence as a function of shuttling velocity.
The main result of the paper given is that for realistic parameters ($T_{2}^*$ times, orbital excitation energies, valley splitting, density of atomic interface steps) of Si/SiGe structures, we predict the existence of an optimal electron velocity $v_{\mathrm{opt}}$ in the CB transfer mode. This $v_{\mathrm{opt}}$ is between five and a few tens of m/s, and we predict that the operation of the \bus{} with this velocity will lead to qubit coherence error below the targeted $10^{-3}$, showing that using of the proposed mid-range link will allow for scalable quantum computing architecture.

In the following, we summarize our considerations on the spin dephasing mechanism in the \bus{} starting from the lowest qubit transfer velocities.  
When the  voltages controlling the \bus{} are varied slowly enough, the qubit transfer should be adiabatic, i.e.~the electron should remain in its lowest-energy orbital/valley state while it is being pushed along the channel. As for the spin degree of freedom, if we assume that the electron does not pass through spin-relaxation hotspots that occur when the spin splitting matches the valley splitting in a given QD \cite{Yang13,Huang14,Petit18,Borjans18, Hollmann20},
the targeted transfer time, $\TP\leq 10\,\mathrm{\mu}$s, is at least three to four orders of magnitude below spin relaxation times in stationary dots in the presence of a magnetic field gradient \cite{Borjans18, Hollmann20}, and the latter are not expected to be lowered significantly due to the quantum dot moving at velocities $\vd \leq 100$\,m/s \cite{Huang13}.
Note that a relatively spatially uniform valley splitting is helpful for choosing a global magnetic field $B$ that leads to such avoidance of the hot-spots. 

In absence of the hot-spots, spin state can then only undergo dephasing due to fluctuations of local values of spin splitting along the channel due to nuclear dynamics \cite{abragam61} and charge noise \cite{Freeman16,Thorgrimmson17,Chan18,Connors19,Chanrion20,Struck20,Kranz20,Stuyck21} modulating the effects of spin-orbit interactions on the spin splitting of an electron (e.g.~through fluctuations of $g$-factors) in a QD at a given location. For a stationary QD, these fluctuations lead to finite $T_{2}^*$ dephasing times \cite{Kawakami14,Yoneda18,Sigillito19,Tanttu19,Struck20}, and the motion of the electron through a channel longer than the correlation length of random contributions to spin splitting enhances the spin dephasing time due to the motional narrowing effect \cite{abragam61}. Making the shuttling velocity $\vd$ higher seems then to be an obvious way to suppress the phase error: the qubit spends less time exposed to perturbations, and their noisy influence is additionally suppressed by fast motion. However, with increasing $\vd$, changes in electrostatic potentials and valley fields experienced by the electron become faster, and the assumption of adiabatic character of the evolution of its orbital and valley degrees of freedom has to become untenable.

When the dynamics of the electron becomes non-adiabatic, motion-induced excitation of the electron into higher-energy states has to be taken into account. Transitions to excited orbital states of the electron in a potential of the moving QD are caused by electrostatic disorder in the channel: in the frame co-moving with the QD the quasi-statically fluctuating disorder turns into dynamic noise coupling the orbital levels. Analogously, atomic-scale interface roughness \cite{Friesen07,Culcer10,Zwanenburg13} that affects the valley splitting \cite{Borselli11,Shi11, Kawakami14,Scarlino17,Zajac15,Mi17,Watson18,Ferdous18,Mi18-2,Borjans18,Hollmann20,Yang13,Petit18,Zhang20,Ciriano-Tejel20} and determines the composition of valley states \cite{Friesen07,Friesen10,Culcer10,Hollmann20,Wuetz21,Mcjunkin21} in a static QD, becomes a time-dependent valley-coupling term for a moving QD, with intervalley excitations appearing as a result. Once the electron starts to occupy excited states, it becomes susceptible to processes of energy relaxation accompanied by emission of phonons. This makes the evolution of the orbital/valley state stochastic: the electron will spend random fractions of shuttling time in various orbital/valley states, thus opening up a new channel for qubit dephasing. Spin-orbit coupling makes the $g$-factor state dependent \cite{Kawakami14,Veldhorst15,Ruskov18}, with a relative variation of electron spin $g$-factor between distinct valley states being $\sim 10^{-3}$ \cite{Kawakami14,Ferdous18}. A similar $g$-factor difference was measured between neighouring QDs \cite{Liu21,Cai23}, which we assume to be an upper bound for g-factor difference between lowest-energy and excited orbital in a single QD. $B$-field independent contribution to spin splitting due to spin-orbit interaction also depends on the valley state \cite{Nestoklon07,Tanttu19}.
Consequently, any randomness in time spent by the electron in distinct valley/orbital states will lead to randomness in qubit phase (and thus dephasing) for any finite external magnetic field. With rms of the phase given by $\delta \phi$, and the probability of excitation out of instantaneous ground state given by $p_e$, the phase error $\delta C$ is given by
\begin{equation}
    \delta C = p_e (1-e^{-\delta\phi^2/2}) \approx p_e \delta\phi^2/2  \,\, \mathrm{when} \, \delta \phi \ll 1 \,\, , \label{eq:dC}
\end{equation}
where in the first formula we have assumed that the distribution of the random contributions to qubit phase is Gaussian, while the second one, relevant in the small-error regime of interest here, does not require this assumption. As $\delta C \! \leq \! p_e$, limiting the probability of orbital and valley excitations, i.e.~keeping $p_e \! < \! 10^{-3}$, is one route towards reaching the targeted level of phase error. When this turns out to be impossible, suppression of $\delta \phi$, e.g.~by making the orbital/valley relaxation faster (thus making the electron spend shorter periods of time in excited states), is the remaining route towards a coherent \bus{}. Quantitative calculation of  both orbital/valley excitations caused by a QD motion, and orbital/valley relaxation, as functions of parameters of the \bus{}, is thus the main topic of the second part of the paper, in which we focus on coherence of the electron transferred in the CB mode.

\begin{figure}
    \centering
    \includegraphics[width=\columnwidth]{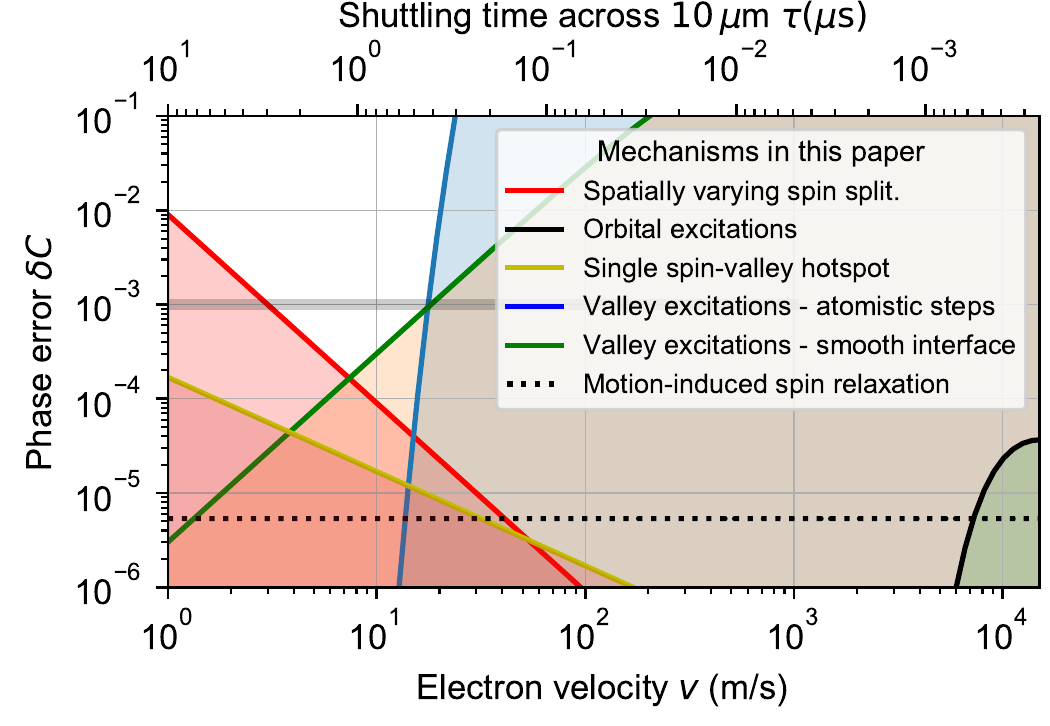}
    \caption{A sketch of the transfer errors of an electron spin-qubit after 10\,$\mu$m shuttling in the CB mode as a function of the shuttling velocity for various decoherence mechanisms discussed in this paper. In Sec.~VIII, we discuss the detailed plots of these dependencies that summarize all the calculations in this paper.}
    \label{fig:intro}
\end{figure}

The key results of our calculations are the following: Already for reliable transfer of the electron, the CB mode of electron-spin shuttling is superior vs. its BB counterpart in terms of scalability, i.e. robustness against potential disorder with only few signal lines independent from the \bus{}-length available (Sec.~\ref{sec:device_idea}). The propagating QD potential required for the CB mode can be generated in a realistic Si/SiGe \bus{} despite having typical density of charged defects at the Si/SiO$_2$ interface (Sec.~\ref{sec:design_CB}). Then, motional narrowing enhances spin coherence of a single electron confined in the propagating QD compared to a static QD, setting a comfortable lower velocity limit to the CB-mode (red line in Fig. \ref{fig:intro}). The state-dependent electron g-factor sets the upper velocity limit in conjunction with diabatic QD motion (Sec.~\ref{sec:nonadiab_dephasing}): Orbital excitations (black line in Fig. \ref{fig:intro}) appear rarely and relax quickly in the propagating QD in our realistic Si/SiGe \bus{}, so they do not limit the spin coherence (Sec.~\ref{sec:orbital_nonadiabaticity}). However, valley excitations last orders of magnitude longer, setting the upper velocity limit of the CB (green and blue lines in Fig. \ref{fig:intro}) within the bounds of our models  of the lateral valley-splitting fluctuations (Sec.~\ref{sec:valley_nonadiabaticity}). Spin relaxation hot-spots have to be passed sufficiently fast, setting another lower velocity limit (yellow line in Fig. \ref{fig:intro}) (Sec.~\ref{sec:valley_nonadiabaticity}). Finally, dotted line shows that motion-induced spin relaxation caused by spin-orbit interaction analyzed previously in \cite{Huang13} (see Sec.~\ref{subsec:spin_relax}) does not pose a threat to the coherence of the shuttled qubit. Taking all spin-decoherence mechanisms into account in Sec.~\ref{sec:discussion}, we predict that CB mode shuttling across a distance of 10~$\mu$m with less than 0.1~\% infidelity is feasible under favorable QD velocity (white area underneath gray line in Fig. \ref{fig:intro}), magnetic field, and \bus{}-geometry.

\section{\bus{} device concept} 
\label{sec:device_idea}

\begin{figure}
\includegraphics[width=\linewidth]{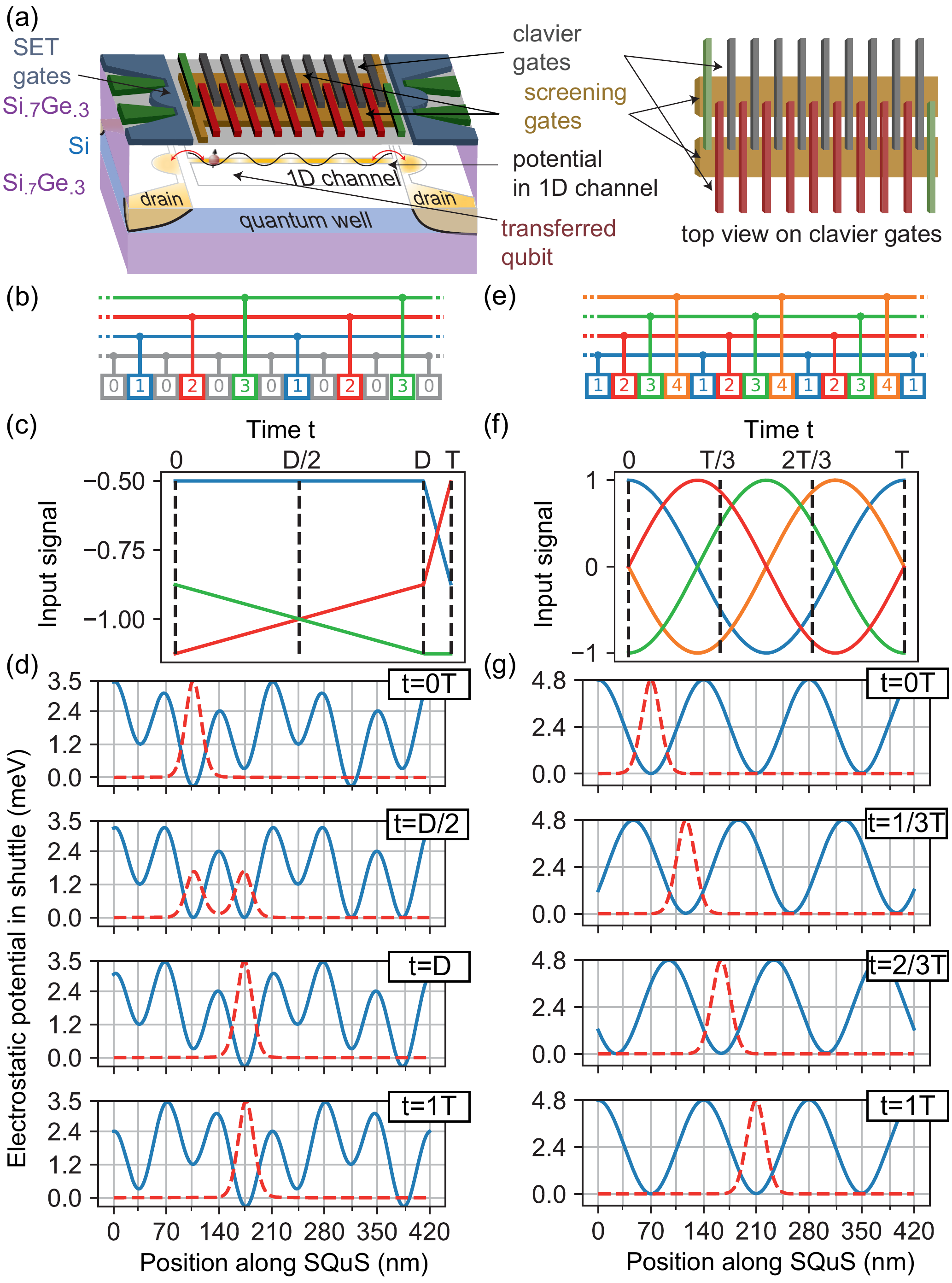}
\caption{\bus{} concept and operation modes. (a) Sketch of a \bus{} device. (b)-(g) Comparison of BB (left panels) and CB (right panels) transfer modes.
(b),(e): Position of the clavier gates and electric connection scheme for the BB mode (b) and CB mode (e), respectively.
(c),(f): Input voltages during a \bus{} step cycle T for the BB mode (c) and CB mode (f), respectively. d,g: Electrical potential (blue) calculated along the center of a \bus{} within the Si/SiGe quantum well for four time frames (dashed lines in panels b and d for each mode). Dashed red line indicates the calculated electron density.}
\label{fig:modes}
\end{figure}

A sketch of the \bus{} device is shown in Fig. \ref{fig:modes}a. It is based on multi-layer electrostatic gates and thus its fabrication is compatible with common QD devices \cite{Mills19,Yang13}, in which qubit manipulation with necessary fidelities was demonstrated, and can be readily adapted by industrial fabrication lines \cite{Zwerver22}. The shuttling process is controlled by an array of metal gates called clavier gates on top of a Si/SiGe heterostructure. Two parallel gates (called screening gates) underneath the clavier gate layer, define a one-dimensional electron channel and screen the electric field of the clavier gates at the edge of the channel. The DC voltage applied to these gates is chosen to deplete this channel. Our main idea is to connect a few control lines to a small number of gate sets (i.e. clavier gates set to the same potential), the number of which is independent of the length of the coherent link \cite{Seidler22}. The constant number of input terminals of the coherent link solves the signal fan-out problem and assures full scalability of our approach. Along the \bus{} channel we require no charge detector. For a \bus{} test device, we suggest two single electron transistors (SET) at each end of the channel, which detects the charge state at the end of the channel (Fig. \ref{fig:modes}a). If the SET is tunnel-coupled to the channel, electrons can be loaded into and unloaded from the channel on demand, controlled by small number of dedicated gates at the end of the \bus{}. For a quantum computing architecture, other approaches for loading/detecting electrons may be found.   

\subsection{Two modes of qubit transfer}
There are two modes of operation for a shuttling which we call ``bucket brigade'' (BB) and ``conveyor belt'' (CB) transfer mode, which differ by the connection scheme of the clavier gates (Fig. \ref{fig:modes}b for BB and \ref{fig:modes}e for CB, respectively). Both modes are explained in more detail below.  

\begin{figure}
\includegraphics[width=1\columnwidth]{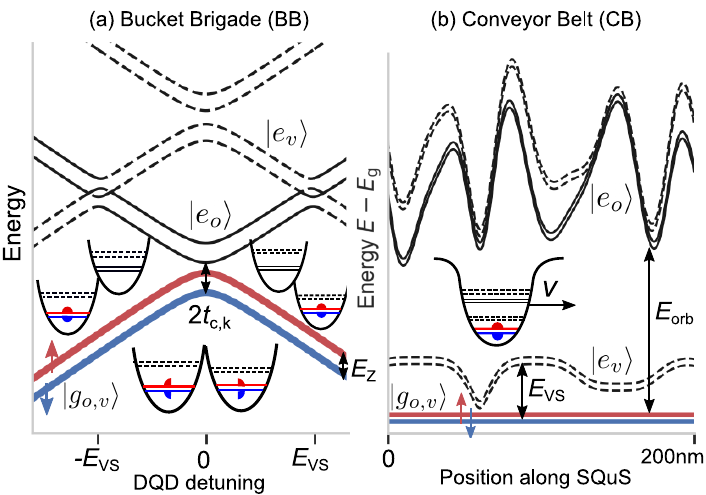}
\caption{Schematic picture of the energy spectrum during transfer of spin superposition in presence of orbital (ground and first excited only), valley and spin degrees of freedom, as a function of: (a) DQDs detuning $\epsilon$ in a single transition between the QDs in Bucket Brigade, (b) position along the \bus{} in the Conveyor Belt. We denoted excited valley states $\ket{e_v}$ using dashed line and the joint ground state of orbital and valley degrees of freedom $\ket{g_{o,v}}$ by thicker: blue (spin down) and red (spin up) line. In the insets, we illustrate mechanism of electron shuttling in each mode, note that contrary to BB mode the charge transfer in CB does not require effectively adiabatic evolution in the lowest energy levels. In the conveyor belt (b), we plotted energy difference between excited and ground orbital levels including expected electrostatic disorder (See \ref{sec:e_disorder}). We also included the modification of energy spectrum due to atomistic steps (left kink) and a smooth interface gradient (right kink), that lead to temporal reduction of valley splitting $E_{\text{VS}}$. }
\label{fig:levels}
\end{figure}

The BB transfer mode (Fig. \ref{fig:modes}, left panel column) requires a linear array of QDs along the \bus{}, which can be pairwise described by the Hamiltonian $H_k=\epsilon_k \sigma_z/2+ t_{c,k} \sigma_x$, where $\epsilon_k$ and $t_{c,k}$ are the interdot energy detuning and interdot tunnel-coupling of the $k$-th double quantum dot (DQD) pair of the linear array, respectively, and $\sigma_x$ and $\sigma_z$ are Pauli-matrices. The electron is transferred by adiabatic Landau-Zener transitions (LZT) between adjacent DQDs: time-dependent voltages applied to adjacent QDs change $\epsilon_k$ from negative to positive, and the energies of states localized in each of the two QDs anti-cross, with the minimum gap given by twice the tunnel coupling $t_{c,k}$ (Fig.~\ref{fig:levels}a). Three clavier gate sets (1,2,3) operating as plunger gates (i.e. controlling mainly the chemical potential of the QD underneath) in Fig. \ref{fig:modes}b trigger the LZTs at time $D/2$ by sweeping the gate voltages of the gates sets as plotted in Fig. \ref{fig:modes}c. The potential at four time frames is sketched in Fig. \ref{fig:modes}d. The channel is initially depleted and only one QD is filled (red wavefunction in Fig. \ref{fig:modes}d) by loading a single electron from one channel end. 

In the CB mode (Fig. \ref{fig:modes}, right panel column), the existence of an array of tunnel-coupled QDs is not needed. Every fourth clavier gate is connected (Fig. \ref{fig:modes}e) and by applying sine-signal with $\pi/2$ phase shift to each gate set as depicted in Fig. \ref{fig:modes}f, a propagating sine-wave potential in the 1D quantum channel is induced (Fig. \ref{fig:modes}g) with only one pocket filled by the electron to be transferred (red wavefunction in Fig. \ref{fig:modes}g). We refer to this pocket as {\it single} moving QD. Tunnel coupling between pockets of the sine function has to be excluded by proper choice of gate pitch and the signal amplitude. We will show that this requirement can easily be fulfilled by a realistic device in Section \ref{sec:design_CB}, for which electron shuttling on a short length scale has been demonstrated already \cite{Seidler22}. A minimum of three gate sets is required to define the direction of the transfer. We choose a 4-gate sets scheme here, since it eases the realisation of a connection scheme of the \bus{} in the CB mode as will be elaborated in Section \ref{sec:design_CB}. The input signals sketched in Fig. \ref{fig:modes}c,f presume a constant transfer velocity. Smooth acceleration can be implemented in CB by sweeping the frequency of the \bus{} signals applied to the clavier gates. Adiabatic reversal of the transfer direction is feasible as well. Such a flexibility is lacking in the surface acoustic wave approach demonstrated in GaAs as the speed of sound is intrinsically fixed. Note that mixtures of the BB and CB modes with more input signals might be beneficial, but we restrict the considerations to these two extreme case of transfer modes. 

\subsection{Challenges in the bucket-brigade mode}
Within our scalable gate-set approach two significant limitations arise in BB transfer mode, both originating from typical potential disorder in Si devices: Firstly, fine-tuning of gate voltages for setting the $t_{c,k}$ close to a common value of $t_c$, and secondly fine-tuning of voltage pulses applied to specific QDs triggering adiabatic LZTs along the chain, are both impossible to achieve, since gates are electrically connected. As potential disorder is unavoidable, we have to deal with a range of $t_{c,k}$, while the sweeping range for $\epsilon_k$ at all LZT has to be large enough to compensate offsets in zero-detuning points among adjacent QDs. The requirement of being able to transfer the qubit with the use of only a few synchronized time-dependent voltage signals (Fig. \ref{fig:modes}c), applied to appropriately designed gates spanning the whole length of the \bus{}, leads to rather stringent requirements on the degree of uniformity of the channel through which the electron is to be sent. In the following we quantitatively explicate on this requirement. Considering a 10\,$\mu$m long \bus{}, an array of $N \! \sim \! 100$ tunnel-coupled QDs is required to span the distance. Then the qubit shuttling is effected by consecutive LZT of the electron between neighboring DQDs, driven by a sweep of the DQD interdot detuning $\epsilon$ (Fig. \ref{fig:levels}a). First, we focus on the probability $Q$ of nonadiabatic LZT evolution given by $Q_k \! =\! \exp( -2\pi t_{c,k}^2 /\hbar v_\epsilon)$, where $v_{\epsilon}$ is the rate of change of detuning \cite{Shevchenko10}. This $Q_k$ is the probability that the electron will fail to transfer adiabatically from QD $k$ to QD $k+1$ when $\epsilon_k$ is driven through the anticrossing of tunnel-coupled states localized in the two QDs. It should be stressed that while in principle the charge transfer could occur in an inelastic way (via phonon-assisted tunneling) in subsequent part of the BB driving cycle in which the energy of electron in $k$-th QD is larger than the energy in $k+1$-th QD (between $t\! =\! D/2$ and $t=D$, see Fig.~\ref{fig:modes}), this process is inefficient in Si/SiGe QDs \cite{Krzywda21}, and it can be neglected on timescales relevant here.

To achieve adiabatic (and thus deterministic) evolution through the whole length of the \bus{} with an error of $<10^{-3}$, each of the $N$ transfers has to fulfill $Q<10^{-5}$. This condition is rather restrictive as can be understood by translating it into a lower limit of all $t_{c,k}$: Let us assume that the transfer time $\LQBL/v$ through the \bus{} is at least 10\,$\mu$s, to avoid limitation of quantum computer clock speed. For $N\approx 100$ this means that the interdot transfer time has to be $\approx \! 100$\,ns.  
In state-of-art Si/SiGe devices, the offsets in zero-detuning points among adjacent QDs have a Gaussian distribution with an rms of about 3\,meV  (cf. simulations in Fig.~\ref{fig:corr_fun}). Thus the potential disorder sets rigorous restrictions on $\epsilon_k(t)$: (I) It has to span a range of $\approx 10$\,meV in order to include all $N-1$ DQD zero-detuning points (at time $D/2$ in Fig. \ref{fig:modes}d), since no individual compensation of disorder is possible in our gate-set approach. (II) $v_{\epsilon}$ has to be constant, since detuning at which tunneling-induced anticrossing of states occurs is unknown due to disorder. (III) $v_{\epsilon}> 100$\,$\mu$eV/ns to pass the \bus{} in less than 10\,$\mu$s. This conditions translate into $t_{c,k}>10$\,$\mu$eV for all $N-1$ DQD-pairs, in order to achieve adiabatic charge transfer across the whole \bus{}. Accordingly, passing the \bus{} in less than 1\,$\mu$s, requires $t_{c,k}>35$\,$\mu$eV. We believe that such a high $t_{c,k}$ is hard to achieve for all $k$ with state-of-art disorder in Si/SiGe devices, if only a common voltage can be applied to all the barrier gates. In particular, for an ensemble of $N\sim 100$ QDs with expected value of $t_c = 35\,\mu$eV, the $35\,\%$ variation of tunnel coupling would on average result in two weakly coupled QD pairs (with $t_c<10\,\mu$eV) for which adiabatic transfer would fail with high probability. 

\begin{figure*}
\includegraphics[width=0.8\linewidth]{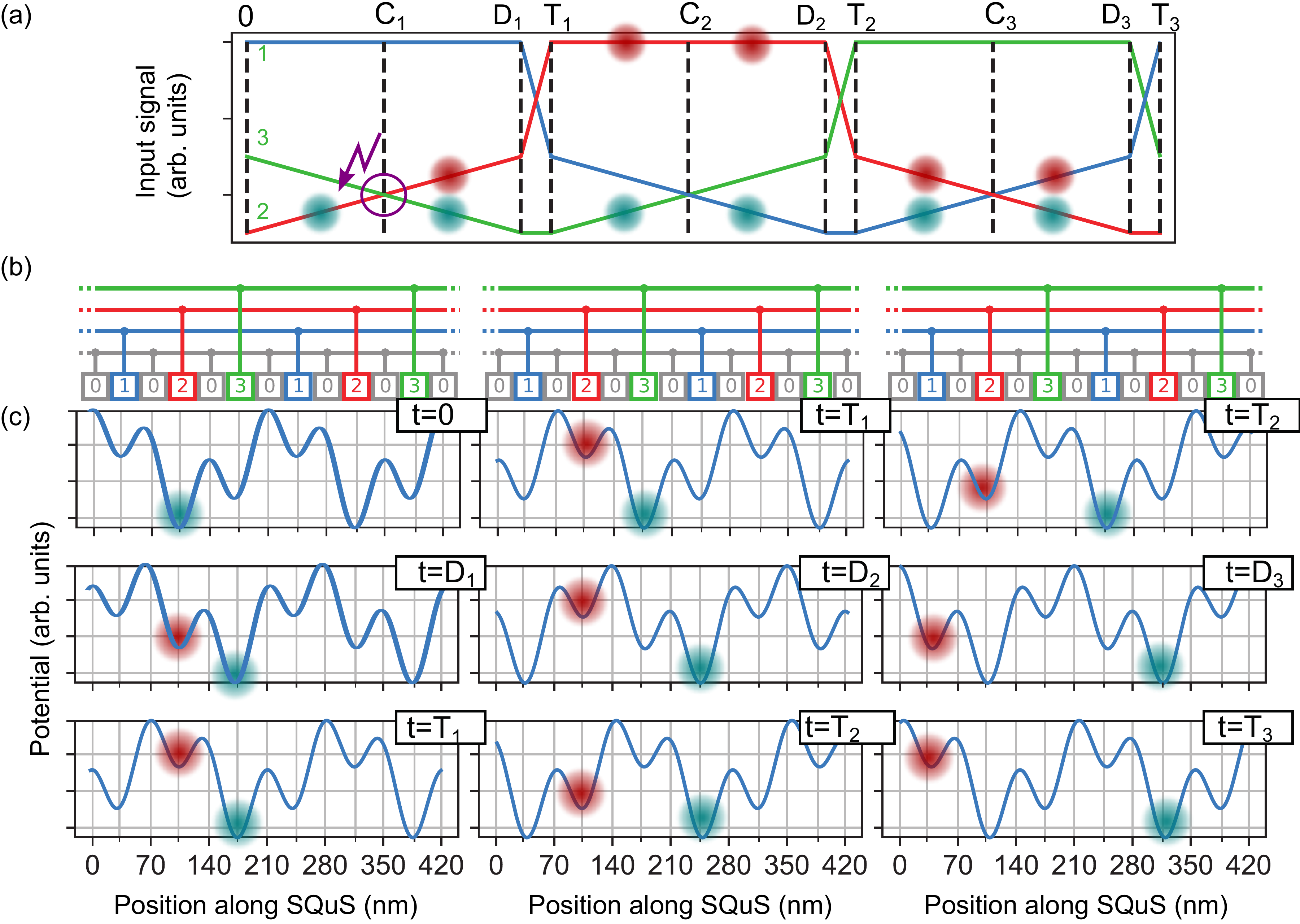}
\caption{Illustration of the propagation of an electron in BB mode for which the QD transfer fails once (red) and comparison with an electron which follows the intended path (blue). (a) Three input signal as a function of time applied to the gate sets labeled 1, 2 and 3 in panel b. (b) Labeling of the clavier gates and electrical connection. Fixed voltages are applied to all barrier gates, labeled 0. (c) Sketch of the potential in the QW along the \bus{} channel aligned with the clavier gates in panel b. The potential plotted in each sub-panel corresponds to a different time $t$ labeled in panel a. Blue circles show the intended position of the shuttled electron and red circles mark the expected electron position in case of a failed LZR transition marked by the circle in panel a.}
\label{fig:backwardspropagation}
\end{figure*}

How does imperfectly adiabatic LZT degrade qubit coherence? Naively, one might think the qubit will arrive just a few BB signal cycles delayed. Besides besides losing track of the exact qubit position this will lead to
qubit dephasing, if the Zeeman energies of all QDs are not exactly equal due to small variations of electron g-factor and/or local magnetic field. In fact, it can become much worse, since only one diabatic LZT might lead to a reversal of the shuttling direction as illustrated in Fig. \ref{fig:backwardspropagation}. If the adiabatic transfer is unsuccessful within the time range 0 to $D_1$ (electron position marked red), then the electron will move backward, if the following LZT within time $D_1$ to $T_1$ is adiabatic. While this following LZT can be to some extent enforced to be diabatic by a fast detuning ramp, this is not possible for the LZT transition at $C_3$. The effect is that the electron starts to shuttle in the opposite direction as can be seen by comparing the electron position marked blue (intended) and red (unintened) in Fig. \ref{fig:backwardspropagation}. As noted before,  electron-phonon coupling in Si is too weak to allow for efficient enough inelastic phonon-assisted tunneling between detuned QDs \cite{Krzywda21} and thus this back-transfer can be persistent for several BB signal cycles. Such an unintentional reversal of shuttling direction is catastrophic to a quantum computing architecture and can be triggered by only one diabatic LZT of two adjacent QDs. This rules out the BB transfer mode, unless ways to mitigate this reversal are found. Certainly, lowering the potential disorder will lead to significant improvements of stability of operation, as it will allow for smaller sweep range (and thus lower $v_\epsilon$ while keeping the same total shuttling time). There are even more challenges of the BB transfer mode: If the LZT evolution is made very slow to ensure low $Q$, not only transfer velocity $v$ and thus clock-speed sets a limit, but processes of electron excitation due to charge noise become significant, and in fact at lower $v$ their presence limits $Q$ from below, as it has been shown in recent work considering coupling of the transferred electron to phonons and sources of both $1/f$ and Johnson-Nyquist charge noise \cite{Krzywda20,Krzywda21}. Additionally, transition between neighbouring sites separated by $50-100 \,\mathrm{nm}$ might lead to valley excitations, caused by spatial variation of valley splitting $\vs$ in typical Si/SiGe heterostructures (Fig. \ref{fig:levels}a) \cite{Dodson22}. The temporal occupation of higher valley and presence of valley-orbit mixing can lead to spin decoherence as discussed in detail in this paper in the context of the CB mode.

We conclude that coherence error below $10^{-3}$ will be difficult to achieve in scalable BB mode without a significant improvement of the uniformity of state of the art devices. The tension between conflicting requirements of small interdot barriers resulting in large tunnel couplings (necessary for deterministic, and consequently coherent spin shuttling) and homogeneity of parameters of 100 QDs in a realistically disordered heterostructure (necessary for scalability) is absent in the CB mode of operation.  In the remaining part of the paper we will thus focus on analysis of that mode of the \bus{}.

\subsection{Larger robustness of the conveyor-belt mode}
In order to start thinking about qubit transfer in CB mode, we only have to require that the depth of the single moving QD is much larger than than a typical variation of electrostatic disorder potential on length-scale of QD size.
Compared to the BB mode, we do not have to worry about tension between the requirement for existence of separate (albeit well-coupled with finite $t_c$) quantum dots, and the need for charge-transfer control with global pulses, {\it both in presence of disorder}. We only need to create a moving QD of a stable shape (Fig. \ref{fig:levels}b). 

Due to this, we expect the CB mode to be more robust to disorder in the channel as will be further discussed in Section \ref{sec:design_CB}. Considering signal generation and bandwidth of signal lines, a maximal input signal frequency of 100\,MHz is convenient, and it results in a sufficiently high transfer velocity of 20\,m/s at a typical gate pitch of 50\,nm. As a disadvantage compared to BB mode, the CB mode requires a higher dynamic range of the input signals, which might cause Ohmic heating, if the clavier gates are not superconducting. It also sets limits on the distance between clavier gates and the moving QD, and thus the depth of the QW and the clavier gate pitch must be well-balanced (Section \ref{sec:design_CB}). Transfer velocity, signal amplitude and other issues such as spatial fluctuations of $\vs$ (shown in Fig. \ref{fig:levels}b) affecting the coherent spin transfer in the CB mode will be discussed in the following sections.

\section{Optimization of gate-design for conveyor belt mode shuttling}
\label{sec:design_CB}

\begin{figure}
\includegraphics[width=\linewidth]{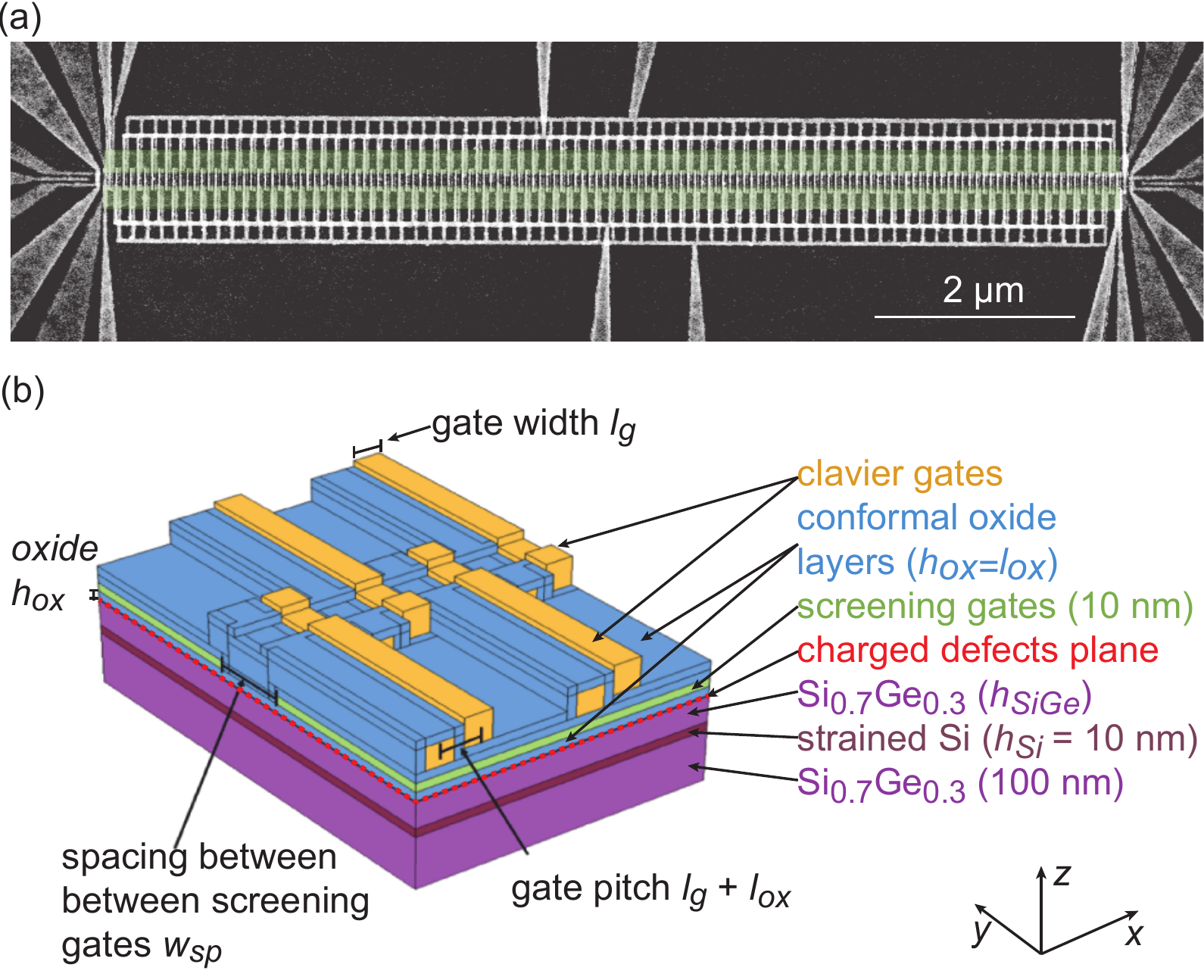}
\caption{Realistic Si/SiGe \bus{} device. (a) Scanning electron micrograph of 10\,$\mu$m long \bus{} showing three gate layers: on top of a screening gate layer (colored in green), two clavier gate layers connected to four gate sets for CB mode. Gates at the ends of the \bus{} are for the formation of single electron transistors. (b) COMSOL model of the \bus{} device from panel a showing two \bus{} unit cells and thus eight clavier gates. The typical thickness of layers is given in brackets.}
\label{Fig1}
\end{figure}

In this Section we elaborate further on the blueprint of a \bus{} operating in conveyer belt (CB) mode and optimize the gate design of a realistic undoped Si/SiGe \bus{} for CB mode,  as the upper SiGe spacer layer keeps charged defects, typical for the semiconductor-oxide interface, at a larger distance and thus reduce its impact on potential disorder \cite{Bluhm19}. In order maximize the robustness of the adiabatic charge shuttling against the potential disorder, we check whether its magnitude in the \bus{} channel is sufficiently low compared to the confinement of the propagating QD and whether the associated correlation length of disorder is sufficiently large not to break the QD apart. As the dominant source for potential fluctuations, we simulate the impact of charged defects at the interface between a thin Si cap layer (on top of the Si/SiGe heterostructure) and the planar SiO$_{2}$ layer. The model employed for finite-element calculations is based on realistic Si/SiGe device with three metallic gate layers (Fig. \ref{Fig1}a). By alignment of two clavier gate layers fabricated by electron-beam lithography, we can achieve an effective minimal gate pitch of $l_\mathrm{g}+l_\mathrm{ox}=$35\,nm, where $l_\mathrm{g}$ and $l_\mathrm{ox}$ are the width of a single clavier gate and the oxide thickness, respectively. The connection scheme of the clavier gates required for CB mode (cf. Fig. \ref{fig:modes}) can be realised within each of the two clavier gate layers by connections on both sides of the \bus{} (Fig. \ref{Fig1}a). We will discuss whether this minimal gate pitch is sufficient for the \bus{} or whether even larger gate pitches are optimal and thus fabrication constraints can be relaxed. For finite element simulations of the \bus{} electrostatics, we use a COMSOL model of the realistic \bus{} device, a part of which is shown in Fig. \ref{Fig1}b. In this model having a total size of 2600 nm by 400 nm, we randomly distribute spatially uncorrelated singly charged defects at the planar Si/SiO$_{2}$ interface having a density of $5 \times 10^{10}$\,cm$^{-2}$. This is a typical defect density for Si/SiO$_{2}$ interface extracted from room-temperature C-V measurements \cite{Klos2019,Thoan11,Campbell05}.

\subsection{Analytical optimization of the device design without defects}
\label{sec:analitical_opt_dev}
\begin{figure}
\includegraphics[width=\linewidth]{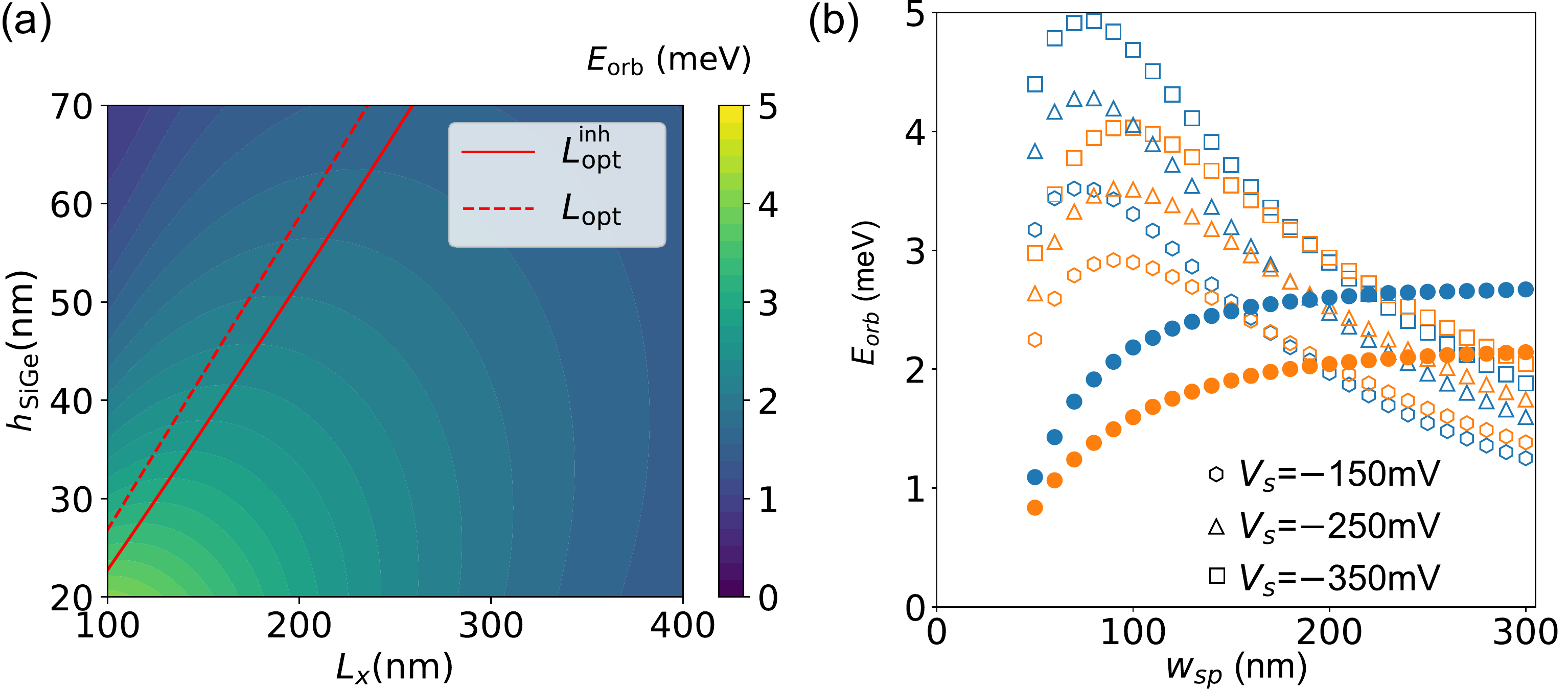}
\caption{Orbital-splitting as a function of the geometric parameters of the \bus{} without disorder. (a) Using $h_\mathrm{ox}=10$\,nm and $V_a=100$\,mV, the minimum orbital splitting $E_\mathrm{orb}^x$ is computed by solving the Schr{\"o}dinger equation on the full-mode periodic potential. The red dashed line indicates $L_{\mathrm{opt}}$ and the red solid line the optimal unit cell length for the case of inhomogeneous dielectics $L_{\mathrm{opt}}^\mathrm{inh}$ (discussed in Appendix \ref{App:electrostaticEstimatesDetails}). (b) Orbital splitting $E_\mathrm{orb}^x$ (filled symbols) and $E_\mathrm{orb}^y$ (open symbols) as a function of the gap between the screening gates $w_\mathrm{sp}$ for an oxide thickness $h_{ox}=5$\,nm (blue symbols) and $h_\mathrm{ox}=10$\,nm (orange symbols) and the voltage $V_S$ applied to the screening gates (see the model in Fig. \ref{Fig1}b). A fixed sin-signal amplitude $V_a=100$\,mV is applied to the clavier gates and $h_\mathrm{SiGe}=45$\,nm.}
\label{Fig5}
\end{figure}

The geometry of the \bus{} is not only given by the width of the clavier gates $l_\mathrm{g}$ and their spacing $l_\mathrm{ox}$, but also by the thicknesses $h_\mathrm{ox}$, $h_\mathrm{SiGe}$ and $h_\mathrm{Si} \approx 10$\,nm of the oxide layer, the Si$_{0.7}$Ge$_{0.3}$ layer and the thickness of the strained Si quantum well (QW) layer, respectively, as well as the spacing between the screening gates $w_\mathrm{sp}$ (Fig. \ref{Fig1}b). As the parameter space for the \bus{} geometry is large, we start our discussion by neglecting the charged defects and investigate the harmonicity of the propagating potential in the center of the quantum well layer as a function of the interplay between the depth of the quantum well and the gate pitch. We aim at maximizing the orbital energy along the \bus{} transfer direction $E_\mathrm{orb}^x(x)$ at a given sine wave voltage amplitude $V_a$ applied to the clavier gate sets at all positions of the propagating QD. For further simplicity, we assume that all four clavier gate sets are at the same height (other than plotted in Fig. \ref{Fig1}b), thus the thickness of the SiO$_{2}$ layer $h_\mathrm{ox}$ underneath each layer is the same. This approximation allows us to argue with the Fourier analysis of the potential formed by the array of clavier gates, in order to find an analytical expression for the relation of the gate pitch to the depth of the QW given by the thickness of the oxide and the thickness $h_\mathrm{SiGe}$ of the SiGe top layer (a 1\,nm thin Si cap layer is neglected). We assume $w_\mathrm{sp}$ to be large enough that we can reduce the problem to the xz-plane. As the \bus{} potential exhibits a periodicity of four gate pitches due to the use of four gate sets, we introduce the unit cell length of the \bus{} $L_x=4 (l_\mathrm{g}+l_\mathrm{ox})$. 
For a homogeneous dielectric, we then find an optimal unit cell length of $L_x=L_{\mathrm{opt}}\equiv\pi(h_\mathrm{SiGe}+h_\mathrm{ox})$, with which the orbital confinement energy along the $x$-direction scales as $E_\mathrm{orb}\propto\frac{1}{L_\mathrm{opt}}$, inversely proportional to the depth of the quantum well. Fig. \ref{Fig5}a compares the estimated optimum $L_\mathrm{opt}$ to a numerically exact solution of the one dimensional potential. Details on the calculations and additional information are given in Appendix \ref{App:electrostaticEstimatesDetails}.

Next, we analyse the effect of $w_\mathrm{sp}$ on the curvature of the QD potential by harmonic fits in $x$- and $y$-direction and express the result as an effective $E_\mathrm{orb}^x$ and $E_\mathrm{orb}^y$, respectively. We plot the minimum of all $E_\mathrm{orb}^x(x)$ and $E_\mathrm{orb}^y(x)$ of all $x$-positions along the \bus{} unit cell for two oxide thicknesses (Fig. \ref{Fig5}\add{b}). As $w_\mathrm{sp}$ is increased, the capacitive coupling of the clavier gates to the QD increase and therefore $E_\mathrm{orb}^x$ increases with $w_\mathrm{sp}$. For larger $w_\mathrm{sp}$, the gain in $E_\mathrm{orb}^x(x)$ decreases. On the other hand, $E_\mathrm{orb}^y(x)$ decreases with increasing $w_\mathrm{sp}$, since with widening of the \bus{} channel the QD becomes more elliptical towards the $y$-direction. The QD confinement is maximized, if the QD remains approximately circular during the shuttling, which is fulfilled at $w_\mathrm{sp} \! \approx \! 200$\,nm here. A negative voltage applied to the screening gates $V_S$ can be used as an additional degree of freedom to enhance $E_\mathrm{orb}^y$ independent from $E_\mathrm{orb}^x$ (Fig. \ref{Fig5}\add{b}). 

\subsection{Numerical simulation of the QD shuttling in presence of interface defects}
\label{sec:e_disorder}

\begin{figure}[ht!]
\includegraphics[width=\linewidth]{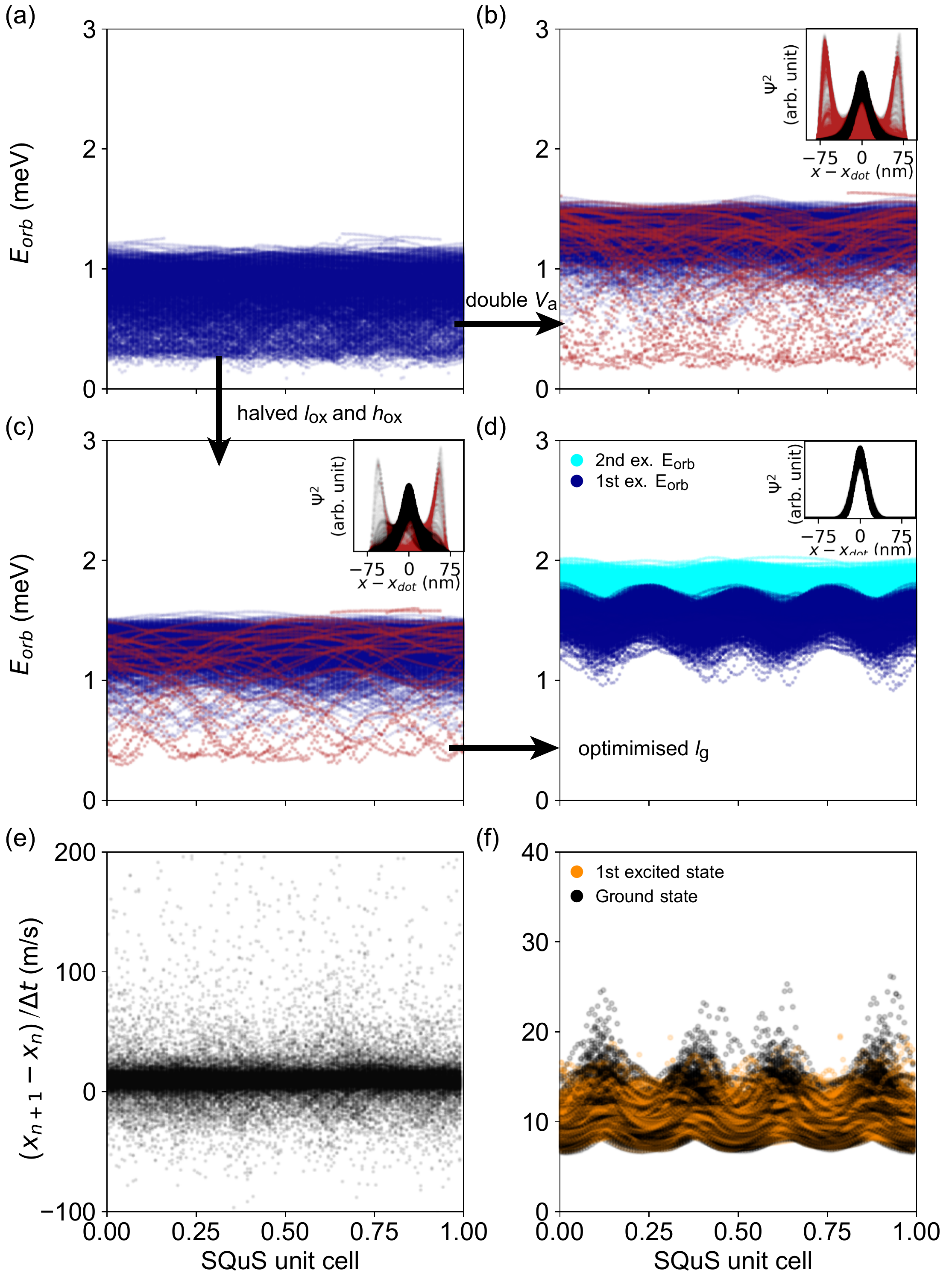}
\caption{Simulation based on Poisson-Schr{\"o}dinger solutions of propagating potential in CB mode for 800 ensembles of randomly distributed charged defects at the Si/SiO$_2$ interface for $h_\mathrm{SiGe}=45$\,nm and $w_\mathrm{sp}=200$\,nm. (a) Orbital energy $E_\mathrm{orb}$ of the first excited state of the propagating QD assuming an oxide thickness  $h_\mathrm{ox} = 10$\,nm, gate width $l_\mathrm{g}=30$\,nm and signal amplitude of $V_{a,1}= 45$\,mV and $V_{a,2}= 100$\,mV applied to the first and second claviature gate layer, respectively. (b) Same as in panel a with parameters ($h_\mathrm{ox} = 10$\,nm, $V_{a,1}= 90$\,mV, $V_{a,2}= 200$\,mV, $l_\mathrm{g}=30$\,nm). Inset: The probability density $\int |\psi(x,y)|^2 dy$ of the ground electron wavefunction is plotted for all expected QD positions $x_{dot}$ and all charge defect ensembles. Wavefunctions deviating from a single QD are marked in red together with their corresponding $E_{orb}$ data point. (c) Same as in panel b with parameters ($h_\mathrm{ox} = 5$\,nm,$V_{a,1}= 60$\,mV, $V_{a,2}= 100$\,mV, $l_\mathrm{g}=30$\,nm) and the probability density in the inset. (d) Same as in panel b and c with parameters ($h_\mathrm{ox} = 5$\,nm, $V_{a,1}= 65$\,mV, $V_{a,2}= 100$\,mV, $l_\mathrm{g}=60$\,nm) and the probability density in the inset. In addition to the first excited orbital energy (blue dots), the second excited orbital energy is plotted (bright blue dots). (e)-(f) The numerically calculate velocity for each ensemble for the ground (black) and first excited state (orange) along the \bus{} direction is plotted corresponding to the parameters from panel a and d, respectively.
}
\label{Fig2}
\end{figure}

We analyse now the impact of charged defects at the Si/SiO$_2$ interface on the \bus{} in CB mode. We numerically calculate the potential within the center of the 10\,nm thick QW plane applying the model shown in Fig. \ref{Fig1}b. The orbital energy of the first excited state of this QD is obtained using a Poisson-Schr{\"o}dinger solver. Since the \bus{} is periodic in $x$-direction, we focus on a unit cell of the \bus{} consisting of four clavier gates and having a length of $L_x=4 (l_\mathrm{g}+l_\mathrm{ox})$. We calculate the potential in each unit cell 800 times with 800 different ensembles of randomly distributed singly charged defects all having on average constant defect density of $5 \times 10^{10}$\,cm$^{-2}$.  We focus a QW depth of $h_\mathrm{SiGe}=45$\,nm here and will optimize the gate pitch $l_g+l_\mathrm{ox}$ and the screening gate spacing $w_\mathrm{sp}$ according to Fig. \ref{Fig5}. We assume the oxide to be conformally deposited and thus $h_\mathrm{ox}\approx l_\mathrm{ox}$. We start close to our minimal clavier gate pitch of 40\,nm and an oxide thickness $h_\mathrm{ox}$ of 10\,nm, hence $l_\mathrm{g}=30$\,nm. The spacing between the screening gates $w_\mathrm{sp}$ is fixed to 200\,nm and $V_S \approx -100$\,mV to obtain an approximately circular propagating QD. For calculation of the electron ground state and the first excited orbital state, we solve the time-independent Schr{\"o}dinger equation for the 2D x,y-potential in the center of the QW at various positions $x_n$ along the center of the QuBus. We solve numerically  within boundaries of $\pm 0.5 L_x$ around $x_n$, in order to exclude the neighbouring minima of the periodic potential.

Applying sine wave signals with an amplitude of $V_{a,1}=45$\,mV to clavier gates on the lower and  $V_{a,2}=100$\,mV to the upper gate layer, the first orbital splitting energy $E_{orb}$ fluctuates considerably along the \bus{} unit cell and among the defect ensembles (Fig. \ref{Fig2}a). For some defect ensembles, the orbital splitting approaches zero. 
If we double the signal voltage amplitude ($V_{a,1}=90$\,mV and $V_{a,2}=200$\,mV)  applied to the clavier gates (Fig. \ref{Fig2}b), the orbital splitting $E_\mathrm{orb}$ of the propagating QD is enlarged and the variance is reduced. Further investigations of the wavefunction (inset of Fig. \ref{Fig2}b) reveal that the propagating QD breaks into a tunnel-coupled double quantum dot (DQD) at the locations (red lines in insert of Fig. \ref{Fig2}b).  The second QD appears either because the propagating QD splits into a DQD at a large potential ripple or it approaches a disorder-induced deep QD. 
The unintentional formation of a second QDs, which is strongly tunnel-coupled to the propagating QD might lead to orbital excitation and therefore must be avoided.
The low $E_\mathrm{orb}$ values in close proximity belong to the same defect ensemble and appear in our simulation when the randomly distributed defects form a charged cluster. Since we have not taken correlation effects for the distribution of charged defects into account, such clusters are expected to be suppressed in realistic devices due their Coulomb repulsion, but cannot be fully excluded here.

Note a large signal amplitude $V_{a,2}$ might lead to other problems such as sample heating. Hence, just increasing the voltage amplitudes is insufficient and geometrical optimization of the \bus{} is desirable as well. Therefore, we reduce the maximum amplitude back to $V_{a,2}=100$\,mV and $V_{a,1}=60$\,mV similar to (Fig. \ref{Fig2}a) and reduce the SiO$_2$ thickness to $l_\mathrm{ox}=$5\,nm. The resulting variance of the orbital splitting energies is reduced (Fig. \ref{Fig2}c) compared to Fig. \ref{Fig2}a mainly due to an enhanced screening of the defects by the metal gates. The metal gates are also closer to the QW, which increase the confinement of the intentional QD slightly. Still we observe the propagating QD breaking into a DQD (red lines in insert of Fig. \ref{Fig2}c) for some defect ensembles.

Next, we enlarge the width of the finger gate to $l_\mathrm{g}=60$\,nm keeping the oxide thickness at $l_\mathrm{ox}=5$\,nm. Such a \bus{} is not only easier to fabricate, but also exhibits a larger orbital splitting staying above 0.93\,meV for all 800 defect ensembles (Fig. \ref{Fig2}d), since the capacitive coupling of the gates is enhanced. The difference of the orbital energies of the first and second exited orbital reveals only a small ellipticity of the propagating QD. Most noteably, the electron wavefunction (insert of Fig. \ref{Fig2}d) shows no trace of breaking into a DQD. Thus, the QD can propagate sufficiently smoothly across the disordered potential (cf.~the results on disorder autocorrelation function in Sec.~\ref{sec:orbital_nonadiabaticity}).
The increased gate pitch also enhances anharmonicity of the propagating potential as expected from Fig. \ref{Fig5}a. It results in  wobbling and breathing of the QD visible by the deterministic increasing and lowering of the orbital splitting with a wavelength given by the gate pitch, the magnitude of which however does not exceed the effect of $E_\mathrm{orb}$ fluctuations due to the potential disorder. 

We have also considered the expectation value of the $x$-position of the propagating QD as a function time for the ground and first excited orbital. In order to follow correlations in the variation of the QD location within one defect ensemble, we calculate the corresponding finite velocity of the QD orbitals (Fig. \ref{Fig2}e,f). The velocity corresponding to the parameters used for Fig. \ref{Fig2}a reveal large variations in the positions of the QD (Fig. \ref{Fig2}e). The optimized \bus{} geometry discussed in (Fig. \ref{Fig2}d), reveals smaller variations in velocity (Fig. \ref{Fig2}f). The deterministic variation of the QD velocity following the gate periodicity, is of the same order of magnitude as variations of the QD orbital splitting and velocity variations due to potential disorder. These simulations show that QD propagation is feasible despite realistic charge disorder at the Si/SiO$_2$ interface, but the shuttling velocity $\vd$ will not be \add{strictly} constant. We always have to consider a variance of shuttling velocities when we calculate the coherent spin transport conditions, see Tab. \ref{tab:parameters}. 

\subsection{Conclusion of device design optimization}
\label{sec:e_disorder_conclusion}
 We have arrived at a realistic \bus{} design for the CB mode. The numerical simulations indicate that state-of-art defect densities at the Si/SiO$_2$ are sufficient to realize a QD transfer. Obviously, disorder can be counteracted by increasing the dynamic voltage $V_a$ on the clavier gates, but this approach is limited by Ohmic heating, which is is expected due to dielectric loss in the oxides and Ohmic dissipation if normal-conductive metallic gates are used. The latter will appear not directly at the \bus{} as clavier gates constitutes an open terminal. Decisive is the optimization of the \bus{} geometry: Using thin oxides with low defect density, the charged defects can be screened and the metal coveraged increased. For each $h_\mathrm{SiGe}$, a clavier gate-pitch can be chosen to maximize $E_\mathrm{orb}$. Increasing the gate-pitch further leads on the one hand to breathing of the QD potential and thus to deterministic oscillations of the orbital splitting and to deterministic variations of the transfer velocity $\vd$. On the other hand, it increases the mean orbital splitting due to an enhanced capacitive coupling. The deterministic breathing effect can be balanced with the stochastic $E_\mathrm{orb}$-variations due to disorder. The QD should be approximately circular, which can be achieved by a proper gap between the screening gates $w_\mathrm{sp}$, and the applied voltage $V_S$. If the QD is elliptical in $y$-direction, $E_\mathrm{orb}$ is limited by a weak confinement in this direction. If the QD is elliptical in $x$-direction, an increase of $w_\mathrm{sp}$ can increase the capacitive coupling of the clavier gates and thus deepen QD potential, hence increasing $E_\mathrm{orb}$.

\begin{table}
\caption{Relevant parameters and their ranges of values, which will be used for calculation of spin dephasing during CB mode type shuttling in the following sections.}
\begin{center}
\begin{tabular}{c|c|c|c}  
\label{tab:parameters}
 Parameter & low & usual & high \\
 \hline
 Orbital energy $E_\mathrm{orb}$ [meV] & 1 & & 3\\ 
 QD size $L$ [nm] & 12 & & 20\\
 Transfer velocity $\vd$ [m/s] & 1..2 &\,10..20 & 100..200\\ 
 Homog. B-field $B$ [mT]& 20 & 100 & 1000\\ 
 Valley splitting $E_\mathrm{VS,0}$ [$\mu$eV]& 100 & 200 & 500 \\
 Valley $\delta g_v/g$ Refs. \cite{Kawakami14, Ferdous18}& & <10$^{-3}$ & \\
 Orbital $\delta g_o/g$ Refs. \cite{Liu21,Cai23}& & <10$^{-3}$ & \\
 Static QD dephasing $T_2^*$ [$\mu$s]& 10 & 20 & 50 \\
 \hline
 
\end{tabular}

\end{center}
\end{table}

The detailed simulation of our blueprint \bus{} device results in ranges of orbital energy $E_\mathrm{orb}$, QD size $L$ and transfer velocity $\vd$ given in Tab.~\ref{tab:parameters}.  
Let us briefly discuss the other parameters from this table.   We consider three different magnetic fields (Tab.~\ref{tab:parameters}), the highest of which allows spin read-out by Zeeman-energy dependent tunneling to a reservoir, while the others suggest Pauli-spin blockade schemes for spin to charge conversion. Finally, we use typical variations of effective electron g-factors $\delta g_o$ and $\delta g_v$ for orbital and valley state variations, respectively. The typical $T_2^*$ time of a quasi-static quantum dot, depends on the degree of $^{29}$Si isotopical purification and might be limited by the presence of gradient fields as pointed out in Refs.~\cite{Yoneda18, Borjans18, Hollmann20} (see discussion in the next Section).
These parameters are used to calculate the transfer infidelity in the remainder of our paper.  

\addt{We conclude by pointing out the significance of above-described the modeling of realistic electrostatic disorder in an optimized device geometry, and its influence on the properties of the moving QD, on the calculations in the subsequent Sections.
The result that the moving QD does not break apart into an unintentional DQD motivates the perturbative treatment of orbital and spin excitations due to dot motion through a disordered channel in Sec.~\ref{sec:orbital_nonadiabaticity}B. Autocorrelation function of electrostatic disorder is calculated in Sec.~\ref{sec:autocorrelation} for the optimized structure design (Fig.~\ref{Fig2}d,f), and using the model of disorder discussed above. This function provides the key input into the calculation of the orbital excitation rate in Sec.~\ref{subsec:orb_exc} and the  motion-induced spin relaxation rate in Sec.~\ref{subsec:spin_relax}. Furthermore, the correlation length obtained from that function determines also the  motional narrowing during spin transfer calculated in Sec.~\ref{sec:quasistatic}. The upper bound for the orbital relaxation rate in Sec.~\ref{sec:orbital_nonadiabaticity}D is calculated with respect to the range of $E_\mathrm{orb}$ from Fig. \ref{Fig2}d. All these rates are then used to estimate the coherent transfer error due to motion through electrostatically disordered channel in Sec.~\ref{sec:orbital_nonadiabaticity}E.
Finally, the variations of the transfer velocity $\vd$ (Fig. \ref{Fig2}f) enter the discussion of the optimal mean transfer velocity in  Sec.~\ref{sec:discussion}. } 
%

\section{Coherent electron spin transfer in Conveyor-Belt without non-adiabatic effects}
\label{sec:quasistatic}
In this Section we discuss issues affecting slow operation of the \bus{} in the CB mode, for which the adiabatic character of the charge transfer can be taken for granted - but spin dephasing mechanisms, leading to finite $T_{2}^{*}$ time for a stationary electron in a QD, have to be considered.

 If all the voltages controlling the \bus{} are varied slowly enough,  the qubit transfer is be adiabatic, and the electron should remains in the lowest-energy orbital and valley state while it is being pushed along the channel. 
 We assume that an external magnetic field is establishing a quantization axis for the spin, and the only effect of Overhauser field due to nuclear spins of the host material, spin-orbit interactions,  or magnetic field gradients, is to make the spin splitting of an electron a position-dependent quantity $\hbar\omega(x)$, where $x$ denotes the position of the QD, understood as the position of the minimum of the confining potential along the the propagation direction. 

A local spin splitting, $\hbar\omega(x)$, has a frozen-in random component, e.g.~due to $g$-factor dependence on the QD position. The influence of such static disorder in $\hbar\omega(x)$ can be calibrated away: it amounts to a fixed shift of the phase of the transferred spin superposition. However, the fluctuations $\hbar\delta\omega(x)$ of $\hbar\omega(x)$ which occur on timescales shorter than that of accurate measurement of the transferred spin's phase, but longer than the timescale of a single qubit transfer, amount to dephasing of the qubit. 
This means that the measured coherence of transferred electron corresponds to a value averaged over a quasi-static  distribution of realizations of $\hbar\delta\omega(x)$.
These fluctuations can be caused by slow nuclear dynamics due to dipolar interaction (for contributions from Overhauser fields), or low-frequency $1/f$ charge noise leading to slow changes of electric fields that lead to fluctuations of spin-orbit interactions affecting the $g$-factor of an electron at given $x$. 

If the dot-confined electron is shuttled with velocity $\vd$, the shuttling takes $\TP \! = \! \LQB/\vd$ and the overall spin phase acquired during a single transfer will be $\phi(\TP)=\int_0^\TP\mathrm{d}t \, \delta \omega(\vd dt)$. We assume $\delta \omega(x)$ to be translationally invariant, and that is has a finite correlation length $\lcspin$, such that the phase variance at the end of the shuttling process becomes:
\begin{align}
\delta\phi^2 = \int_{0}^{\tau} \text{d}t\int_{0}^{\tau} &\text{d}t'\langle \delta \omega(0) \delta \omega(v(t-t'))\rangle =  \\&= \delta\omega_0^2 \int_{0}^{\tau} \text{d}t \int_{0}^{\tau}\text{d}t' \exp(-\frac{v\abs{t-t'}}{\lcspin})
\nonumber 
\end{align}
where $\langle \ldots \rangle$ denotes the averaging over realizations of $\delta \omega(x)$, $\delta \omega_0^2$ is the variance in Larmor precession frequency of spni in a stationary QD, and we have assumed an exponential decay of the autocorrelation function of $\delta\omega$. In the expected case of shuttling distance being larger than the correlation length, $\LQBL\gg \lcspin$, we obtain:
\begin{align}
\delta \phi^2\big|_{\LQB \gg \lcspin} &\approx \TP \frac{2 \lcspin}{\vd}\delta{\omega}_0^2 =  2 \left( \frac{\TP}{T_{2}^{*}}\right)^2 \frac{ 2 \lcspin}{\LQB} \,\, \label{eq:T2star}.
\end{align}
In the above we have identified $\sqrt{2}/\delta \omega_0$ with spin dephasing time $T_{2}^*$ observed for stationary spin affected by relevant sources of quasi-static noise in its spin splitting. The phase variation in the moving dot is suppressed in comparison to the case of stationary QD for which $\delta \phi_{v=0}^2 = 2(\tau/T_2^*)^2$, and scales linearly with $\tau$ when $\LQB \! =\! \vd \tau$. This is the well-known effect of motional narrowing of inhomogeneous broadening.
If $\LQBL\gg \lcspin$ and $\vd$ is large enough for $\delta \phi^2 \! \ll \! 1$, the loss of spin coherence introduced in Eq.~(\ref{eq:dC}) reads
\begin{equation}
\delta C \! \approx \! \frac{\delta\phi^2}{2} \! = 2 \frac{\lcspin \LQB}{(\vd T_{2}^{*})^2} \,\, . \label{eq:dCML}
\end{equation}
Making $\vd$ larger suppresses dephasing - but this dependence will cease to hold once $\vd$ becomes too large for the assumption of adiabaticity of electron transfer to hold. 

Let us now use the above formulas to calculate the expected spin dephasing during an adiabatic evolution using parameters from ranges given in Tab.~\ref{tab:parameters}. 
We are focusing on transfer across $\LQB\! =\! 10$ $\mu$m with velocities $\vd \in [1,200]$ m$/$s. 
These lead to  passage times of $\TP \!  \in \! [0.05,10]$\,$\mu$s. 

\begin{figure}[tb]
    \centering
    \includegraphics[width=\columnwidth]{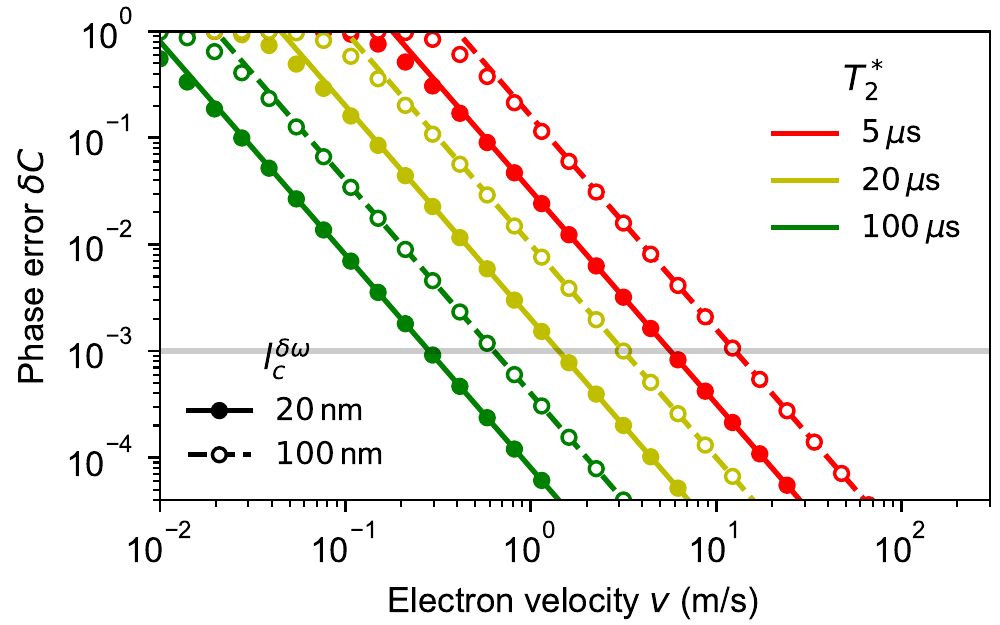}
    \caption{Improvement of spin coherence of an electron transferred adiabatically along a 10 $\mu$m long channel with its velocity. We plot the coherence error due to quasistatic noise as a function of velocity for dephasing times of static spin $T_{2}^{*} \! =\! 5$, $20$, $100$\,$\mu$s, and two values of spatial autocorrelation length of quasi-static disorder in spin splitting: $\lcspin  \! =\! 20$ nm (corresponding to typical QD size) and $\lcspin =  100$\,nm $\approx \lcorb $ (corresponding to typical electrostatic disorder autocorrelation length, see Fig.~\ref{fig:corr_fun}). \add{Lines correspond to Eq.~\eqref{eq:dCML} that holds when $\delta \phi \! \ll \! 1$, while circles correspond to the formula, $\delta C \! = \! 1-e^{-\delta \phi^2/2}$, which holds for Gaussian statistics of random phase fluctuations.}}
    \label{fig:T2*}
\end{figure}

 For Si/SiGe quantum QDs the observed values of $T_{2}^{*}$ are in $\approx \! 20$ $\mu$s range for isotopically purified silicon. In the presence of magnetic field gradients this dephasing is caused by charge noise leading to slow variations in electron's position along the gradient \cite{Yoneda18,Struck20}. For quasistatic charge noise its correlation length should be limited by correlation length of static disorder. We use the numerically calculated static potential disorder from Sec.~\ref{sec:e_disorder} to fit the correlation length (see Fig. \ref{fig:corr_fun} in Sec.~\ref{sec:orbital_nonadiabaticity}A) to be $\lcgg \approx \! 100$\,nm. In the worst-case scenario of magnetic field gradient being constant  and equal to value used in single-dot spin coherence experiments \cite{Yoneda18,Struck20}, we should thus use $T_{2}^{*} \approx 20$ $\mu$s and $\lcspin\! \approx \! 100$\,nm. As shown in Fig.~\ref{fig:T2*} this leads to $\delta C < 10^{-3}$ for $\vd \! \gtrsim \! 5$\,m/s.
With a more realistic assumption that the gradient is sizable only near the ends of the channel (close to the registers of stationary qubits that need to be manipulated), taking only 1\,$\mu$m as the length of the region in which charge noise and gradient dominate $T_{2}^{*}$, we get contributions to phase error that are 10 times smaller, and lead to tolerable phase error in the whole range of velocities that we consider.

 With Si containing 60\,ppm of spinful $^{29}$Si isotope, the $T_{2}^*$ resulting from interaction with very slowly evolving Overhauser field of the nuclei in the quantum well is expected to be  $\approx \! 10 $\,$\mu$s \cite{Assali11,Struck20},  and in fact the nuclear-induced dephasing could be dominated by interaction with a few $^{29}$Si and $^{73}$Ge nuclei in the barrier for which the predicted $T_{2}^{*}$ could be as low as $\sim \! 1$\,$\mu$s, if the wavefunction overlaps with a few $^{73}$Ge atoms \cite{Struck20}. It is however unclear to what extent these spins are frozen out, i.e.~if the dynamics of Overhauser field generated by them can be considered ergodic on timescales relevant for quantum computation \cite{Struck20,Madzik20}. Note that $T_{2}^{*}\! \approx\! 20$\,$\mu$s was observed in Si/SiGe with about 800\,ppm of $^{29}$Si \cite{Yoneda21}, consistent with non-ergodic nuclear dynamics.
 We are however considering the worst-case scenario of nuclei-induced $T_{2}^{*} \! \approx \! 5$ $\mu$s, while making a natural assumption  of lack of spatial correlations between polarization of nuclei, leading to correlation length given by typical size of the QD, $\lcspin\! \approx \! 20 \, \mathrm{nm}.$ This gives us the phase errors $\delta C \! \leq\! 10^{-3}$ for $v\! \geq\! 1 \, \mathrm{m/s}.$
 
 Finally, in the absence of a gradient, and with even more strongly isotopically purified samples (or with nuclear dynamics too slow to be relevant on timescale on which we want to operate our quantum registers), we are left with a mechanism in which charge noise leads to fluctuations of electron $g$-factors. We use the parameters for quasi-static g-factor noise from $T_2^*$-measurements carried out in MOS devices \cite{Chan18} without magnetic field gradients. Assuming a Gaussian distribution of quasi-static fluctuations $g$ factors with rms $\sigma_g$, one obtains $T_{2}^* = \sqrt{2}\hbar/\mu_B B \sigma_g$.
The measured values of $T_2^* = 30\,\mathrm{\mu s}$ at applied magnetic field $B = 1.4\, \mathrm{T}$ implies $T_{2}^{*} > 100$\,$\mu$s in at most a few hundred of mT range that we consider here.
The correlation length is again the QD size, $\lcspin  \! \approx \! 20$\,nm, as the effect of electric fields on $g$-factor relies to a large degree on presence of atomic length-scale interface roughness.
The phase errors are consequently three orders of magnitude smaller than the ones given above for the case of Overhauser field noise, see Fig.~\ref{fig:T2*}.

Summarizing, the phase error $\delta C$ due to spatial dependence of the spin splitting of an electron that is transferred adiabatically is smaller than the targeted benchmark of $10^{-3}$ for all of  the above-discussed values of $T_{2}^{*}$ and $l_c$ when $\vd \! > \!20$ m/s, and for most of them it is enough for $\vd$ to be larger than a few m/s, i.e.~all the values of $\vd$ from the range considered in Tab. \ref{tab:parameters} are admissible then. The error is the largest in the case of  $T_{2}^{*} \! =\! 5$\,$\mu$s and  $l_c^{\delta \omega} \approx 20$\,nm, relevant for dephasing due to natural concentration of $^{29}$Si, or due to coupling to $^{73}$Ge nuclei that are dynamic on the timescale of experiment. In Fig.~\ref{fig:T2*}, we also show the case of even larger error for $T_{2}^{*} \! =\! 5$\,$\mu$s and  $l_c^{\delta \omega} \approx 100$\,nm, relevant for dephasing due to charge noise in a  very large gradient of magnetic field (about 4 times larger than the ones used in \cite{Struck20,Yoneda18}). These cases can be avoided by isotopic purification and proper design of the gradient magnetic fields.

\section{Spin dephasing due to nonadiabaticity of electron dynamics at larger velocities}  \label{sec:nonadiab_dephasing}
We now move to the regime of larger $\vd$, in which non-adiabatic character of the charge transfer has to be seriously considered. We will discuss how the presence of orbital/valley level dependent spin precession together with not-fully adiabatic character of charge transfer, open up a highly relevant channel of spin dephasing. 

The adiabaticity of the \bus{} operation in CB mode, i.e.~lack of excitation of the electron out of its instantaneous lowest energy state $\ket{g_{o,v}}$, is not necessary for the charge transfer to occur: the electron will traverse the channel unless the excitations cause the electron to become ``lost'' by being excited from the pre-existing/moving QDs into the continuum of high-energy states, or by being trapped in an unintentional deep potential well. Device optimization presented in Section \ref{sec:design_CB} makes the probability of such events low. However,  even without them, the non-adiabaticity leads to randomness in state trajectory. 
    Transitions between ground and excited {\it orbital} states of the moving QD happen due to electrostatic disorder (Fig.~\ref{fig:levels}) that acts as a time-dependent perturbation of the confinement potential in the frame of reference moving with the electron-confining QD. 
Furthermore, when the wavefunction of the electron confined in the moving QD overlaps with defects in the interface such as atomistic steps or interface gradients, orbit-valley coupling is activated \cite{Culcer10,Zwanenburg13,Gamble13,Boross16}, and its time-dependence excites the electron into the higher-energy valley state.

These processes lead to finite probability of excitation into a higher energy orbital/valley state. After such an excitation, relaxation back to the ground state will occur on timescale of $\tau_r$. When this $\tau_r$ is smaller than the total shuttling time $\TP$, or when the excitation occurs at a random place along the channel, the electron will spend a randomly distributed time in an excited state. Spin precession frequency $\omega$ is  expected to depend on the orbital/valley state. Random time spent in a state characterized by  random $\omega$ results in random contribution to spin phase characterized by rms $\delta \phi$. For a probability of excitation given by $p_e$, the phase error is then given by Eq.~(\ref{eq:dC}). A more detailed formal derivation of that formula is given in Appendix \ref{app:formal_dephasing}.

Let us discuss the physical mechanisms leading to state-dependent precession rates.
The first mechanism is the state-dependence of the effective $g$-factor of the electron: while $g\! \approx \! 2$ in silicon, its value is exhibiting relative variation of $\delta g_o/g \! \lesssim \! 10^{-3}$ when we compare electrons in states localized in two distinct QDs, 
or ground and excited states of a single QD, 
and similarly $\delta g_v/g \! \lesssim \! 10^{-3}$ in each of the two valley states \cite{Kawakami14, Ferdous18}. This g-factor variation is caused by finite spin-orbit coupling and electrostatic/interface disorder \cite{Ruskov18,Tanttu19}. If the electron spends the time $\tau_e$ in an excited state, the additional spin phase acquired by it compared to the case of  adiabatic evolution is
\begin{equation}
\delta\phi_g \approx 2\frac{\delta g}{g} \frac{\mu_B B \tau_e}{\hbar} \approx 350 \frac{\delta g}{g}  B[\mathrm{T}] \tau_e[\mathrm{ns}] \,\, .\label{eq:phi}  
\end{equation}
For the typical value of $\delta g/g \approx 10^{-3}$, this yields $\delta\phi_g \! \approx \! 0.35 B[\mathrm{T}] \tau_e[\mathrm{ns}]$. \add{Note that $\tau_e$ is given by the relaxation time $\tau_r$ if $\tau_r \! \ll \! \TP$, and it is $\tau_e \! \lesssim \! \TP $ when relaxation is too slow, and an electron excited at some position along the channel will stay in the higher-energy state for the rest of the shuttling time.
}

The second mechanism is the state-dependence of spin-orbit coupling for the two different valley-states. An analogous mechanism for orbital excitation is neglected here, because orbital excitations relax quickly via phonon emission, $\tau_e\! =\! \tau_r \! < \! 1$\,ns for the $E_\mathbf{orb}$ range found in Fig. \ref{Fig2}d, as will be discussed Sec.~\ref{sec:orbital_nonadiabaticity}.
After a Galilean boost of the spin-orbit interaction Hamiltonian into the frame co-moving with the QD, we obtain
\begin{equation}
    H_\mathrm{SO,i_v}=\alpha_{+,i_v}m^*v_{[1\overline{1}0]}\sigma_{[110]}+\alpha_{-,i_v}m^*v_{[110]}\sigma_{[1\overline{1}0]},
\end{equation}
where $i_v\in\{v_+,v_-\}$ labels the lowest-energy valley states, $\alpha_{\pm,i_v}=(\alpha_{i_v}\pm\beta_{i_v})$ 
is the valley dependent spin-orbit strength due to Rashba and (effective) Dresselhaus 2D effects in the Si/SiGe system, and $v_{[1\overline{1}0]}$ and $v_{[110]}$ are the instantaneous QD velocities along along the $[1\overline{1}0]$ and $[110]$ crystal axes.
The first order correction to the valley-dependent spin precession frequency is then
\begin{equation}
    \delta \omega_{\mathrm{SO},i_v}=\frac{m^*}{\hbar}\left(\alpha_{+,i_v}v_{[1\overline{1}0]}\sin(\phi_\mathrm{B})+\alpha_{-,i_v}v_{[110]}\cos(\phi_\mathrm{B})\right),
\end{equation}
with $\phi_\mathrm{B}$ being the angle between the ${[1\overline{1}0]}$ crystal axis and the external magnetic field $\boldsymbol{B}$. A suitable choice of electron velocity $\boldsymbol{v}$ and magnetic field direction parameterized by $\phi_\mathrm{B}$ may  reduce the impact of this effect. In particular, choosing $\boldsymbol{v}\parallel \boldsymbol{B}$ (i.e. $\phi_\mathrm{B}=0$)  being oriented along either $[1\overline{1}0]$ or $[110]$ eliminates the above first order contribution. {However, to realize a two dimensional spin qubit architecture, binding the shuttling direction to the external magnetic field orientation is undesirable.}

Assuming that the spin-orbit coefficients are simply opposite in sign in different valleys \cite{Nestoklon07}, and using an estimate from SiMOS device $|\alpha| \! \approx \! 50$ nm/ns \cite{Tanttu19}, we find 
\begin{equation}
    \delta\phi_\mathrm{SO} \approx  1.6 \cdot 10^{-4} \vd  \left[\frac{\mathrm{nm}}{\mathrm{ns}}\right] \tau_e[\mathrm{ns}] \,\, . \label{eq:dphiso50}
\end{equation}
Here, $|\alpha|$ has been introduced as the general strength of spin-orbit coupling. As the spin-orbit interaction is expected to be weaker in Si/SiGe nanostructure \cite{Hollmann20}, the above gives an upper bound on the spin-orbit induced phase rotation in the considered \bus{}.  The second-order contributions to precession frequency due to transverse fields will give corrections of the order of $|\alpha|m^{*}v/ E_Z \! \ll \! 1$ to this formula. 

We have singled out the secular spin-valley interaction term (of type $\sigma_z\tau_z$), since both spin and valley precession frequencies -- as well as their difference -- are large compared to the strength of the interaction ($E_\mathrm{Z}/h>0.1\,\mathrm{GHz}$ and $E_\mathrm{VS}\gg E_\mathrm{Z}$, while $|\alpha| m^*v/h\leq 1\,\mathrm{MHz}$), justifying a secular approximation. An estimate for the error due to the transverse components is the misalignment of the spin quantization axes in the two valley subspaces. Again assuming spin-orbit coefficients of opposite signs, this amounts to $(|\alpha| m^* v/E_\mathrm{Z})^2 <10^{-4}$ for magnetic fields larger than $5\,\mathrm{mT}$ and shuttling speeds ($v<100\,\mathrm{m/s}$).
There is the  remaining effect that spatial variations in spin-orbit strengths are rendered dynamical due to the motion of the QD. \add{As shown in  \cite{Huang13}, spin relaxation time due to such effectively dynamical transverse spin-orbit fields is $T_1 \!>\! 1$ ms for $\vd \! \leq \! 100$ m/s.  This corresponds to bit flip error $<10^{-4}$ for a qubit moved by $10$ $\mu$m (see alse section VIC). }

Before moving to calculations of $p_e$ and $\tau_r$ in the subsequent sections, let us check  in which parameter ranges we expect the phase error \addt{$\delta C$} to be below our targeted threshold of $10^{-3}$. Since $\delta C \! \leq \! p_e$ according to Eq.~(\ref{eq:dC}),  when $ p_e \! <\! 10^{-3}$, the phase error is guaranteed to be below the targeted threshold no matter what $\tau_e$, and thus $\delta \phi$, is. Only for $p_e \! > \! 10^{-3}$  we have to rely on $\delta \phi \! \ll \! 1$ to find ourselves in the regime in which $\delta C \! \approx \! p_e \delta\phi^2 /2 \! < \! 10^{-3}.$ 
The spin-orbit phase contributions from Eq.~(\ref{eq:dphiso50}) are $\delta \phi_{\mathrm{SO}} < 1$ when $\vd [\mathrm{m/s}] \tau_e[\mathrm{ns}] \! < \! 10^4$, so even for $\vd \! \approx \! 100$\,m/s, relaxation times below 100 ns make $\delta C \! \ll\! p_e$. On the other hand, having no relaxation during shuttling time $\tau$ makes $\delta C \! \approx \! p_e$, as in that case $\tau_e\! \approx \! \tau $ and $\delta \phi_{\mathrm{SO}} \! \approx \! 1.6$.
For phase variations due to randomness in $g$-factors, from Eq.~\eqref{eq:phi}with $\delta g/g\! =\! 10^{-3}$, we have $\delta \phi_g \! < \! 1$, if $\tau_e \! < \! 140 \, $\,ns at $B\! = \! 20$\,mT ($\tau_e \! < \! 14 \, $\,ns at $B\! = \! 200$\,mT). Note that when $\tau_e\! \approx \tau$, we have $\delta \phi_g \! \approx \! 3.5\cdot(10^1-10^3)B[T]$ for $\vd\in[1,100]$ m/s range. In this case, $\delta \phi_g \! < \! 1$ only for lowest magnetic fields of $\approx 20 $\,mT compatible with ESR control of spin qubits, and for the highest considered $\vd \! \approx \! 100$\,m/s (corresponding to $\tau\! \approx \! 100$ ns). 
Note that in this case the electron will finally have to undergo relaxation to its ground valley state at its final destination, and the stochastic nature of this relaxation will lead to additional dephasing.

Clearly, working at lowest $B$ fields compatible with qubit control is beneficial if we cannot suppress orbital/valley excitations so that $p_e$ is below the error threshold: having $B \! \approx \! 20$ mT strengthens the phase error suppression when $\tau_r \! \ll\! 100$ ns, and it is necessary for any such suppression when $\tau_r \! > \! \tau \approx 100$ ns at the highest considered $\vd$.

\section{Coherent electron transfer in Conveyor-Belt in presence of orbital non-adiabaticity and motion-induced spin relaxation}  \label{sec:orbital_nonadiabaticity}
\addt{
Movement of the quantum dot turns the spatial electrostatic disorder into dynamical charge noise in the frame co-moving with the dot. This noise can then cause transitions between orbital states, and it can also couple to spin degree of freedom by spin-orbit interaction. The latter mechanism was first considered in \cite{Huang13}. Here  we use numerical simulations of electrostatic disorder from Sec.~\ref{sec:design_CB} to  estimate the orbital excitation rate.
Then we calculate the orbital relaxation rate due to phonon emission, and combine both orbital transition rates to estimate the spin dephasing caused by mechanism  discussed in previous Section. Furthermore, using the autocorrelation function of disorder calculated here as an input into the theory from \cite{Huang13}, we prove that spin relaxation along the channel is not an obstacle for reaching the targeted qubit transfer fidelities.}

\subsection{Autocorrelation function of electrostatic disorder} \label{sec:autocorrelation}
We start by determining the correlation length $l_c$, and variance $\delta V^2$, of the potential disorder along the channel for the optimized geometry (Fig. \ref{Fig2}d,f). We calculate the autocorrelation function of the matrix element $M_{\text{gg}}(x_0) = \bra{g_o}  \delta V(x) \ket{g_o}$, which quantifies the effects of the electrostatic disorder on the electron occupying the ground state of the QD localized at $x_0$. We compute this function, defined as 
\begin{align}
\label{eq:kgg}
K_{\text{gg}}(x_0,x_0+x) &= \langle M_{\text{gg}}(x_0) M_{\text{gg}}(x_0+x) \rangle \nonumber\\
&- \langle M_{\text{gg}}(x_0) \rangle \langle M_{\text{gg}}(x_0+x) \rangle
\end{align}
by fixing $x_0$ in a simulated device having length of 10 \bus{} unit cells, and averaging over 80 realizations of disorder.
 To create an effectively 1D potential we additionally integrate over the y axis, i.e. $M_{\text{gg}}(x_0) = \int |\psi_{g}(x'-x_0,y')|^2 V(x',y') \text{d}y'\text{d}x'$. At each position $x_0$ of the QD, and for each disorder realization, we solve the Schr{\"o}dinger equation and fit ground-state wavefunctions (Gausssian shapes with variable width) to the ansatz $\psi_{g}(x',y')=\psi_{g}^{(x)}(x')\cdot\psi_{g}^{(y)}(y')$ in both directions x and y.

\begin{figure}[tb!]
    \centering
    \includegraphics[width = \columnwidth]{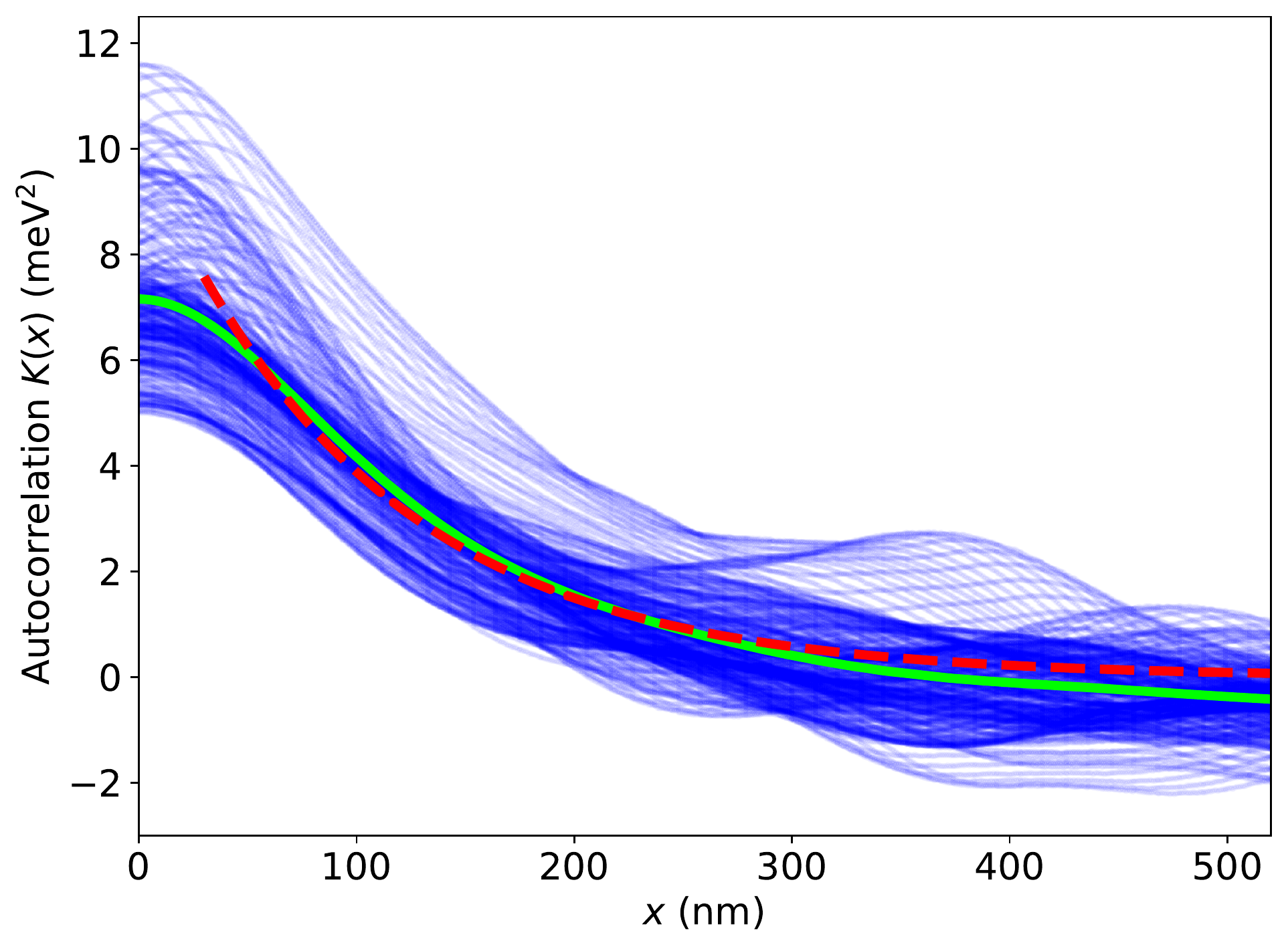}
    \caption{Autocorrelation function $K_{\text{gg}}(x_0,x_0+x)$
    of the matrix element $ \delta M_{\text{gg}}(x_0)=\bra{g_{x_0}} \delta V(x)\ket{g_{x_0}}$, defined in Eq.~(\ref{eq:kgg}), calculated for device geometry and voltages used in Fig.~\ref{Fig2}d,f, and averaged over 80 disorder realizations. Blue lines correspond to 250 distinct values of $x_0$. 
    We plot averaged autocorrelation (solid green line) and an exponential fit (dashed red line) for $x\! > \! \ldot$ of the  $K_{\text{gg}}(\ldot) e^{-\abs{x-\ldot}/\lcgg}$ form. The fitted values $\sqrt{K_{\text{gg}}(0)}\! =\! 2.8$\,meV and $\lcgg\! =\! 90$\,nm indicate typical amplitude and correlation length of effective 1D disorder $\delta V(x)$. The long correlation length $l_c^{\delta V}\gg \ldot$ is an indicator of stability of the moving QD against potential disorder.}
    \label{fig:corr_fun}
\end{figure}

We perform this calculation for 250 values of $x_0$, and as shown in Fig.~\ref{fig:corr_fun}, $K_{\text{gg}}(x,x_0)$ is only slightly dependent on the choice of $x_0$, which confirms its approximate stationarity, i.e.~$K_{\text{gg}}(x_0,x_0+x) \approx K_{\text{gg}}(x)$, \add{and justifies further averaging of the results over $x_0$ (solid green line).} 
For small $x$, the correlation function flattens due to finite size of electron wavefunction $\psi_g(x)$, which effectively filters out the disorder fluctuations on lengthscales below the QD size, $\ldot$.
An exponential fit of $x\! > \! \ldot$ of the $K_{\text{gg}}(\ldot) e^{-|x-\ldot|/\lcgg}$ form, gives an estimate of the correlation length $\lcgg \! =\!  90$\,nm  while the standard deviation of the disorder-induced energy shift of the QD ground state is $\sqrt{\delta V^2} \approx \sqrt{K_{\text{gg}}(0)} \! = \! 2.8$\,meV. 
Note that $\lcgg \! > \! \ldot$ confirms that the disorder in the optimized design does not vary fast enough to break the QD apart, in agreement with other calculations from Sec.~\ref{sec:design_CB}.
The good fit at $x \! > \! \ldot$ means that the correlation length of the matrix element $M_{\text{gg}}$ is a good indicator of the correlation length of the 1D potential, i.e.~we can approximate the autocorrelation of $\delta V(x)$ itself, $K_{\delta V}(x)$, by an exponential with correlation length $l_c^{\delta V}  \! \approx \! l_c^{gg}\gg \ldot$. 

\subsection{Orbital excitations due to electrostatic disorder}
\label{subsec:orb_exc}   
\addt{Motion of the QD converts the electrostatic  disorder into a time-dependent electric field in the frame of the QD.} Due to $1/f$ character of charge noise, electrostatic disorder can be treated as static during a single realization of electron transfer, but it varies between consecutive realizations of the shuttling protocol, \addt{and thus in order to calculate the probability of orbital excitation of the electron, one should average over realizations of electrostatic disorder, i.e.~average over realizations of electric noise experienced by the electron confined in the moving QD. This electric noise will}
cause transitions from ground to excited orbital states, if it has appreciable spectral power at the frequency close the $E_{\text{orb}}/\hbar$.  Such transitions are not captured by solving the time-independent Poisson-Schr{\"o}dinger equation in Sec. \ref{sec:design_CB}. There, only the $E_{\text{orb}}$ and velocity variations of the moving QD have been calculated (Fig. \ref{Fig2}d,f).

In a  single realisation, the electron,treated here as a two-level system, with Hilbert space spanned by ground and excited orbital states, effectively feels a transverse time-dependent field, which in the frame of reference of the moving QD is defined as:
\begin{equation}
\label{eq:mat_elem}
    \delta M_{\text{eg}}(x_0) = \bra{e_o} \delta V(x+x_0) \ket{g_o},
\end{equation}
where $x_0 = vt$, $\delta V(x)$ is a random contribution to the effective 1D potential and $\ket{g_o}, \ket{e_o}$ are ground and first excited orbital state of the moving harmonic potential, respectively. We calculate the excitation probability in the first order of perturbation theory by averaging over the realizations of quasistatic noise in $\delta V(x)$, which translates into averaging over realizations of $\delta M_{\text{eg}}(x_0)$. We follow Refs. \cite{Schoelkopf03,Clerk10} and relate the excitation rate with the spectral density of the dynamical noise at the frequency corresponding to the gap, i.e. 
\begin{equation}
\label{eq:gamo_meg}
    \Gamma_{+,o} = \frac{1}{\hbar^2} \int \text{d}(\Delta t) \langle \delta M_{\text{eg}}(v \Delta t)\delta M_{\text{eg}}(0) \rangle e^{-iE_\text{orb} \Delta t/\hbar},
\end{equation}
where $\langle \ldots \rangle$ denotes averaging over realizations of electrostatic disorder $\delta V(x)$. Using the autocorrelation of $\delta V(x)$ of the form $K_{\delta V} = \delta V^2 e^{-|x|/\lcorb}$, one obtains (see Appendix.~\ref{app:excitation} for details) a result that holds in the regime of $v\! \ll \! E_{\mathrm{orb}} \lcorb /\hbar \! \sim \! 10^5$ m/s:
\begin{equation}
\label{eq:gam_po}
    \Gamma_{+,o} = \frac{\delta V^2 }{\hbar^2 v} \frac{\ldot^2}{\lcorb}\exp(-\frac{1}{2}\Big[\frac{E_\text{orb}\ldot}{\hbar \vd}\Big]^2).
\end{equation}
We can see that for $v/\ldot \ll E_\text{orb}/\hbar$ ($v \! \ll \! 10^4$ m/s) the rate is suppressed by a Gaussian factor. 

\subsection{Spin relaxation}
\label{subsec:spin_relax}
\addt{An analogous calculation can be (and in fact was, \cite{Huang13}) performed for transitions between two spin states of the electron due to spin-orbit coupling being time-dependent in the frame co-moving with the QD across an electrostatic disorder, with the latter modulating the local value of spin-orbit coupling term.}
\addt{ 
As calculated in \cite{Huang13}, using the model of the electrostatic disorder with exponentially decaying correlations that we have employed above, the spin-relaxation rate is given by:
\begin{equation}
    \Gamma_{-,\text{spin}} = \left(\frac{2\hbar \delta V}{l_c^{\delta V} E_\text{orb}^2}\right)^2 \left(\frac{v \, \omega^2 l_c^{\delta V}}{v^2 + (\omega l_c^{\delta V})^2}\right) \alpha^2,
\end{equation}
where $\alpha $ is the spin-orbit coupling 
and $\hbar \omega$ is the Zeeman splitting. Using the above, the estimated probability of spin flip after the transfer follows from $\delta p_\uparrow = \Gamma_{-,\text{spin}} L_s/v$, and it is largest in the limit of small velocities, $v \ll \omega l_c^{\delta V}/\hbar$. We obtain then
\begin{equation}
\label{eq:spin_relax}
    \delta p_\uparrow \leqslant 4 \hbar^2 \frac{\delta V^2 L_s }{(l_c^{\delta V})^3 E_\text{orb}^4} \alpha^2 \approx \frac{4 \times 10^{-4}}{(E_\text{orb}[\text{meV}])^4} \,\, ,
\end{equation}
for parameters obtained from the numerical simulations of electrostatic disorder, i.e. for $\delta V = 3\,\mathrm{meV}$ and $l_c^{\delta V} = 100$ nm, and using $\alpha \! = \! 50$ m$/$s. Since the latter is almost certainly a generous overestimate for Si/SiGe, as discussed previously in Sec.~(\ref{sec:nonadiab_dephasing}), and
we are targeting the orbital excitation gaps $E_{\mathrm{orb}} \! \geq \! 1$ meV, this  result shows that effects of spin relaxation due to electrostatic disorder and spin-orbit coupling in CB are not endangering the goal of keeping the error rate below $10^{-3}$ threshold. In Fig.~1 we have shown the result from this Equation for $\alpha \! =\! 5$ m$/$s taken from \cite{Huang13} and $E_{\mathrm{orb}}\! =\! 1$ meV as a dashed line.
}

\subsection{Orbital relaxation due to electron-phonon coupling}
\label{sec:orbital_relax}
As the relevant relaxation mechanism, we only consider phonon mediated relaxation. Other relaxation mechanisms may dominate at low orbital energies in stationary QDs, located close to charge fluctuators such as electron reservoirs \cite{Hollmann20}, but we neglect these effects when considering a shuttled QD far away from such regions, and assume that high frequency noise on the gates may be sufficiently suppressed in the experiment.

\begin{figure}[tb!]
	\includegraphics[scale=0.5]{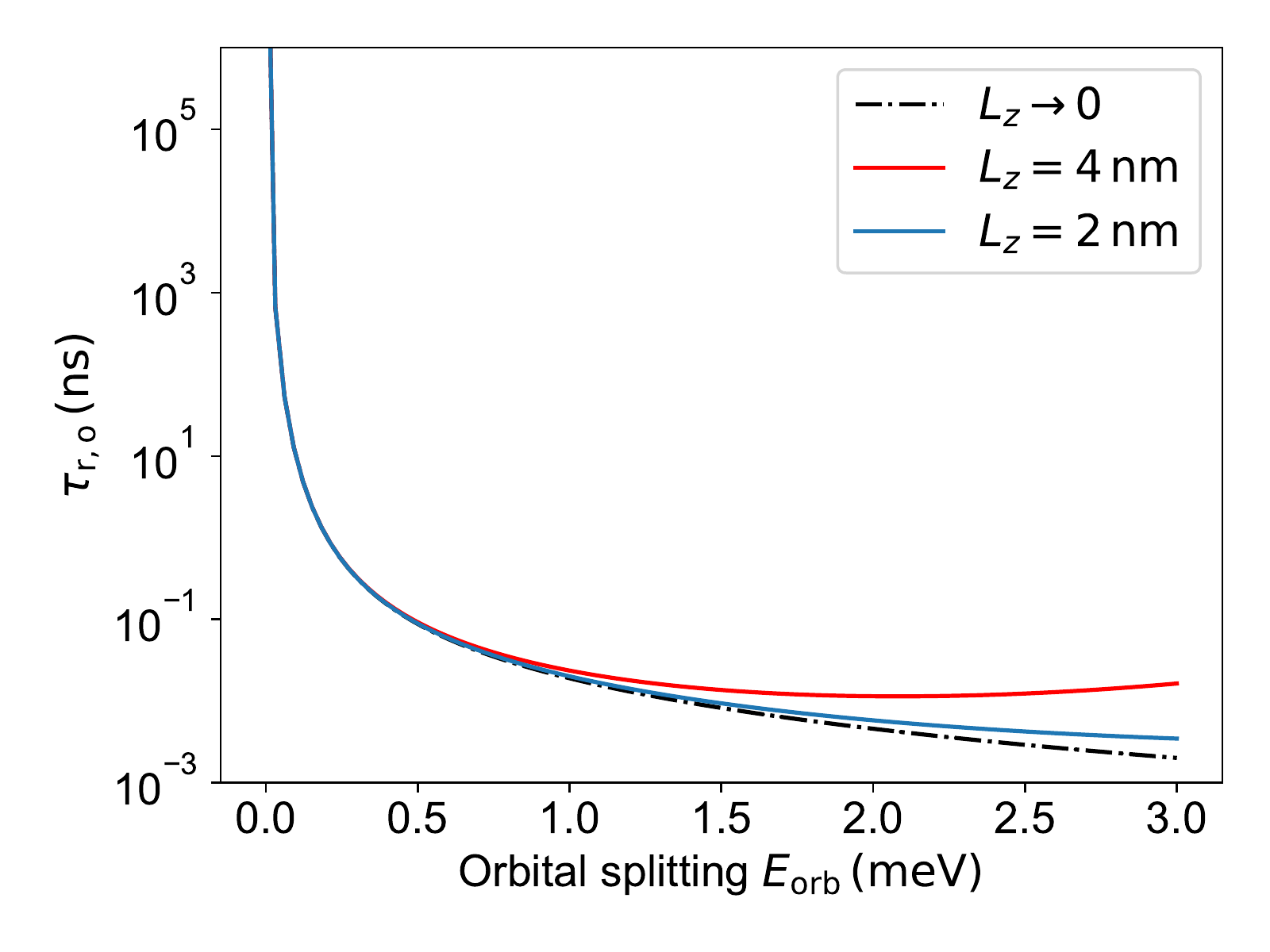}
	\caption{Orbital relaxation time in a circular QD ($L_x=L_y$), with $L_z$ as the remaining parameter. $L_z\rightarrow 0$ illustrates the fictious case of a completely 2D electron (where phonon-bottlenecking along the $z$-direction is absent), while $L_z= 4\,\mathrm{nm}$ is a reasonable upper bound for the localisation of the electron wavefuntion in z-direction.}
	\label{fig:relaxationrates}
\end{figure}

To compute the orbital relaxation due to phonons, we employ Fermi's Golden Rule in the zero-temperature limit \cite{Yu10,Boross16}, as $k_\mathrm{B}T \! \ll\! E_{\mathrm{orb}}$ for  $T \! \approx \! 20\,\mathrm{mK}$, so that temperature enters only as slight modification factor in the relaxation rate. The relevant orbital-phonon coupling will only be due to deformation potential, in contrast to GaAs heterostructures where also piezoelectric coupling is present \cite{Krzywda21,Srinivasa13}. Employing the Herring-Vogt deformation potential Hamiltonian to model this interaction \cite{Yu10}, and taking into account that only the $[001]$-valleys are occupied due to the strain in the QW, we arrive at a relaxation rate of
\begin{equation}
\frac{1}{\tau_{r,\text{o}}} = \frac{E_\mathrm{orb}^3}{8\pi^2\hbar^4 \rho}\left(\frac{\Xi_\mathrm{d}^2I_0+2\Xi_\mathrm{d}\Xi_\mathrm{u}I_2+\Xi_\mathrm{u}^2I_4}{v_l^5}+\frac{\Xi_\mathrm{u}^2J}{v_t^5}\right)
\end{equation}
 written in terms of dilatation and shear deformation potentials $\Xi_d$, $\Xi_u$, transverse and longitudinal speeds of sound $v_t$ and $v_l$, and the mass density of silicon $\rho$, respectively \cite{Boross16}. 
The relevant integrals are given by
\begin{equation}I_n=\int_0^{2\pi}\mathrm{d}\phi_k\int_0^{2\pi}\mathrm{d}\theta_k\sin(\theta_k)\cos^n(\theta_k)|\langle g_o | e^{i\boldsymbol{k}_t\cdot\boldsymbol{\hat{r}}}| e_o  \rangle|^2\label{eq:sphericalIntegral_In}\end{equation}
\begin{equation}J=\int_0^{2\pi}\mathrm{d}\phi_k\int_0^{2\pi}\mathrm{d}\theta_k\sin^3(\theta_k)\cos^2(\theta_k)|\langle g_o | e^{i\boldsymbol{k}_l\cdot\boldsymbol{\hat{r}}}| e_o \rangle|^2\label{eq:sphericalIntegral_J}\end{equation}
with $\boldsymbol{k}$ being the angular wave vector of the phonons matched to the orbital energy splitting $\boldsymbol{k} = [E_{\mathrm{orb}}/(\hbar v_\lambda)]\boldsymbol{e}_{\boldsymbol{k}} \equiv k_0\boldsymbol{e}_{\boldsymbol{k}}$ with $\boldsymbol{e}_{\boldsymbol{k}}$ the unit vector along the direction of $\boldsymbol{k}$.
$| g_o \rangle$ and $| e_o \rangle$ are the ground and first excited orbital state, respectively. Making a harmonic confinement ansatz for the three spatial directions with length scales $L_x$, $L_y$ and $L_z$ allows us to perform the $\phi_k$ integration analytically: 
\begin{align}
\int_0^{2\pi} &\mathrm{d}\phi_k|\langle g_o | e^{i\boldsymbol{k}_\lambda\cdot\boldsymbol{\hat{r}}}| e_o \rangle|^2=2\pi (k_0\ldot)^2\sin^2(\theta_k)\nonumber \\
&\times \exp(-\frac{k_0^2}{2}\big[L_z^2\cos^2(\theta_k)+\ldot^2\sin^2(\theta_k)\big]).
\end{align}
In the above, we assumed that the in-plane confinement is roughly isotropic ($L_x\approx L_y = \ldot$), which is a sensible approximation for the considered QD shapes, cf. Fig.~\ref{Fig5}\add{b} and \ref{Fig2}\add{d}. The $\theta_k$ integration is then carried out numerically and the results are plotted in Fig.~\ref{fig:relaxationrates}. \add{Most importantly, the orbital relaxation time $\tau_{r,o}$ is shorter than 100\,ps for the considered orbital energies from 1 to 3\,meV (Tab. \ref{tab:parameters}). Such an confinement can be achieved for the moving QD across the \bus{} according to Fig. \ref{Fig2}d. Thus, orbital relaxation is sufficiently fast to relax back to the orbital ground-state before a significant difference in spin-phase can be accumulated in the moving QD.} 

A single QD exhibits strong orbital relaxation, attributed to the large orbital splitting  $\sim \! 1 \,\mathrm{meV}$, matching a large phonon density of states at these frequencies, and the strong orbital-phonon coupling, which allows phonons to siphon excess orbital energy on below $100 \,\mathrm{ps}$ timescales. As phonons in Si are relatively slow, and therefore have short wavelengths approaching the characteristic QD dimension at the relevant frequencies, bottlenecking effects where orbital-phonon coupling becomes inefficient due to matching/exceeding of phonon wavelengths to the QD size (suppression of the coupling elements $\langle g_o | e^{i\boldsymbol{k}\cdot\boldsymbol{\hat{r}}}| e_o \rangle$) become relevant. However, as the characteristic QD size scales with $1/\sqrt{E_\mathrm{orb}}$, the relevant parameter is $k_0L_\mathrm{dot}\propto\sqrt{E_\mathrm{orb}}$. This is sufficient to soften the effect of bottlenecking in the relevant energy regime $1-3 \,\mathrm{meV}$. 

While the effect of phonon bottlenecking is of significance at characteristic orbital energy scales, its effects are reduced in two ways. Firstly, the small parameter for bottlenecking for the in-plane components $x$ and $y$ is $k_0L_x=\sqrt{E_\mathrm{orb}/(m^*v_\lambda^2)}$ and only takes on moderate values ($\approx 5$ at $E_\mathrm{orb}=1\,\mathrm{meV}$) for the relevant energies, as discussed above. Secondly, the much stronger confinement along the growth direction ($z$-direction) leads to an additional suppression of these effects ($k_0L_z=\frac{E_\mathrm{orb}}{\hbar v_\lambda}L_z$), which is $\approx 1$ for $L_z\approx 4\,\mathrm{nm}$ and $E_\mathrm{orb}=1\,\mathrm{meV}$.

\subsection{Transfer infidelity due to orbital non-adiabaticity } \label{sec:fidelity_charge}
Let us first estimate the final occupation of the excited orbital state $p_{e,o}$ for the typical case, in which the shuttling time is much longer that the relaxation time $\TP \! \gg\! \tau_r$. We can then estimate $\overline p_{\text{e,o}} \! \approx  \! \Gamma_{+,o} \tau_{\text{r,o}}$ from the steady-state value. 
For typical $\tau_{r,o} \sim 10$\,ps, the final occupation of the excited state is always below $\overline p_{e,o}\leqslant 10^{-3}$ as long as $\Gamma_+ \leq 0.1\,\mathrm{ns}^{-1}$. In our design, the excitation rate due to static disorder computed in Eq.~\eqref{eq:gam_po} gives non-negligible $\Gamma_+ \sim 10^{-4}\mathrm{ns}^{-1}$ only at $\vd \sim 4000$\,m/s ($E_\text{orb} = 1\,\mathrm{meV}$, $\sqrt{\langle \delta V^2 \rangle}\! = \! 3$\,meV and $\lcorb = 100$\,nm), and it is suppressed by a Gaussian factor at lower $\vd$.

Now, let us estimate the spin dephasing caused by repeated processes of orbital excitation followed by relaxation that occur during the shuttling.
According to the model used in Eq.~(\ref{eq:phi}), for time spent in the excited state given by typical orbital relaxation time, $\tau_{e,o} \approx 10 \, \mathrm{ps}$, the variance of random phase acquired after each relaxation event is given by:
\begin{equation}
    \delta \phi_g^2 \approx (\delta \omega \,\tau_{e,o} )^2 \approx 10^{-4} \big(\hbar\delta\omega[\mu \text{eV}]\big)^2,
\end{equation}
where $\delta\omega =\delta g_o \mu_B B/\hbar$ is the difference of Larmor frequencies between ground and excited orbital state. As the relaxation is expected to be orders of magnitude faster than the excitation, we estimate total coherence error as the phase error per relaxation event $\delta \phi_g^2$ /2 times the number of transitions from ground to excited state, which depends on the excitation rate and the shuttling time $N_e \! \approx \! \Gamma_{+,o}\TP$. In this way we estimate total phase error due to temporal occupation of excited orbital state during the shuttling as
\begin{equation}
    \delta C \approx \Gamma_{+,o}\,\TP\,\delta \phi_g^2/2  = \frac{(\hbar\delta \omega[\mu \text{eV}])^2 \Gamma_+[\text{ns}^{-1}]}{2 v[m/s]},
\end{equation}
which can be used to define a tolerable level of excitation rate. Using parameters from Tab.~\ref{tab:parameters}, in the non-optimal (for spin coherence) regime of large magnetic field $B = 1$\,T, g-factor difference $\delta g_o/g = 10^{-3}$ and velocity of $v = 10$\,m/s (transfer time $\TP \sim 1\,\mu \mathrm{s}$), the phase error below the threshold $\delta C \! =\! 10^{-3}$  requires the excitation rate  $\Gamma_{+,o} \! < \! 1 \, [\mathrm{ns}^{-1}]$. This value is orders of magnitude larger than the above-estimated transition rate due to electrostatic disorder.
Even if sources of orbital excitations other than charge disorder simulated in Sec.~\ref{sec:design_CB}, e.g.~charged defects in the channel or threading dislocations, are relevant, the excitation rate associated with them would have to be $\gtrsim  1 \, \mathrm{ns}^{-1}$ for dephasing to become dangerous due to orbtial excitation.
 
We hence come to the important conclusion that spin dephasing due to orbital non-adiabatic effects \addt{(and also qubit state error due to motion-induced spin relaxtion)}, should not pose a limitation for coherent electron transfer in the CB mode. \addt{See Figs.~\ref{fig:intro} and \ref{fig:final_fig} for the comparison to other, more relevant mechanisms considered throughout the paper. }

\section{Coherent electron transfer in Conveyor-Belt in presence of valley degree of freedom}
\label{sec:valley_nonadiabaticity}
After analyzing dephasing due to quasistatic noise in Sec.~\ref{sec:quasistatic} and due to temporal occupation of higher orbital state in Sec.~\ref{sec:orbital_nonadiabaticity}, we finally analyze the phase error resulting from non-adiabatic evolution of the valley degree of freedom.

\subsection{Model of instantaneous valley states}
\begin{figure}[tb]
	\includegraphics[width=\columnwidth]{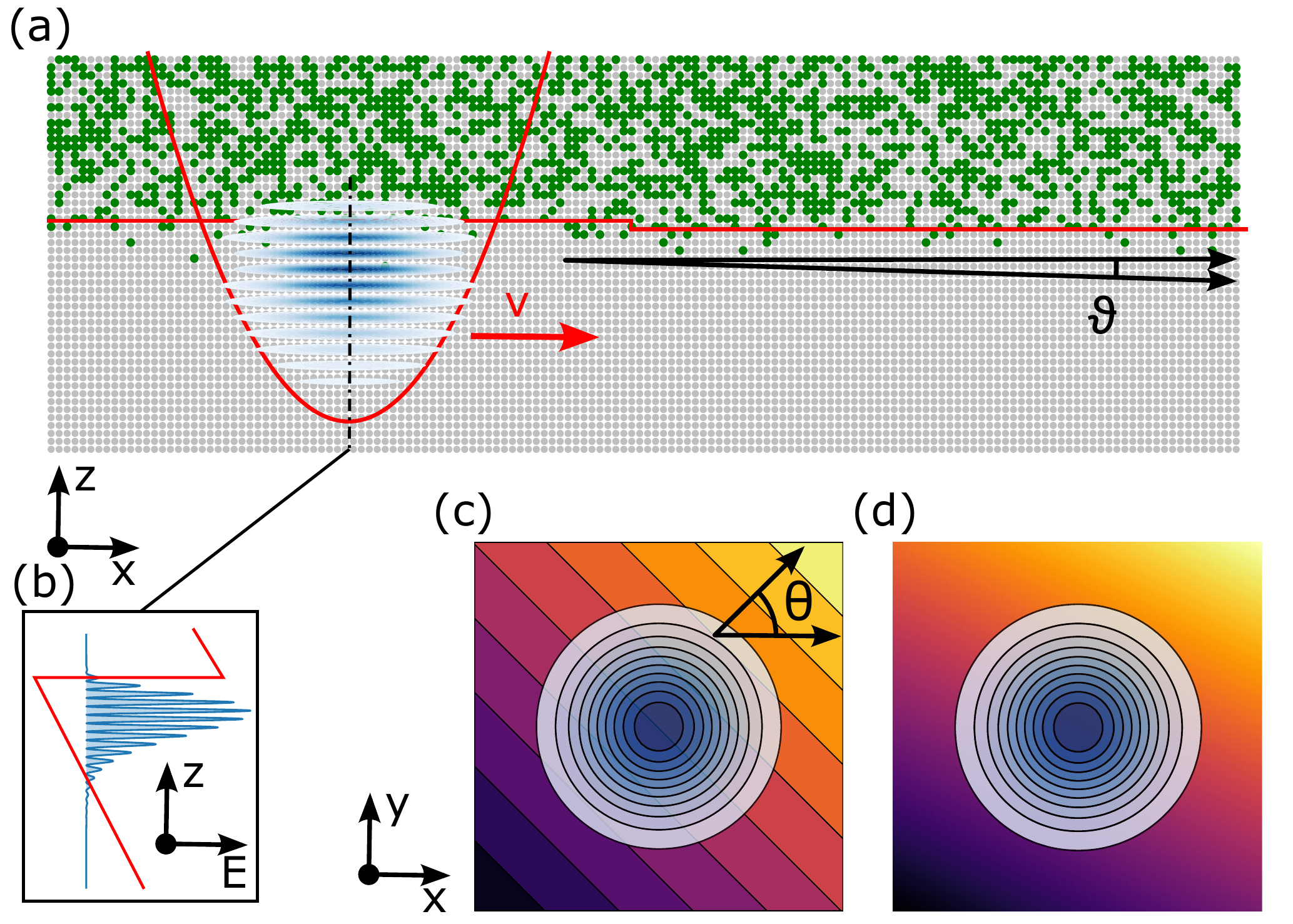}
	\caption{Modelling of the local valley splitting across the \bus{} channel. \textbf{(a)} A 2D cross-section through a Si/SiGe interface region with a gate definied quantum dot containing the electron. A single atomistic step is indicated at the center, with Ge-alloy discorder smearing out the location of the interface. Grey dots represent Si atoms in the strained Si-layer and QW barrier region, respectively. Dark green dots represent Ge atoms. The angle $\vartheta$ indicates the average effective misscut. \textbf{(b)} A 1D linecut along the growth direction which is the model containing all short-range disorder discussed in the main text. The models describing long range disorder in the 2D-plane perpendicular to the growth-direction are sketched in \textbf{(c)} (step model) and \textbf{(d)} (gradient model), with the color coding indicating the continuous and discrete spatial variation in valley phase respectively. The angle $\theta$ in \textbf{(c)} is the angle between step normals and shuttling direction.}

	\label{fig:sketchinterfacedisorder}
\end{figure}

The relevant degree of freedom with low energy splitting in the Si/SiGe quantum wells discussed in this paper are the conduction band minima (valleys) along the growth direction in $k$ space (labelled as $|k_{[001]}\rangle$ and $|k_{[00\overline{1}]}\rangle$), with the in-plane valleys ($|k_{[100]}\rangle$, $|k_{[\overline{1}00]}\rangle$, $|k_{[010]}\rangle$ and $|k_{[0\overline{1}0]}\rangle$) being much higher in energy due to strain \cite{Zwanenburg13}. In gate-defined Si/SiGe QDs, reported values of the splitting between the valleys $|k_{[001]}\rangle$ and $|k_{[00\overline{1}]}\rangle$ vary between $10$ and $\approx 200$\,$\mu$eV \cite{Borselli11,Shi11, Kawakami14,Scarlino17,Zajac15,Mi17,Watson18,Ferdous18,Mi18-2,Borjans18,Hollmann20}. It can exceed 500\,$\mu$eV in MOS structures \cite{Yang13,Petit18,Zhang20,Ciriano-Tejel20}. Crucially, the valley splitting $\vs$ in Si/SiGe is a local property of the heterostructure depending on atomic steps and Ge segregation at the Si/SiGe interface \cite{Dyck17,Wuetz21}. Variations of electric field and crystal compositions that are spatially smooth on the scale of the lattice constant affect the local value of $\vs$, but they do not lead to valley-orbit mixing that would couple the valley degree of freedom with the electron motion. Perturbations on a length scale of the Si lattice spacing e.g.~atomic steps at interface of the quantum well \cite{Voigtlander01}, Ge segregation and SiGe alloy disorder\cite{Dyck17} (see Fig. \ref{fig:sketchinterfacedisorder} (a)), however, severely affect not only $\vs$, but also the composition of valley eigenstates \cite{Friesen07,Friesen10,Culcer10,Hollmann20,Wuetz21,Mcjunkin21}. We parameterize the inhomogeneous valley splitting $\vs$ using an effective valley parameter termed bare valley splitting $\vsz$, which is assumed to be approximately homogeneous on a length scale of the \bus{} device. This $\vsz$ is the expected valley splitting in absence of the models of interface perturbations discussed below (interface step and smooth interface gradient) and therefore the theoretical maximum value of $\vs$ observed in a QD located somewhere along the channel. It includes the influence of alloy disorder, the dependence on the electric field along the QW confinement direction, the Ge content of the barrier, the thickness of the QW layer and the mean length-scale of Ge segregation at the interface. 

While a complete description of atomistic interface disorder would require 3D modelling, we restrict ourselves to a 2D model as will be specified below \cite{Friesen10,Tariq19, Gamble13,Boross16}. 
As the orbital and valley degrees of freedom are only weakly coupled, we may, to a good approximation, find an effective valley Hamiltonian by averaging over the two-dimensional spatial probability density $\rho(x,y)$.
Introducing the valley operators $\hat{\tau}_x=|k_{[001]}\rangle\langle k_{[00\overline{1}]}|+\mathrm{h.c.},$\;$\hat{\tau}_y=-i|k_{[001]}\rangle\langle k_{[00\overline{1}]}|+\mathrm{h.c.},$ we can write the averaged Hamiltonian as 
\begin{align}
\label{eq:Hv_int}
\hat H_{\mathrm{v}}(x_0)=\frac{E_\mathrm{VS,0}}{2} \iint &\rho(x-x_0,y)\left[\cos\left(\varphi_{\mathrm{VS}}(x,y)\right)\hat \tau_x\right.\nonumber\\ &\left.+\sin\left(\varphi_{\mathrm{VS}}(x,y)\right)\hat \tau_y\right] \mathrm{d}x\,
\mathrm{d}y,
\end{align}
where $\rho(x-x_0,y)$ is the electron probability density centered around $(x_0,y_0)$, $y_0$ represents a constant position in the direction perpendicular to shuttling, and $\varphi_{\mathrm{VS}}(x,y)$ is the spatially dependent valley-field. We will consider two models of $\varphi_{\mathrm{VS}}(x,y)$, describing the extreme cases of its gradual and instantaneous change. Both model can be expressed in terms of an effective Hamiltonian,
\begin{equation}
    \hat H_{\mathrm{v}}(x_0) = \frac{E_{\text{VS}}(x_0)}{2} \Big[\cos( \tilde \varphi_{\text{VS}}(x_0)) \hat \tau_x + \sin( \tilde \varphi_{\text{VS}}(x_0)) \hat \tau_y\Big], \label{eq:gradientmodel_Hv}
\end{equation}
written in terms of local valley splitting $E_{\text{VS}}(x_0)$ and the local valley phase $\tilde \varphi_{\text{VS}}(x_0)$. 

The first is a linear gradient model (or smoothly tilted interface from \cite{Friesen07}) depicted in Fig. \ref{fig:sketchinterfacedisorder}d, in which the valley phase is given by $\varphi_{\text{VS}}(x,y) = a_x x + a_y y$, where $a_x$ and $a_y$ denote gradients along and perpendicular to the \bus{} respectively. When substituted to Eq.~\eqref{eq:Hv_int}, the parameters of effective gradient Hamiltonian $\hat H_\mathrm{v,g}(x_0)$ are given by:
\begin{equation}
\label{eq:evs_grad}
    E_{\text{VS}}(x_0) = E_{\text{VS},0} \exp[-\left(\frac{a_x L_x}{2}\right)^2],\,\, \tilde \varphi_\text{VS}(x_0) = a_x x_0,
\end{equation}
where $L_x\approx \ldot$ is the size of the QD in the x-direction. The gradient in y-direction can be incorporated into definition of bare valley splitting $E_{\text{VS},0} \equiv E_{\text{VS},0}'e^{-a_y^2L_y^2/4}$, where $L_y \approx \ldot$ is the size of the QD in the y-direction.

In the second model we consider the  modification of valley field $\varphi_{\text{VS}}(x,y)$ caused by an atomistic step at the interface. We consider regions $A_n$ having piecewise constant  $\varphi_{\mathrm{VS},n}$, see Fig.~\ref{fig:sketchinterfacedisorder}c.  The Hamiltonian reads then:
\begin{equation}
    \hat H_\mathrm{v,s}(x_0) = \frac{E_{\text{VS},0}}{2}\sum_n p_n(x_0)\Big[\cos(\varphi_{\text{VS},n})\hat \tau_x + \sin(\varphi_{\text{VS},n})\hat \tau_y\Big]  \label{eq:stepmodel_Hv}
\end{equation}
with the $p_n(x_0)=\int_{A_n}\mathrm{d}A\,\rho(x-x_0,y)$ being the probability of the electron occupying region $A_n$. As a result, the valley dynamics can be expressed in terms of the effective Hamiltonian \eqref{eq:gradientmodel_Hv}, where the
$E_{\text{VS}}(x_0)$ and $\tilde \varphi_{\text{VS}}(x_0)$ are indirectly defined via the equation:
\begin{equation}
    E_{\text{VS}}(x_0)e^{i\tilde \varphi_{\text{VS}}(x_0)} \equiv \vsz \sum_n p_n(x_0) e^{i\varphi_{\text{VS},n}},
\end{equation}
which mathematically represents the sum of the complex numbers, each corresponding to $A_n$ region with respective modulus $p_n(x_0)$ and argument $\varphi_{\text{VS},n}$. We refer to this model as the step model. When the electron travels over a single atomistic step the phase rotates by $\varphi_1 \equiv \varphi_{\text{VS},n+1}-\varphi_{\text{VS},n} \approx 0.85\pi$, and hence in the limit of separated steps, traveling electron will experience local dips of valley splitting \cite{Friesen07,Friesen10,Tariq19}. If the regions are all separated by parallel steps, the y-confinement direction may be integrated out and the model becomes effectively one dimensional with the missalignment of the steps w.r.t. the shuttling direction $x$ entering as an effective reduction of the shuttling velocity  $\tilde{v}=v/\cos(\theta)$ with $\theta$ being the angle between the shuttling direction and the step normal, see Fig.~\ref{fig:sketchinterfacedisorder}c. 

In order to parameterize the density of atomistic steps, the average tilt of the Si/SiGe interface can serve as a reference parameter \cite{Zandvliet00-2, Swartzentruber93}, originating from the miscut of the silicon wafer on which the heterstructures is grown. A typical miscut angle of $\vartheta < 1^\circ{}$ translates to an average gradient of $a_x \approx 0.85\pi \vartheta/h$ or an average step separation $\overline{d}\approx h/\vartheta$ with single atomic layer height $h = a_\mathrm{Si}/4 \sim 0.136$\,nm. We note that the local gradients and step densities may differ significantly from the global average (e.g. due to step bunching or outliers in alloy disorder profile), and hence the miscut is only taken as an indicator of order of magnitude of these effects. 

\subsection{Excitation in the linear gradient model}
\label{sec:gradient}
To compute effects of non-zero gradient \add{(Fig. \ref{fig:sketchinterfacedisorder}d)} from the model Hamiltonian given in Eq.~\eqref{eq:gradientmodel_Hv}, we move to an adiabatic frame using a time-dependent operator $\hat R(t)$ that diagonalizes the valley Hamiltonian $\hat R(t)\hat H_\mathrm{v,g}(t)\hat R^\dagger(t) = \frac{1}{2}\vs(t)\hat \tau_x$ at every instant of time \cite{Krzywda20}. When substituted into the Schr{\"o}dinger equation, it produces an effective Hamiltonian in the adiabatic basis $\ket{e_v},\ket{g_v}$, which we assume to be the eigenstates of $\hat \tau_x$ Pauli operator. Due to time-dependence of $\hat R(t)$ the total effective Hamiltonian includes also the coupling between instantaneous levels, which together reads:
\begin{equation}
\label{eq:had}
    \hat{\mathcal{H}}_\mathrm{v,g}(t) = \hat R \hat H_\mathrm{v,g}(t) \hat R^\dagger -i\hbar \hat R \dot{\hat{R}}^\dagger =\frac{1}{2} E_{\text{VS}}\, \hat \tau_x + \frac{1}{2} \hbar \dot {\tilde \varphi}_{\text{VS}}\,\hat \tau_z,
\end{equation}
where for each region of constant gradient, $E_{\text{VS}} = E_{\text{VS},0} \exp(-[a_x \ldot]^2/4)$ and $\dot {\tilde \varphi}_{\text{VS}} = a_x v$. As a result, the occupation of the excited valley state is:
\begin{align}
\label{eq:Q_grad}
    &p_{e,v}(t) = |\bra{e_v} e^{-i \hat{\mathcal{H}}_{\mathrm{v,g}} t/\hbar} \ket{g_v}|^2  \nonumber \\&\quad= \frac{(\hbar a_x v)^2}{E_{\text{VS}}^2 + (\hbar a_x v)^2}  \sin^{2}\left(\sqrt{E_{\text{VS}}^2 + \hbar^2 a_x^2 v^2 }\,\frac{t}{2\hbar}\right),
\end{align}
which is the well known result of Rabi oscillations in the rotating frame. In reality, instead of coherent oscillation, one should expect some time-averaging due to inevitable fluctuation of electron velocity (see Fig.~\ref{Fig2}), or valley-orbit coupling that allows for relaxation via phonon emission. Thus, for every region of constant gradient we estimate typical occupation of excited state as:
\begin{equation}
    \overline p_{e,v} \approx \frac{1}{2}\frac{(\hbar a_x v)^2}{E_{\text{VS}}^2+(\hbar a_x v)^2} \approx \frac{1}{2}\frac{(\hbar a_x v)^2}{ E_{\text{VS,0}}^2} \exp(\frac{(a_x \ldot)^2}2),
\end{equation}
where the last approximation is valid if $\hbar a_x v \ll E_{\text{VS}}$, which is fulfilled for typical gradients $a_x$ and parameters from Tab.~\ref{tab:parameters}.

\subsection{Excitation caused by sharp atomistic steps}
\label{sec:single_step}
Now, let us investigate the excitations caused by abrupt changes of $\tilde \varphi_\text{VS}(x_0)$ due to parallel atomistic steps at the interface \add{(Fig. \ref{fig:sketchinterfacedisorder}c)}. 
For concreteness, we assume that the orientation of the steps is perpendicular to the \bus{} direction. The value of local valley splitting reads $E_{\text{VS}}(x_0) = E_{\text{VS},0}\abs{\sum_n p_n(x_0) e^{i n \varphi_1}}$, where $p_n$ is the probability of electron to be found in the $n$th interstep region. 

We start by computing the probability of the valley excitation on a single atomistic step located at $x=0$. Assuming ground state electron density of the form $\rho(x-vt) \propto e^{-(x-vt)^2/\ldot ^2}$, we can write probability of occupying regions on two sides of the step as $  p_{\text{L}(\text{R})}(t) = \frac{1}{2}(1-(+)\erf[vt/L])$, where erf($x$) is the error function, using which we obtain
\begin{equation}
E_{\text{VS}}(vt) = E_{\text{VS},0} \sqrt{\erf^2(vt/L) \sin^2\tfrac{\varphi_1}{2} + \cos^2\tfrac{\varphi_1}{2}}.
\end{equation}
\noindent In order to estimate the excitation probability on an isolated step, we use a Landau-Zener model of non-adiabatic transition $Q = \exp(-\pi \Delta^2/2\hbar \dot \epsilon)$ \cite{Shevchenko10}, in which we identify time-independent valley coupling as $\Delta = E_{\text{VS},0}\cos\tfrac{\varphi_1}{2}$, and linearize time-dependent part $\epsilon(t) \equiv  E_{\text{VS},0} \erf(\tfrac{vt}{L})\sin\tfrac{\varphi_1}{2} \approx \frac{2 vt E_{\text{VS},0}}{L\sqrt{\pi}} \sin\tfrac{\varphi_1}{2}$. The probability of occupying a higher valley state after single step passage can then be written as
\begin{equation}
\label{eq:qstep}
Q_1  = \exp{-\pi^{3/2} \frac{ E_{\text{VS},0}\,\ldot}{4\hbar  v} \frac{\cos^2\tfrac{\varphi_1}{2}}{\sin\tfrac{\varphi_1}{2}}} = 10^{-0.03/\eta}.
\end{equation}
In the last approximate expression, we have substituted the phase shift corresponding to a single step,  $\varphi_1 = 0.85\pi$, and combined the remaining quantities into the shuttling parameter
$\eta = \hbar v/E_{\mathrm{VS},0}\ldot $. For a typical range of parameters $v = 10 \,(50)$\,m/s, $E_{\text{VS},0} = 200\,\mu$eV, $\ldot= 20$\,nm, the value of shuttling parameter $\eta \sim 0.002 (0.01)$ corresponds to excitation probabilities of the order of $Q_1 \leq 10^{-17}\, (10^{-4})$. In Appendix \ref{app:LZ_step} we prove validity of L-Z approximation by a direct comparison against numerical simulation.

\begin{figure}[tb!]
	\includegraphics[width=\columnwidth]{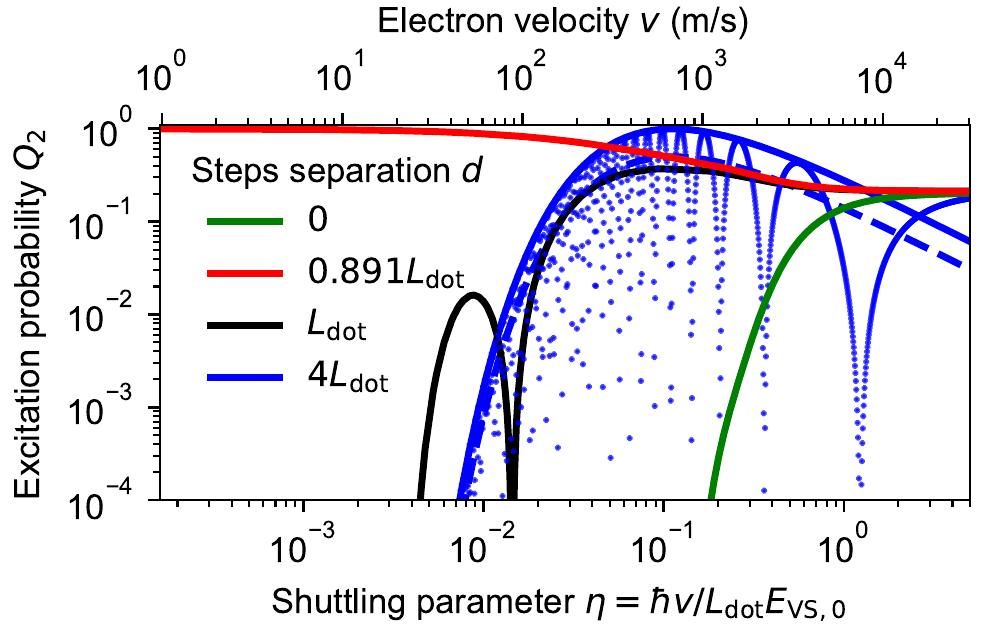}
\caption{Probability of occupying excited valley state, after two step passage $Q_2$ for four different step separations $d$ corresponding to double height step $d=0$ (green), valley collapse $d \approx 0.891\ldot $ (red) and interference $d\gg \ldot $ (blue) as a function of shuttling parameter $\eta$ (lower scale) and electron velocity $\vd$ (upper scale) for fixed $E_{\text{VS},0}=200\mu$\,eV and $\ldot = 20$\,nm ($E_\text{orb} = 1$\,meV). We added the intermediate case of $d=\ldot $ (black) to show that $d\geqslant \ldot $ distances between the steps should be sufficient to avoid the valley collapse. In the interference regime (blue) we compare analytical formulas for constructive interference $Q_2^{\text{max}} = 4Q_1$ (solid line) and the average $\langle Q_2 \rangle = 2Q_1$ (dashed line) against the results of numerical simulation (dots).}
	\label{fig:twostep}
\end{figure}

In the case of two parallel steps, located at $x = \pm d/2$, the $\vs$ can identically vanish if the lateral distance between two steps fulfills $E_{\text{VS}}(d)/E_{\text{VS},0} = \erf(d/2\ldot ) (1-\cos\varphi_1) + \cos\varphi_1=0$, which is numerically solved by $d \approx 0.891\ldot$. Using a numerical simulation of evolution in Fig.~\ref{fig:twostep} we show that the probability of occupying the higher valley state after the two-step passage, $Q_2$, for the step separation $d = 0.891\ldot $ is close to unity even at small $\vd$. This is caused by the above-mentioned a decrease of the energy gap due to the competition between three regions of different valley phase. In contrast, this is no longer true for slightly larger step separation $d = \ldot $ (black line), where the $Q_2$ starts to show signatures of interference (note e.g. the dip around $\eta = 10^{-2}$). 
The presence of interference is even more visible at larger step separation i.e. $d > \ldot $, for which the example result with $d = 4\ldot $ is shown in Fig.~\ref{fig:twostep} (blue line). In this regime, the electron is expected to undergo consecutive transitions through isolated atomistic steps, which coherently mix the two valley states. Thus the whole process resembles Landau-Zener-Stückelberg interferometry \cite{Shevchenko10}. In the case of two distant enough steps, the value of $Q_2 \sim |a_1+a_2|^2$ results from the coherent sum of probability amplitudes, which correspond to excitation on the first $a_1 = \sqrt{Q_1}\sqrt{1-Q_1}$, and second step $a_2 = \sqrt{1-Q_1}\sqrt{Q_1}e^{i\Phi_{12}}$, where phase difference between the two paths $\Phi_{12}$ gives:
\begin{equation}
\label{eq:q2_phi}
    Q_2 = 4(1-Q_1)Q_1\cos^2(\Phi_{12}/2) \approx  4 Q_1\cos^2(\Phi_{12}/2).
\end{equation}
The last approximation works in the typical limit of $Q_1 \ll 1$, see Eq.~\eqref{eq:qstep}. For estimation purposes one can consider the worst case scenario of constructive interference ($\cos^2(\Phi_{12}/2) \to 1$), where $Q_2^{\text{max}} \approx 4 Q_1$. However, the typical occupation of the excited valley state is expected to be closer to an average $\langle Q_2\rangle \approx 2 Q_1$, due to inevitable fluctuations of parameters that randomize the phase $\Phi_{12}$ between the realizations ($\langle\cos^2(\Phi_{12}/2)\rangle \to 1/2$). In Fig.~\ref{fig:twostep}, we show an agreement between analytical predictions (solid and dashed lines, respectively) and numerical results (dots). The interference pattern is highly sensitive to small variations of the electron velocity, and as such prone to averaging in presence of expected velocity fluctuations (see Fig.~\ref{Fig2}).

Finally, let us generalize these results to the case of multiple interface steps. Following \cite{Zandvliet00-2, Swartzentruber93}, the variation of the interstep distance is expected to be proportional to the average distance between them, $\sigma_d\sim \overline d$. Higher density of steps ($\overline d \ll \ldot $) is expected to give more regular alignment, which in presence of Ge interdiffusion at the interface \cite{Dyck17} should resemble the linear gradient model with relatively large $a_x = 0.85\pi/\overline d$ (see the previous Section). In the opposite limit of one or two atomistic steps per QD size ($\overline d \geqslant\ldot$), their random alignment can lead to both unfavourable valley-splitting collapse or favourable bunching of two atomistic steps into one double-layer step. For the two-step case the bunching can be modeled by Eq.~\eqref{eq:qstep} with $\varphi_1 = 1.7\pi$, and as shown using green line in Fig.~\ref{fig:twostep} it allows for adiabatic transfer even at relatively high electron velocities. 

We assume that the distance between all the steps is larger then the QD size, $\overline d>\ldot $, which corresponds to the previously defined interference regime. In analogy to the double step case, the probability of occupying higher valley state after $N$-step passage, where $N = \LQB/\overline d$ can be written as:
\begin{equation}
\label{eq:Q_n}
    p_{e,v} \approx  Q_1 \left|\sum_{n=1}^{N} e^{i\Phi_{n}} \right|^2 \approx \frac{\LQB}{\overline d} Q_1.
\end{equation}
where the approximation is valid for $N Q_1 \leq 1$, and in the presence of parameter fluctuations which result in uncorrelated sum of phases $\sum_n e^{i\Phi_n}= \sqrt{N}$. The obtained result for multistep passage will is tested against numerical solution in Appendix~\ref{app:LZ_step}.

\subsection{Valley relaxation due to electron-phonon coupling}
\label{sec:valley_relax}
The relaxation of the valley state in a gate defined QD may be treated analogously to the spin relaxation in gate confined quantum dot systems \cite{Hollmann20,Huang14,Golovach08}, in that hybridization of valley and orbital leads to the dominant relaxation mechanism \cite{Tahan14,Penthorn20}.
However, compared to the spin-orbit mediated spin relaxation, the valley relaxation rate is higher  due to both the stronger valley-orbit coupling and the lack of Van-Vleck cancellation \cite{VanVleck40,Abrahams57,Khaetskii01,Hanson07}.

As orbit-valley hybridization yields the dominant relaxation channel for valleys, we may obtain the valley relaxation rates directly by substituting a perturbative ansatz for the hybridized ground/excited valley states $| \widetilde{g}_v \rangle$ and $| \widetilde{e}_v \rangle$, replacing $| {g}_o \rangle$ and $| {e}_o \rangle$ in Eq.~(\ref{eq:sphericalIntegral_In}) and (\ref{eq:sphericalIntegral_J}). This results in the coupling matrix element \add{$|\langle \widetilde{g}_v | e^{i\boldsymbol{k}_\lambda\cdot\boldsymbol{\hat{r}}}| \widetilde{e}_v \rangle|^2\approx \left|\frac{\langle {e_o g_v}|H_\mathrm{v}|{g_o e_v}\rangle}{E_\mathrm{orb}}\right|^2|\langle g_o | e^{i\boldsymbol{k}_\lambda\cdot\boldsymbol{\hat{r}}}| e_o \rangle|^2$} ($g_o$ and $e_o$ and $g_v$ and $e_v$ labelling the ground/(first) exited orbital and ground/exited valley states of the effective valley Hamiltonian $H_\mathrm{v}$ from Eq.~\eqref{eq:gradientmodel_Hv}). Neglecting some correction factor at maximum of the order of unity and assuming $E_\mathrm{VS}\ll E_\mathrm{orb}$, we obtain: 
\begin{equation}\frac{1}{\tau_\mathrm{r,v}}=\left|\frac{\langle {e_o g_v}|H_\mathrm{v}|{g_o e_v}\rangle}{E_\mathrm{orb}}\right|^2F^2\left.\frac{1}{\tau_\mathrm{r,o}}\right|_{E_\mathrm{orb}\rightarrow E_\mathrm{VS}},
\end{equation}
with $F=1+\frac{E^2_\mathrm{VS}}{E^2_\mathrm{orb}}+\mathcal{O}\left(\frac{E^4_\mathrm{VS}}{E^4_\mathrm{orb}}\right)\approx 1$, \add{and with $\tau_\mathrm{r,o}$ being the orbital relaxation time calculated in Sec.~\ref{sec:valley_nonadiabaticity}.}
Neglecting correction factors, the coupling matrix element is proportional to $E_\mathrm{VS}$, so that \add{the relaxation rate} scales as $(E_\mathrm{VS}/E_\mathrm{orb})^2$. The orbital relaxation rate evaluated at the valley splitting scales as $\left.1/\tau_\mathrm{r,o}\right|_{E_\mathrm{orb}\rightarrow E_\mathrm{VS}} \propto (E_\mathrm{VS})^5/E_\mathrm{orb}$ when neglecting bottlenecking effects, which is approximately justified at the low valley splitting energies of order $100\,\mathrm{\mu eV}$, for which in-plane bottlenecking will just start to become relevant and out of plane bottlenecking is sufficiently well suppressed ($k_0L_z=\frac{E_\mathrm{VS}}{\hbar v_\lambda}L_z< 1$). With $E_{\mathrm{orb}}$ roughly one order of magnitude higher than the valley splitting, we would expect a $(E_\mathrm{VS}/E_\mathrm{orb})^7\approx 10^{-7}$ ratio of valley to orbital relaxation rates, if bottlenecking of the orbital relaxation rates were negligible.
However, as discussed in Sec. \ref{sec:orbital_relax}, phonon bottlenecking has an significant effect on the orbital relaxation rates, so that the ratio of valley to orbital relaxation rates is of the order $10^{-5}-10^{-3}$ for relevant values. 
We note that the overall valley relaxation rate scales as $E_\mathrm{orb}^{-3}$, so in principle sacrificing QD confinement, while still ensuring QD stability could lead to an order of magnitude increase in valley relaxation rate for realistic QD confinement energies. 

In order to estimate the orbit-valley hybridization we  again employ the step and gradient models.
The gradient model \eqref{eq:gradientmodel_Hv} yields a matrix coupling element of 
\begin{align*}
\langle g_v|H_\mathrm{v,g}|e_v\rangle = \frac{a_x\ldot\, }{4\sqrt{\pi}}E_\mathrm{VS,0}e^{-(a_x\ldot)^2/4}
\end{align*}
while the step model (\ref{eq:stepmodel_Hv}) yields
\begin{align*}\langle g_v|H_\mathrm{v,s}| e_v\rangle &= E_\mathrm{VS,0}\frac{\sin(\tfrac{\varphi_1}{2})e^{-\frac{\Delta x^2}{L^2}}}{\sqrt{1+(p_\mathrm{R}-p_\mathrm{L})^2\tan^2\left(\frac{\varphi_1}{2}\right)}}
\end{align*}
where $\Delta x$ is the displacement from the QD center to the step position, and $p_L$ ($p_R$) are the probabilities of the electron being to the left (right) hand side of the step defined in Eq.~\eqref{eq:stepmodel_Hv}.

\begin{figure}[tbh!]
	\includegraphics[scale=0.5]{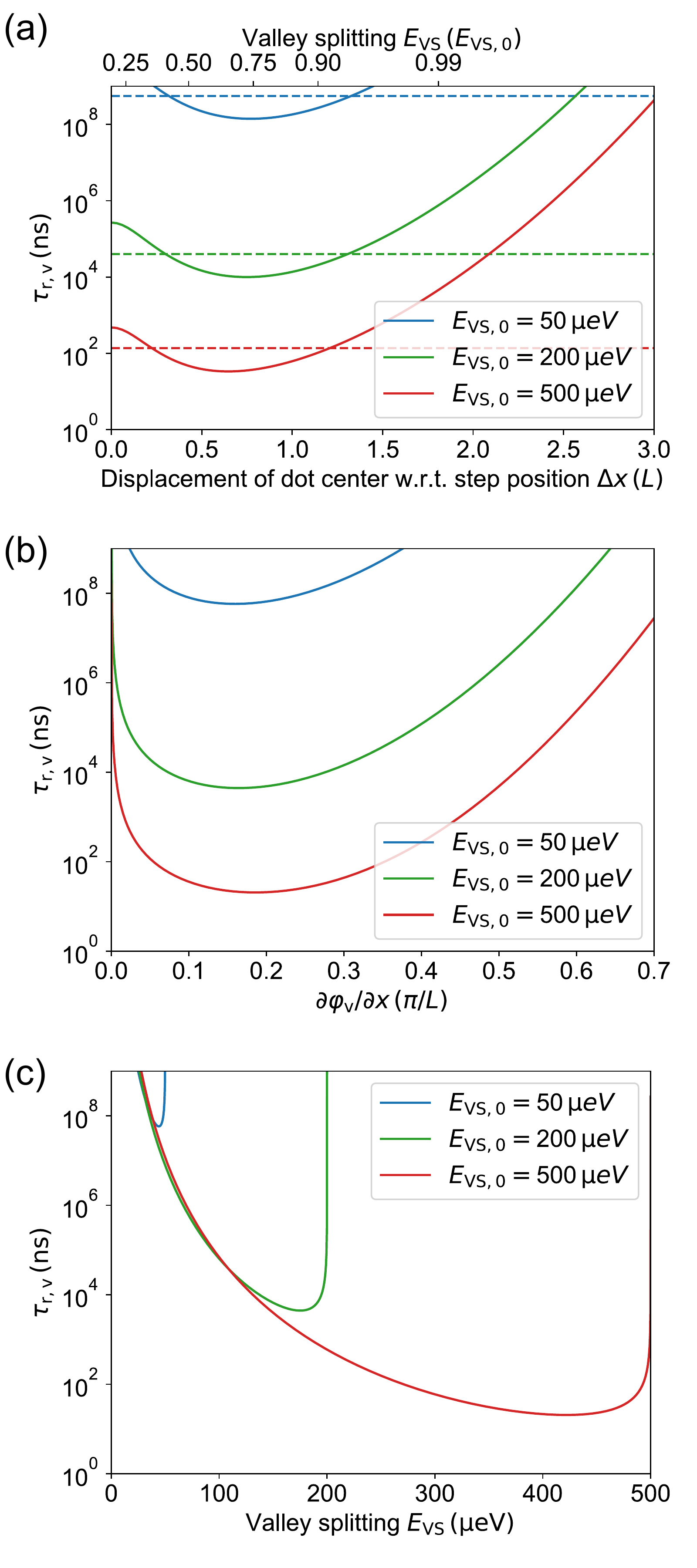}
	\caption{Valley lifetimes for the stepmodel (a) and linear valley phase model (b,c). The dashed lines in (a) are the inverses of the relaxation rate averaged over the displayed region. A QD confinement energy of $E_\mathrm{orb}=1\,\mathrm{meV}$ is assumed for all cases.} 
	\label{fig:valleyT1_step_grad}
\end{figure}

We use these two models to estimate $\tau_{\mathrm{r,v}}$, the results for which are plotted in Fig. \ref{fig:valleyT1_step_grad}. For the step model, see Fig. \ref{fig:valleyT1_step_grad}a, the relaxation rate is the largest when the QD center is between half to one QD length scale ($\ldot/2$ to $\ldot$) away from the position of the step. While valley-orbit coupling is maximized if the QD is centered around the step, the collapse in valley splitting significantly reduces the available phonon modes at this frequency. The optimum for phonon assisted valley relaxation is therefore slightly displaced. By moving further away from the atomistic step, the overlap with the step exponentially decreases for a Gaussian wavepacket, and gives rise to  decrease in valley-orbit coupling, and an effective suppression of this relaxation mechanism.
A similar trade-off is observed in the linear gradient model, see Fig. \ref{fig:valleyT1_step_grad}b: strong interface disorder suppresses the valley splitting, while a lack of interface disorder prevents valley-orbit hybridization, and therefore diminishes the relaxation. The maximum relaxation rate occurs for gradients which change the valley phase by $(0.1-0.2)\pi$ over one QD lengthscale $\ldot$. In terms of the valley splitting, the maximum relaxation rate is achieved for valley splittings close to the bare one ($E_\mathrm{VS,0}$), see Fig. \ref{fig:valleyT1_step_grad}c. 

Overall, the achievable valley relaxation rate is highly dependent on the bare valley splitting $E_\mathrm{VS,0}$.  For our chosen ranges of parameters, averaged valley relaxation times generally lie above $100 \, \mathrm{ns}$, which, as we have discussed in Sec.~\ref{sec:nonadiab_dephasing}, leads to randomization of the phase of the spin qubit after an excitation event if the difference in spin-precession frequencies between the different valley states leads to a significant buildup of phase difference on this timescale. In conclusion, in order to avoid detrimental effects due to the valley excitation, an increase in valley splitting both suppresses non-adiabatic excitation as well as it increases the valley relaxation rate due to phonons.

\subsection{Spin-valley hotspots}
\label{sec:hotspot}
While shuttling the QD across a wide range of interface regions, the Zeeman splitting may temporarily match the valley splitting $E_Z\approx \vs$. In such cases, spin relaxation is dominated by the spin-valley hybridization, referred to as a spin-relaxation hotspot in the literature \cite{Yang13,Zhang20,Hollmann20}. 
As such a relaxation channel leads to both spin-up decay ($\delta p_\uparrow \equiv  p_\uparrow(0)- \Tr{\ketbra{\uparrow}\hat \varrho(\tau)}$) as well as dephasing ($\delta C \sim \delta p_\uparrow$), but we will be simply interested in the overall order of magnitude of the error $\delta p_\uparrow$ in the following. 
Fortunately, this detrimental effect of spin-valley hotspot can be completely
avoided, if the Zeeman splitting is kept below the global minimum or above the global maximum of the local valley splitting. In particular at $B = 20$\,mT, the Zeeman splitting of $E_z \approx 2\,\mu$eV is well below the smallest value of local valley splitting reported in the Si/SiGe quantum well, $\vs \approx 10$\,$\mu$eV \cite{Kawakami14}. Thus, 
the use of small global magnetic fields is not only advantageous for suppressing spin dephasing (Sec. \ref{sec:quasistatic}), but also avoids spin relaxation at the spin-valley hotspot.

\begin{figure}
    \centering
    \includegraphics[width=0.9\columnwidth]{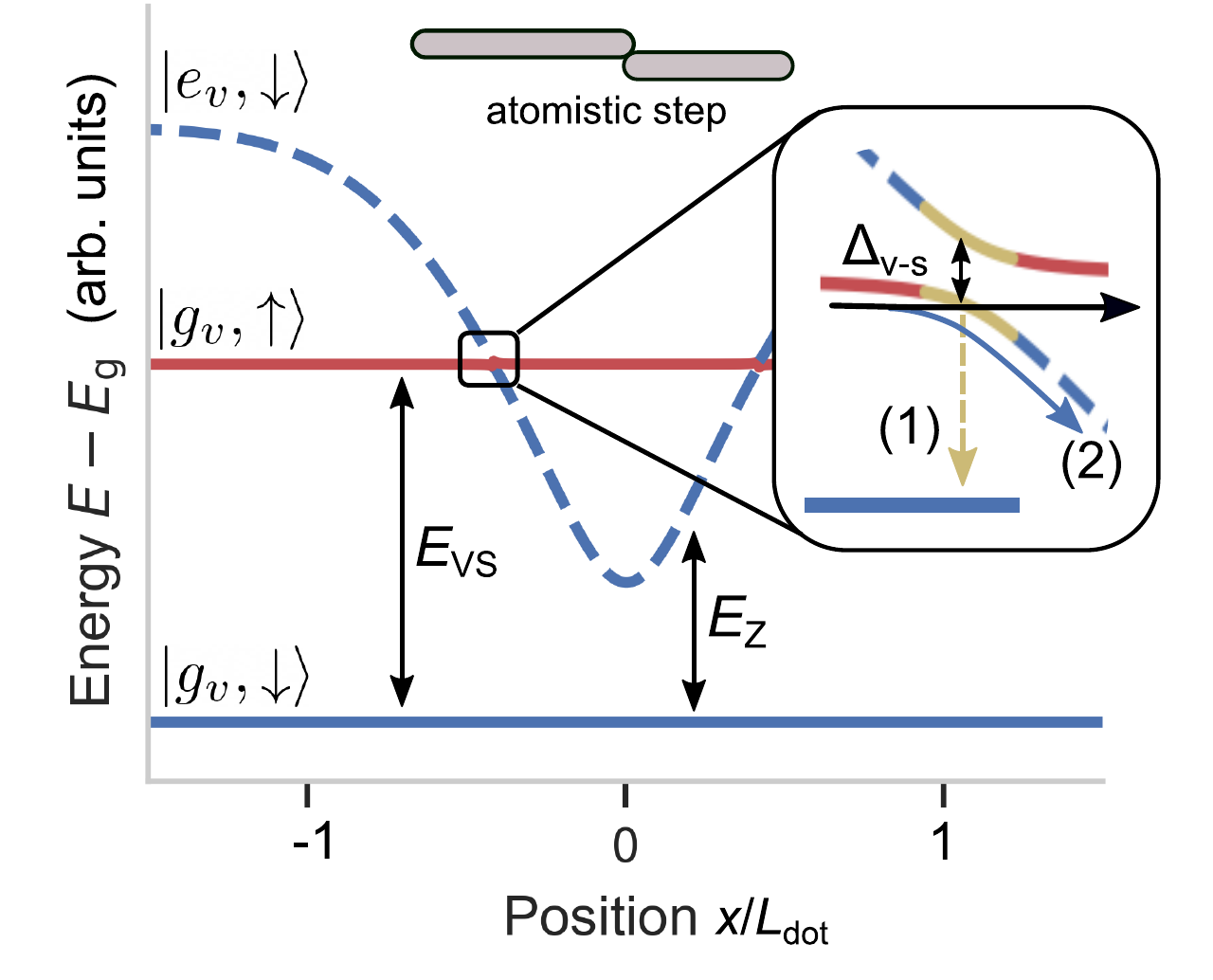}
    \caption{Sketch of energy difference between the following
states: ground valley spin-up state $\ket{g_v,\uparrow}$, excited valley spin-down state $\ket{e_v,\downarrow}$ and the lowest-lying level $\ket{g_v,\downarrow}$ with energy $E_g$ at the spin-valley hotspot caused by the atomistic step. Inset: zoom into avoided crossing between $\ket{e_v,\downarrow}$ and $\ket{g_v,\uparrow}$ states, where we identify two mechanisms responsible for the spin-flip error $\delta p_\uparrow$: (1) the relaxation in vicinity of avoided crossing, due to hybridisation between spin and valley states (yellow arrow) and (2) adiabatic transition that converts spin up to spin down, followed by subsequent relaxation, dephasing, and possibly more non-adiabatic transitions at the subsequent level crossing (blue arrow). For illustration purposes we have used arbitrary units.}
    \label{fig:hotspot}
\end{figure}

    However, if the presence of the hotspot cannot be prevented (e.g. due to local collapse of valley splitting caused by unfavourable step separation or relatively steep gradient), it will add to $\delta p_\uparrow$ and consequently to $\delta C$. The hotspot  is related to the avoided crossing between the states
    $\ket{e_v,\downarrow}$ and $\ket{g_v,\uparrow}$, where $g_v$ and $e_v$ label the ground and excited valley states, respectively. Their energies are sketched in Fig.~\ref{fig:hotspot} together with the ground-state $\ket{g_v,\downarrow}$. There are two mechanisms contributing to $\delta p_\uparrow$: (1) relaxation of the temporarily hybridized $\ket{g_v,\uparrow}$ and  $\ket{e_v,\downarrow}$ states to the ground state $\ket{g_v,\downarrow}$ which has the character of valley relaxation at the hotspot, while effectively resulting in spin relaxation, (2) transfer of the occupation from $\ket{g_v,\uparrow}$ to $\ket{e_v,\downarrow}$ state.  The latter process converts the spin qubit into a valley qubit, as it converts superposition of spin states into that of valley states, provided that the passage through the avoided crossing occurs via an adiabatic path. When this happens, the quantum information carried by the qubit can be destroyed by processes of relaxation to the valley ground state, and uncontrolled (due to fluctuations of velocity and electric field noise) interference effects that arise after the passing through the second anticrossing (see Fig.~\ref{fig:hotspot}). For a conservative estimate of the error we assume that taking an adiabatic path at the first anticrossing results in complete decoherence of qubit state.

For this conservative estimate, we concentrate on the adiabatic conversion labeled as mechanism (2), \addt{since the contribution from (1) is about an order of magnitude weaker, see Appendix.~\ref{app:hot_spot} for details.} We use the Landau-Zener model in the limit of fast passage, where the probability of going adiabatically through the avoided crossing, that corresponds to $\ket{g_v,\uparrow} \to \ket{e_v,\downarrow}$ transition, reads:
\begin{equation}
\label{eq:hot_spot_2}
\delta p_\uparrow^{(2,\text{art})} \leqslant 1-Q_{\text{LZ},\text{s-v}} \approx \frac{\pi}{2}\frac{\Delta_{\text{s-v}}^2}{\hbar a} \to  \frac{10^{-4}}{v[\mathrm{m/s}]},
\end{equation}
where we have used typical value of spin-valley mixing due to artificial-SOC as $\Delta_{\text{s-v}}^{\text{art}} \sim 20$\,neV, that have been estimated for the gradient of $(\Delta B/\Delta x)= 0.1\mathrm{mT/nm}$ (see Appendix.~\ref{app:hot_spot}), and $a = d( E_{VS}(t) - E_Z)/dt$. Larger gradients can significantly increase the error since $\Delta_{\text{s-v}}^{\text{art}}\propto (\Delta B/\Delta x)$, while for smaller ones the 
spin flip probability becomes dominated by intrinsic-SOC mechanism of the similar order of magnitude $\delta p_\uparrow^{(2,\text{int})} \!\leqslant\! 10^{-4}/v\mathrm{[m/s]}$. Naturally the gradient-dominated hotspots are most likely to occur in vicinity of SQuS terminals, close to stationary qubits or in possible manipulation regions, where most coherent control would take place. This shows that avoiding strong gradients of magnetic field or reducing the value of constant magnetic field to relatively hotspot-free $B=20$\,mT is sufficient to avoid shuttling error due to spin-valley hotspot.

\subsection{Transfer infidelity due to valley non-adiabatic effects}
We conclude this section by combining the above calculation of valley excitation and relaxation rates into an estimate for spin dephasing.
We initially consider relatively small bare valley splitting $E_{\text{VS},0} \leq 200$\,$\mu$eV,  which gives $\tau_{e,v} > \tau$ according to values from Fig.~\ref{fig:valleyT1_step_grad} and consequently falls into regime of no relaxation, in which $\delta C \approx p_{e,v}$, see the discussion in Sec.~\ref{sec:nonadiab_dephasing}.
In the smooth gradient model the final occupation $p_{e,v}$ can be estimated by considering the largest gradient along the \bus{}. Assuming that $\hbar a_x v \ll E_{\text{VS},0}$, we have:
\begin{equation}
\label{eq:pev_grad}
p_{e,v}^{\text{grad}} \sim 10\left(\frac{a_x[\tfrac{1}{\text{nm}}] v[\tfrac{\text{m}}{\text{s}}]}{E_{\text{VS},0}[\mu \text{eV}]}\right)^2 \exp(\frac{1}{2}(a_x\big[\tfrac{1}{\text{nm}}\big]\ldot[\text{nm}])^2) \,\, .
\end{equation}
As the upper bound of the linear gradient, we take $a_x= 0.05\pi/$\,nm $ \approx \pi/\ldot$, which would correspond to a miscut angle of $\vartheta = 0.5^\circ{}$ and measurable local valley splitting $\vs = 0.1 \vsz \in [10,20] \, \mathrm{\mu eV}$. The most pessimistic $\vsz=100$\,$\mu$eV gives $p_{e,v}^{\text{grad}} \sim 10^{-4}\,
(10^{-2})$ for $v = 10 \,(100)$\,m/s, respectively, which shows that the probability of valley excitations for the gradient values $a_x \sim \pi/\ldot$ is below the threshold at $v \sim 10$\,m/s. Note that in presence of disordered interface, the considered value of $a_x = \pi/\ldot$ can also emulate atomistic steps with average separation $\overline d \approx 0.85\ldot$ (2-3 steps per QD size).

In contrast, the step model describes the valley dynamics for lower densities of steps, $\overline d > \ldot$ (less then a step per QD size). The local valley splitting $\vs$ is then affected by a single step at a time, and its value is not smaller then $\vs \geqslant\cos(0.85\pi)\vsz \sim 0.2\vsz \in [20,40]\mu$eV, which is in the lower range of commonly measured local valley splittings. In this interference regime, we use Eq.~\eqref{eq:Q_n} to estimate:
\begin{equation}
\label{eq:pev_step}
p_{e,v}^{\text{steps}} \approx   \frac{10^{4}}{\overline d[\text{nm}]} \exp(-\frac{E_{\text{VS},0}[\mu \text{eV}] \ldot[\text{nm}]}{10\, v[\text{m}/\text{s}]}).
\end{equation}
For illustration, we take $\overline d = 30$\,nm (miscut angle $\vartheta  \approx 0.25^\circ{}$), $E_{\text{VS},0} \! = \! 100$ $\mu$\,eV and $\ldot= 20$\,nm, which result in $p_{e,v}^{\text{steps}}\sim 10^{-7}\, (10^{-2})$ for $v = 10\,(20)$\,m/s, respectively. The estimate reveals exponential sensitivity of the excitation probability on a single step $Q_1$ to the shuttling parameter $\hbar v/\ldot E_{\text{VS,0}}$, and in particular to the electron velocity $v$.

Next, we consider the case of large bare valley splitting $\vs$, illustrated by the value of $E_{\text{VS},0} = 500\,\mu\mathrm{eV}$, at which the phonon relaxation becomes non-negligible. For the constant gradient model, phonon relaxation is much slower than the rotation of the valley field and the corresponding coherent oscillation. Hence, it mostly assists in reaching average occupation of Eq.~\eqref{eq:Q_grad}. However, we highlight that larger values of $E_{\text{VS},0}$ reduce the amplitude of oscillation and in this way limit the phase error, i.e. $\delta C^{\text{grad}} \leqslant p_{e,v}^{\text{grad}} \approx E_{\text{VS},0}^{-2}$. For the isolated step model, Fig.~\ref{fig:valleyT1_step_grad} shows an estimate of the shortest valley relaxation time $\tau_{e,v} \approx 100$\,ns, which is only on the verge of reducing the phase error $\delta C$ (see Sec.~\ref{sec:nonadiab_dephasing}). Indeed, for considered difference in valley-dependent Larmor frequencies $\delta \omega_v/h \geqslant 0.5 $\,MHz, where the equality holds for $B =20$\,mT and $\delta g_v =10^{-3}$, one can estimate the possible improvement in $\delta C$, i.e. the lower bound of the error, as:
\begin{equation}
   \delta C \approx   p_{e,v} \frac{\phi_v^2}{2}  = \left( \frac{\delta \omega_v}{\hbar}\tau_{e,v}\right)^2  p_{e,v}\geqslant 0.01 p_{e,v}.
\end{equation} 
This means that the presence of valley relaxation can realistically reduce phase error by at most two orders of magnitude. Further improvement requires shorter relaxation times (larger $E_{\text{VS},0}$ or stronger valley orbit coupling), weaker magnetic fields, or smaller difference in valley-dependent g-factors. We stress that in the step model, for $v \leq 40$\,m/s and $\vsz = 500$\,$\mu$eV, the valley relaxation is not really needed for coherent operation of the \bus{}, since corresponding probability of excitation on a single step is negligibly small, $Q_{1} \sim 10^{-8}$, even for the smallest considered QD size of $\ldot\sim 12$\,nm ($E_{\text{orb}} = 3$\,meV).

\section{Discussion and Conclusion}
\label{sec:discussion}
In summary, we have provided a blueprint of a scalable Spin Qubit Shuttle (\bus{}) for quantum computers based on Si/SiGe quantum dots. The \bus{}  allows for coherent communication over $\sim\!10$ $\mu$m distance. We have argued that this task can be achieved by moving a single electron spin qubit, using one of two methods: consecutive charge transfers between pre-defined tunnel-coupled QDs forming a chain - the Bucket Brigade (BB) mode - or keeping the electron trapped in a moving quantum dot - the Conveyor Belt (CB) mode. The scalability of both modes is provided by clavier gates with four-point electric connection as proposed here, which avoid the signal fanout problems, see Sec.~\ref{sec:device_idea} for details. In the same Section, we have argued that the BB mode requires higher level of fine tuning and device uniformity, and for this reason, in our modeling we have focused on the CB mode, for which these limitations are less severe. We have predicted a relatively wide range of parameters needed for stable operation of the CB \bus{}, and proved its robustness against simulated electrostatic disorder of realistic amplitude, see Tab.~\ref{tab:parameters} and Sec.~\ref{sec:design_CB} for discussion. 

We have discussed two mechanisms of spin dephasing of the spin qubit shuttled in the CB mode: one due to position-dependent spin precession frequency during adiabatic shuttling, and another due to level-dependent spin precession frequency. The latter mechanism is activated by non-adiabatic transitions (see Sec.~\ref{sec:nonadiab_dephasing} for general discussion) between the orbital and valley degrees of freedom, analyzed in detail in Sec.~\ref{sec:orbital_nonadiabaticity} and Sec.~\ref{sec:valley_nonadiabaticity}, respectively. The dephasing due to the first mechanism is suppressed by increasing shuttling velocity $\vd$, which decreases the time the electron spends in the \bus{}, \add{and activates the motional narrowing effect}. The second mechanism becomes more dangerous with increasing $\vd$ due to larger probability of the electron straying away from adiabatic evolution trajectory. 

\begin{figure}
	\includegraphics[width=0.98\columnwidth]{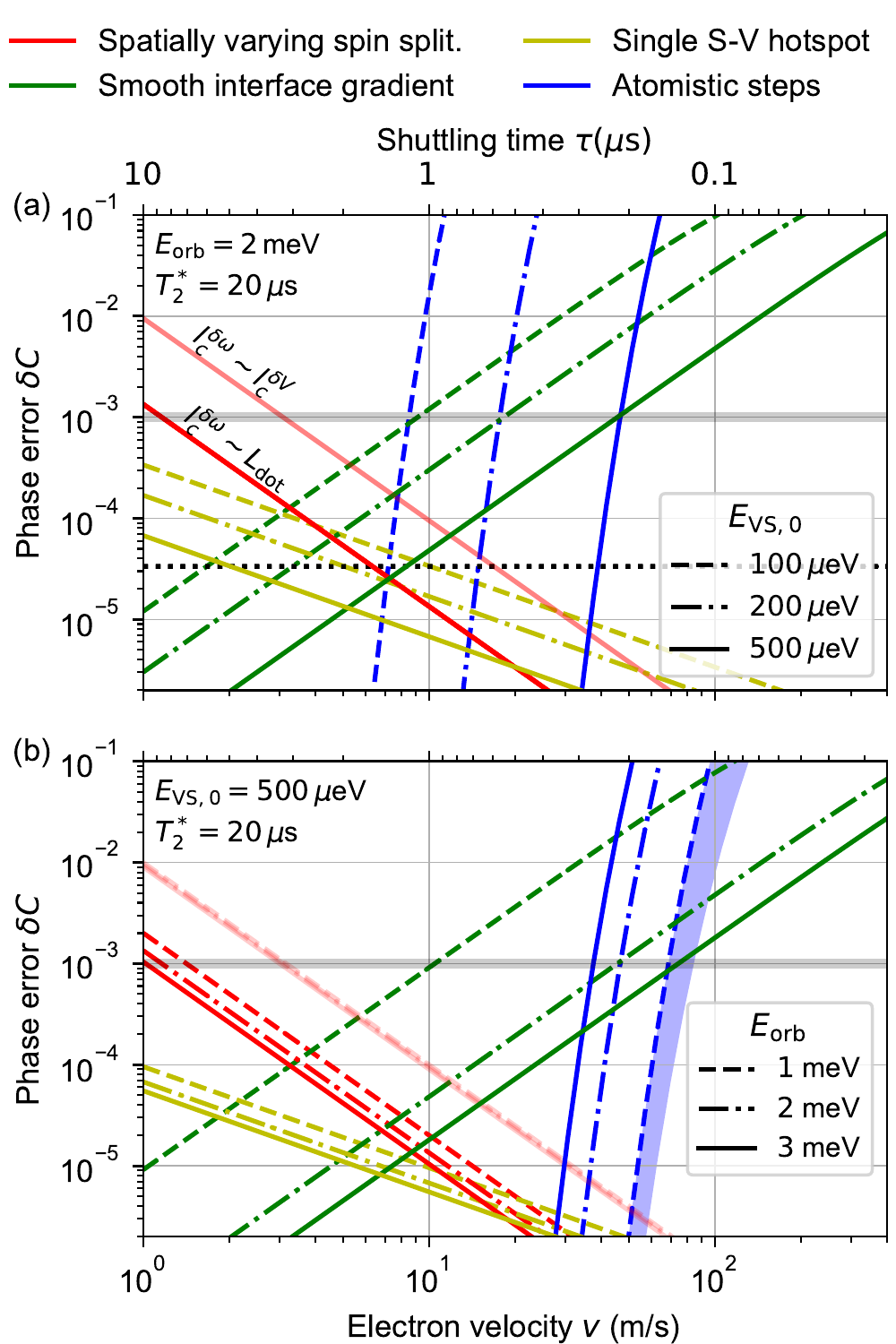}
	\caption{Expected loss of spin coherence $\delta C$  experienced by an electron shuttled through $10$-$\mu$m long SQuS as a function of electron velocity $v$ (lower axes) and shuttling time $\tau$ (upper axes).
	We compare various contribution to such an error: spatial variation of precession frequency (red lines, Sec.~\ref{sec:quasistatic}), single spin-valley hotspot due to intrinsic spin-orbit coupling (yellow, Sec.~\ref{sec:hotspot}) and non-adiabatic effects in valley degrees of freedom caused by: smooth gradient (green, Sec.~\ref{sec:gradient}) and sharp atomistic steps (blue, Sec.~\ref{sec:single_step}). We distinguish between the magnetic noise due to nuclear spins (darker red line) and the charge noise (lighter red line). We highlight that hotspot contribution is the only one caused by spin relaxation process, for which $\delta p_\uparrow \sim \delta C$. The corresponding error should be multiplied by number of hotspots in the SQuS (i.e. no hotspots for $E_Z\ll \vs$). (a) $\delta C$ for relevant values of effective valley splitting $E_{\text{VS},0} = 100,200,500\,\mu \mathrm{eV}$ (dashed, dashed-dotted and solid lines respectively) for fixed $E_{\text{orb}} = 2\,$meV and $T_2^* = 20\,\mu$s. (b) $\delta C$, but for different values of orbital energy $E_{\text{orb}} = 1,2,3$\,meV (dashed, dashed-dotted and solid lines respectively) for fixed values of $E_{\text{VS},0}= 500\,\mathrm{\mu eV}$ and $T_2^* = 20$\,$\mu$s. The results are drawn for the gradient of $a_x \approx 0.05\pi/$\,nm (corresponding to intentional miscut angle $\vartheta \approx 0.5^\circ{}$), average step separation $\overline d = 30$\,nm ($\vartheta \approx 0.25^\circ$) and magnetic field of $B = 100$\,mT. The effect of decreasing magnetic field to $B = 20$\,mT is illustrated by a shaded regions. For plotted range of velocities $v \ll \ldot E_\text{orb}/\hbar$, $\delta C$ due to orbital non-adiabaticity is exponentially small. \addt{For comparison in a) we use black, dotted line to plot probability of spin-relaxation during the transfer from \cite{Huang13} and Eq.~\eqref{eq:spin_relax}} }
	\label{fig:final_fig}
\end{figure}

The summary of the results of the last four Sections is given in Fig.~\ref{fig:final_fig}, where we focus on spin coherence of the qubit transferred using the CB mode.   
In this Figure, the phase error, $\delta C$, is plotted against the electron velocity $\vd$ for the relevant ranges of remaining parameters: typical coherence times $T_2^*$ in stationary QD, ranges of orbital splitting $E_{\text{orb}}$, and the bare valley splitting $E_{\text{VS},0}$. The last parameter corresponds to typical value of valley splitting averaged over alloy disorder, {\it in absence} of additional source of interface inhomogeneity (either atomistic steps or a smooth gradient of interface position that we have considered here). It means that the $E_{\text{VS},0}$ is an {\it upper bound} for $E_{\text{VS}}$ that could be measured in a QD localized at some position. 

The most important qualitative result shown in Fig.~\ref{fig:final_fig}, is the prediction of a relatively wide velocity range for the operation of the SQuS with coherence error $\delta C$ below the targeted threshold of $10^{-3}$. The presence of optimal $\vd$, which leads to smallest $\delta C$, results from existence of the two above-mentioned dephasing mechanisms with opposite dependence of resulting error on $\vd$. This range extends over at least one order of magnitude of velocities,  and it includes $\vd \! =\! 10 \, \mathrm{m/s}$ value, for almost all combinations of parameters. Let us recall that simulations of the QD velocity from Fig.~\ref{Fig2}f have shown that for a nominal $\vd$ of 10\,m/s the velocity might randomly increase up to $\approx \! 20$\,m/s. Such a range of $\vd$ does not endanger the targeted $\delta C \! < \! 10^{-3}$, \addt{if the valley splitting is sufficiently large, i.e. $E_\text{VS,0}>200\,\mu$eV}. 

The velocity of $10 \, \mathrm{m/s}$ lies close to the boundary below which the shuttling time becomes larger than typical qubit manipulation time (let us recall that for shuttling distance of $\LQBL=10\,\mu$s, shuttling time of $\tau \! \approx \! 1 \, \mu\mathrm{s}$ requires $v\approx 10 \, \mathrm{m/s}$). 
If slower shuttling could be accepted, larger values $T_{2}^*$ would allow for $\delta C \! < \! 10^{-3}$ for lower $\vd$: in Fig.~\ref{fig:final_fig}, we see that for the $T_2^* \! = \! 20 \,\mu\mathrm{s}$ the phase error due to quasistatic spin splitting noise grows above the $10^{-3}$ threshold only for velocities below $1 \, \mathrm{m/s}$. Another lower bound on the \bus{} electron velocity is the spin relaxation at the spin-valley hotspots. A single hotspot leads to phase error indicated by the yellow line in Fig.~\ref{fig:final_fig} in the absence of a strong magnetic field gradient. This error is the only one \add{considered here that is} resulting from the spin relaxation process, \add{as the spin-orbit induced relaxation caused by electron motion has much weaker effect \cite{Huang13}.}

On the other hand, the upper bound on $\vd$ is given by the dephasing caused by intervalley excitations. According to Fig.~\ref{fig:final_fig}, it is given by $v_\text{max}\! \approx \! 10 \, \mathrm{m/s}$, in the most pessimistic scenario of $E_{\text{VS},0} \!=\!  100\,\mu\mathrm{eV}$, and can be as large as $v_\text{max} \approx 50$\,m/s for $E_{\text{VS},0} \!=\!  500 \, \mu\mathrm{eV}$. In order to distinguish between the cases relevant for gradient or steps models, we have considereded various average tilts of the quantum well interface, which result in the gradient of $a_x^\text{max} \!= \! \pi/L$ (corresponding to miscut angle $\vartheta \!= \!0.5^\circ{}$) and the step separation $\overline d \!= \!30\, \mathrm{nm}$ ($\vartheta \!=\! 0.25^\circ{}$). With those values, both models reproduce reported values of local valley splittings $\vs \! \approx  \! (0.1-0.2) \vsz \in (10-100)\mu$eV. The upper bound on the electron velocity $v_\text{max}$ will significantly increase for smaller gradients and lower density of steps. In contrast, lateral distance $d$ among the atomic steps being smaller than the QD size $\overline d < L$ may result in unfavourable collapse of the valley splitting and strongly nonadiabatic evolution, see Sec.~\ref{sec:valley_nonadiabaticity} for details.

We have considered charged defects at the oxide interface as the dominant source of potential disorder in Si/SiGe. The impact of other crystalline imperfections of the Si/SiGe heterostructure e.g. charged defects close to the QW and the SQuS channel, or threading dislocations, may be \add{relevant and need to be investigated in the future}. The more localized imperfections might more easily lead to the orbital excitation of the shuttled electron. However, we have shown that even extremely large orbital excitation rates, as high as $100\,\mathrm{MHz}$, are unproblematic due to the fast orbital relaxation rate.

From our analysis it is clear that operating at small magnetic field is beneficial for coherent transfer. It minimizes both the phase error during non-adiabatic transitions, and the probability of the occurrence of a spin-valley hotspot at which $E_Z \! = \! \vs$. In Sec.~\ref{sec:hotspot}, we have shown that the presence of the hotspot can be avoided with magnetic field below $B \leqslant 100\,\mathrm{mT}$ ($E_z \leqslant 12\,\mu \mathrm{eV}$), provided the valley splitting along the \bus{} is relatively stable, i.e.~$\vs>20\,\mu$eV. Otherwise, presence of the hotspots along the \bus{} introduces shuttling error due to spin relaxation. We have shown that for the error threshold of $10^{-3}$, a typical error at the hotspot due to intrinsic spin-orbit mechanism allows for crossing a few of them. In contrast, strong magnetic field gradients (artificial/synthetic SOI) used for fast EDSR qubit manipulation, may introduce spin-flip error above the threshold even at a single hotspot. Such large gradients are not required in a shuttling based architecture, since qubits can be transported to dedicated manipulation zones, in which the displacement amplitude of the electron is larger during EDSR. Hence, orders of magnitude smaller gradients may be sufficient to reach the same EDSR Rabi frequency. Notably, in the absence of hotspot occurrences, the remaining discussed  relevant error processes discussed at length in the paper,  are of pure dephasing character. This may be utilized in tailoring quantum error correction schemes for architectures using \bus{} devices as coherent links.

We have found that for multiple reasons, increasing the effective valley splitting $\vsz$ strongly improves operation of the \bus{}. Firstly, the number of spin-valley hotspots discussed above can be reduced, if the valley splitting is sufficiently large and uniform along the \bus{}, i.e. $\vs\gg E_z$. Secondly, large $\vsz$ significantly limits dephasing caused by valley non-adiabaticity: on one hand it limits excitation probability in both considered interface models (Fig.~\ref{fig:final_fig}), and on the other hand it enhances valley relaxation via phonon emission (Fig~\ref{fig:valleyT1_step_grad}). Recently, a few methods of increasing valley splitting $\vsz$ in stationary QDs have been investigated. One of them suggests back-gates for controlling the electric field across the QD independent from the QD filling \cite{Hosseinkhani20}. Another approach uses engineering of the Ge profile across the Si/SiGe heterostructure \cite{Mcjunkin21,McJunkinPRB21}. 
A third method relies on increasing random fluctuations of alloy composition (Ge concentration in Si QW), which statistically increases average $\vsz$ \cite{Wuetz21}, but at the cost of larger variance. Thus, from the perspective of the \bus{}, it might increase the probability of unfavourable regions with small valley splitting, and also increase spatial variation of valley field, which could be translated to larger gradients and less regular steps.

The results presented here show that even for the smallest considered $E_{\mathrm{VS,0}}\! =\! 100 \, \mu\mathrm{eV}$ it should be possible to transfer the electron spin qubit over $10 \, \mu\mathrm{m}$ in about $1 \, \mu\mathrm{s}$ with phase error of $10^{-3}$. However, having a wide range of parameters allowing for such coherent transfer is clearly desirable. Let us now discuss a few methods to make the operation of the SQuS become more robust and faster.

The effects of the electric noise can be diminished to some extent by populating the SQuS with multiple electrons, creating an effective screening effect \cite{Higginbotham14,Barnes11}. For example, filling the shuttled QD with three electrons, the ground valley states are fully occupied by a spin singlet state. In the opposite regime of low velocities, the noise might be dominated by the magnetic noise either from nuclear spins, or spatial variation of g-factor. In such case, a similar effect can be achieved by shuttling the spin qubit in the singlet/unpolarized triplet  basis, which in principle should be less sensitive to correlated noise, and requires filling the moving quantum QD with two electrons.

Naturally, there are numbers of applicable methods if the \bus{} is relatively short, and thus a certain level of active tuning is feasible: The limitation of compensating potential disorder imposed by our gate sets approach can be circumvented by tuning the voltages of a \bus{} in the time domain. This strategy assumes that only one qubit is shuttled in the \bus{} at a time. Since its position is deterministic, disorder induced potential offsets can be compensated by tuning the voltage signals applied to the \bus{} in real time. Also, the electron velocity can be adapted to disorder/noise in certain regions of the \bus{}. In particular, the possibility of individual tuning of the gates could make the coupling between the predefined QDs more uniform, which in principle could enhance the feasibility of the bucket-brigade mode shuttling. Besides the larger effort of tuning, this approach requires much higher complexity of the control signals, the generation of which is more involved and particularly limited by the space for memory of on-chip control electronics. 

Let us finish with stressing that the presented analysis of sources of errors caused by shuttling of electron in Si/SiGe structures highlights two areas for future research. As the main decoherence mechanism at higher shuttling velocities is due to valley excitations caused by atomic disorder at Si/SiGe interface in presence of valley-dependence of electron spin $g$-factors, further research into interface disorder and $g$-factor physics in Si/SiGe structures is needed. With new insights on the influence of material growth and nanostructuring on atomic disorder at the interface, strategies to decrease both the probability of valley excitations, and the valley-dependence of $g$-factors, may emerge.
 
 Finally, we note that a similar architecture can be used in other semiconductor-based quantum devices, with the relevant example of GaAs-based nanostructures, in which the absence of valley degree of freedom may allow for faster operation of the \bus{} that reduces the detrimental influence of the nuclear spins. Other systems, such as hole spins in Ge/SiGe \cite{Scappucci21} avoid challenges of both coupling to nuclear spins and relevance of the valley degeneracy, but the stronger spin-orbit interaction of hole qubits need to be considered.

\section{Acknowledgements}
We thank Hendrik Bluhm and Lieven Vandersypen for valuable discussions and David DiVincenzo for comments on the manuscript. This work has been funded by the National Science Centre (NCN), Poland under QuantERA program, Grant No.~2017/25/Z/ST3/03044, \add{and under the ETIUDA doctoral scholarship, Grant No.
2020/36/T/ST3/00569 (J.A.K.),} and by the Federal Ministry of Education and Research under Contract No. FKZ: 13N14778 as well as by the German Research Foundation (DFG) within the projects BO 3140/4-1, and under Germany's Excellence Strategy - Cluster of Excellence Matter and Light for Quantum Computing" (ML4Q) EXC 2004/1 - 390534769289786932. Project Si-QuBus has received funding from the QuantERA ERA-NET Cofund in Quantum Technologies implemented within the European Union’s Horizon 2020 program.

V.L. and J.A.K. contributed equally to all the theory presented in this work. N.F. did numerical device simulations and I.S. fabricated the described device. L.R.S. and {\L}.C. equally supervised the work. All authors extensively discussed all sections presented in this work and jointly wrote the manuscript.

\begin{appendix}
\section{Details on geometry estimates for the QuBus gate design}
\label{App:electrostaticEstimatesDetails}
\add{Here, we provide a Fourier mode analysis to arrive at the scaling relations presented in Sec.~\ref{sec:analitical_opt_dev}.
Choosing the lowest non-trivial mode of the confinement potential along the $x$-direction (a sine wave with angular wave number $K_{x,1}=2\pi/L_x$) as the relevant one, the curvature of the QD potential $c_\mathrm{conf}$ will simply be proportional to $K_{x,1}^2$. However, the overall amplitude of the mode is given by the mode decay when propagating the mode from the clavier gate region down into the quantum well. We solve a 2D Laplace equation to account for the propagation of the amplitude of the Fourier modes:
\begin{equation}
\left(\frac{\partial^2}{\partial z^2}-K_{x,n}^2\right)c_n(z)=0
\end{equation}
For a homogeneous medium, the mode amplitude then straightforwardly obeys an exponential decay with the corresponding angular wave number $K_{x,n}\equiv 2\pi n/L_x$:
\begin{equation}
c_n(-h) = c_n(0)e^{-K_{x,n}h},
\end{equation}
\noindent where $h=h_\mathrm{SiGe}+h_\textbf{}{ox}$ is the distance of the considered QW-plane to the lower face of the clavier gates and $c_n$ are the Fourier coefficients. We may then introduce a confinement coefficient by perform a harmonic approximation at the minimum of the first harmonic $n=1$, which then scales as $$c_\mathrm{conf}\propto\frac{1}{L_x^2} c_1(z=-h_\mathrm{SiGe}).$$
Assuming the depth of the QW to be fixed and varying only the conveyor unit cell length $L_x$, the maximum of this confinement coefficient then occurs at
$L_x=L_{\mathrm{opt}}\equiv\pi(h_\mathrm{SiGe}+h_\mathrm{ox})$ in the case of homogeneous dielectrics. Note that the possible depth of the QW is constrained by the electrostatic requirements for qubit manipulation thus the distance to the charged defect is limited.
Choosing this optimal \bus{} cell length, the confinement coefficient scales as $$c_\mathrm{conf}\propto\frac{1}{L_\mathrm{opt}^2} c_1(z=-h_\mathrm{SiGe})\big|_{L_x=L_\mathrm{opt}}.$$ With $c_1(z=-h_\mathrm{SiGe})\big|_{L_x=L_\mathrm{opt}}= c_n(0)e^{-2}= \mathrm{const}$, only the inverse quadratic scaling $c_\mathrm{conf}\propto\frac{1}{L_\mathrm{opt}^2}$ remains, and with $E_\mathrm{orb}\propto \sqrt{c_\mathrm{conf}}$ in a harmonic potential approximation, $E_\mathrm{orb}\propto\frac{1}{L_\mathrm{opt}}$ will be approximately inversely proportional to the depth of the quantum well.}

\add{To be more precise, one may} take the different dielectric constants in the oxide (3.9 for SiO$_{2}$) and semiconductor regions (13.2 averaged over alloy composition for Si$_{0.7}$Ge$_{0.3}$) into account. \add{For that}, we need to choose a slightly more involved exponential ansatz and work the dielectric constants into the constraint of continuity of electric flux density at the boundary between the two media. For convenience, we choose $z=0$ as the position of the semiconductor-oxide interface. This yields
\begin{equation}
c_n(z=-h_\mathrm{SiGe}) = c_n(h_\mathrm{ox})\frac{e^{-K_{x,n}h_\mathrm{SiGe}}(1+r)}{e^{+K_{x,n}h_\mathrm{ox}}+re^{-K_{x,n}h_\mathrm{ox}}}
\end{equation}
with $r \equiv (\varepsilon_{r,\mathrm{ox}}-\varepsilon_{r,\mathrm{SiGe}})/(\varepsilon_{r,\mathrm{ox}}+\varepsilon_{r,\mathrm{SiGe}})$.
Now we discuss the maximization of the confinement coefficient $$c_\mathrm{conf}\propto\frac{1}{L_x^2} c_1(z=-h_\mathrm{SiGe}).$$ We will first assume the depth of the QW to be fixed and vary only the conveyor unit cell length $L_x$. The maximum then occurs at
$L_\mathrm{opt,0}=\pi(h_\mathrm{SiGe}+h_\mathrm{ox})$ for the homogeneous case. Taking the different dielectric constants of the oxide and semiconductor into account, we have to solve the nonlinear equation
\begin{equation}
L^\mathrm{inh}_\mathrm{opt} = \pi(h_\mathrm{SiGe}+h_\mathrm{ox}f(2\pi/L^\mathrm{inh}_\mathrm{opt})) \label{eq:L_opt_nonlin_eq}
\end{equation}
with $$f(K_x) = \frac{e^{+K_{x}h_\mathrm{ox}}-re^{-K_{x}h_\mathrm{ox}}}{e^{+K_{x}h_\mathrm{ox}}+re^{-K_{x}h_\mathrm{ox}}}.$$
Using the homogeneous solution as the argument for the right hand side of (\ref{eq:L_opt_nonlin_eq}), already gives a very good approximation of the optimal period length
\begin{equation}
L^\mathrm{inh}_\mathrm{opt} \approx \pi(h_\mathrm{SiGe}+h_\mathrm{ox}f(2\pi/L_\mathrm{opt})) \label{eq:L_opt_nonlin_eq2}
\end{equation}
for our discussion.

\begin{figure}
\includegraphics[width=\linewidth]{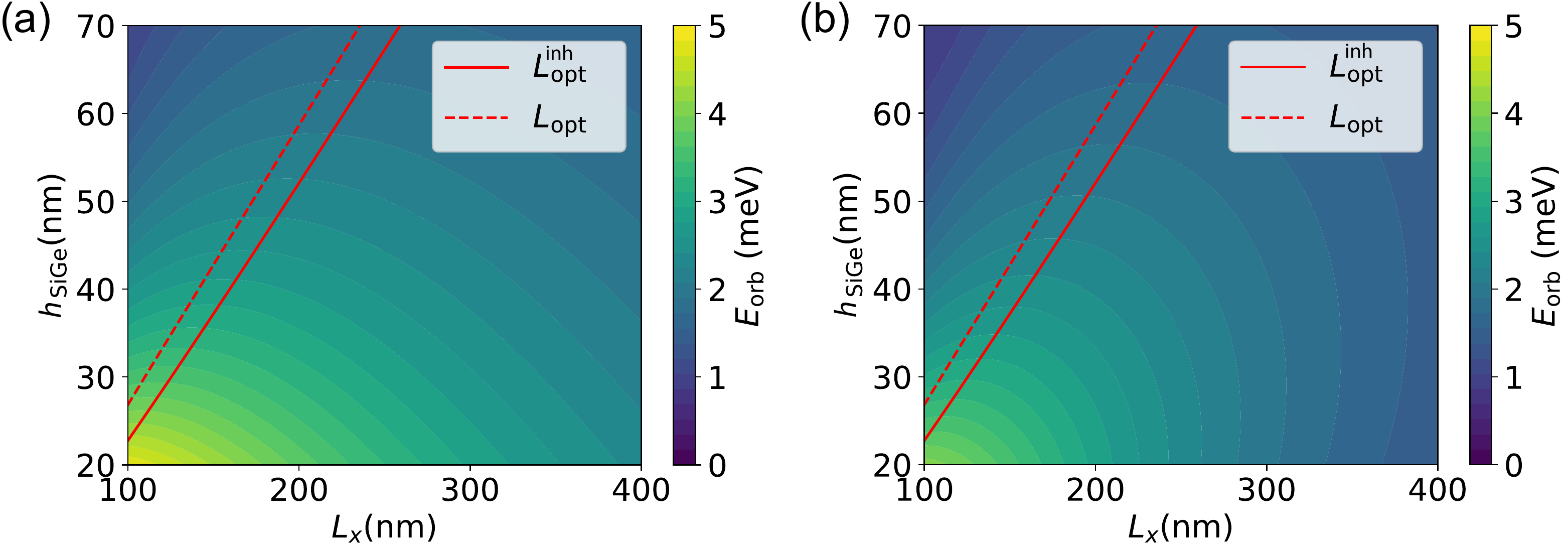}
\caption{Orbital-splitting as a function of the geometric parameters of the \bus{} without disorder. (a) Minimum orbital splitting $E_\mathrm{orb}^x$ extracted from the first mode of the Fourier analysis as a function of the \bus{} unit cell length $L_x$ and the thickness of the SiGe layer ($h_\mathrm{SiGe}$). (b) Calculation of the minimum orbital splitting $E_{orb}^x$ from solving the Schr{\"o}dinger equation on the full-mode periodic potential. In panels a to b, $h_{ox}=10$\,nm and $V_a=100$\,mV is fixed and the red line indicates $L_{\mathrm{opt}}^\mathrm{inh}$ and the red dashed line $L_{\mathrm{opt}}$.}
\label{Fig5app}
\end{figure}

The behavior of the orbital confinement is plotted in Fig. \ref{Fig5app}, illustrating the degree of deviation of $L_{\mathrm{opt}}$ from the inhomogeneous estimate $L_{\mathrm{opt}}^\mathrm{inh}$ and the appropriateness of the first harmonic approximation Fig. \ref{Fig5app}a when compared to the exact solution in Fig. \ref{Fig5app}b. We note that $L^\mathrm{inh}_\mathrm{opt}$ scales approximately linearly with the QW depth parametrized by $h_\mathrm{SiGe}$, mostly shifting the solution for the homogeneous case (cf. Fig. \ref{Fig5}a and \ref{Fig5}b).

\section{Formal treatment of spin dephasing due to orbital/valley transitions} \label{app:formal_dephasing}
Here we extend the analysis of electron dephasing due to relaxation from an excited to the ground orbital/valley state, which has been effectively introduced in Sec.~\ref{sec:nonadiab_dephasing}. For simplicity, we assume that we are dealing with two adiabatic orbital/valley states, $\ket{g}$ and $\ket{e}$, each with a constant $g$-factor. We assume that at an initial time $t_e$ an event leading to finite population of $\ket{e}$ state happens. Such an event is associated with a momentary decrease of the energy gap between these two states, and/or increase of motion-induced coupling between them.
We look at the subsequent evolution of the qubit, while  taking the temperature of the environment as much smaller than the post-excitation event energy difference between $\ket{e}$ and $\ket{g}$ states, so that the environment can only cause transitions from $\ket{e}$ to $\ket{g}$ state. 

We describe the electron spin dynamics in a reference frame rotating with Larmor precession of spin of electron in $\ket{g}$ state, so that there is no spin dynamics in this state, and the spin of electron in $\ket{e}$ is rotating about $z$ axis according to the Zeeman Hamiltonian given by
\begin{equation}
 \hat{H}_Z = \ket{e}\bra{e}\otimes \hbar\delta \omega \hat\sigma_z/2 \,\, ,   \label{eq:HZ}
\end{equation}
where $\hbar\delta \omega = 2(\delta g/g) \mu_B B$.
We assume that the electron had been prepared in the past in $\ket{g}$ state, with its spin in a superposition state $\ket{\Psi_s} = a_{\uparrow}\ket{\uparrow} + a_{\downarrow}\ket{\downarrow}$, so that the initial coherence is $C(-\infty)\! =\! a_\uparrow a^{*}_\downarrow$.
The process of excitation at time $t_e$ is treated for simplicity as point-like in time. It can be coherent, due to Landau-Zener transition  that is  not fully adiabatic,  leading to a creation of superposition of electron being in ground $\ket{g}$ and excited $\ket{e}$ states. It can also be incoherent, due to charge disorder acting as noise in reference frame co-moving with the QD, leading only to creation of occupation of higher-energy state $\ket{e}$. In the first case, the electron at time $t_e$ is in a pure state, 
\begin{equation*}
    \ket{\Phi(t_e)} = (a_g\ket{g} + a_e\ket{e}) \otimes (a_{\uparrow}\ket{\uparrow} + a_{\downarrow}\ket{\downarrow}) \,\, ,
\end{equation*}
and the corresponding density operator  $\ket{\Phi(t_e)}\bra{\Phi(t_e)}$ has all its matrix elements finite (assuming none of $a_i$ amplitudes is zero). In the second case the density operator corresponding to the created partially excited state is
\begin{equation}
\hat\rho(t_e) = \big[(1-p_e)\ket{g}\bra{g} + p_e\ket{e}\bra{e} \big] \otimes \ket{\Psi_s(t_e)}\bra{\Psi_s(t_e)} \,\, ,
\end{equation}
which has zero $e$-$g$ coherences, i.e. $\bra{e} \hat\rho(t_e) \ket{g} \! =\! 0$. In both cases the spin coherence in $\ket{g}$ state is diminished compared to pre-excitation value, as it is multiplied by $|a_g|^2$ or $1-p_e$ factors, respectively.
The ``missing'' part of spin coherence is in the $\ket{e}$ state, and we want to see if subsequent evolution, involving energy relaxation into the environment, can bring back this missing part while returning the electron into the adiabatic ground state $\ket{g}$.

The evolution for $t \! > \! t_e$ is then due to $\hat{H}_Z$ from Eq.~(\ref{eq:HZ}) and to transitions from $\ket{e}$ to $\ket{g}$ state caused by interaction with the environment. The first process is unitary, and for $\delta \omega \! \neq \! 0$ it correlates the spin degrees of freedom with $e/g$ orbital or valley degrees of freedom. The second process in nonunitary, and we describe it using the Born-Markov approximation (which is definitely appropriate for treating phonon-induced orbital and valley relaxation in quantum dots), which leads to Lindblad equation for evolution of density matrix:
\begin{align}
\frac{\mathrm{d} \hat\rho(t)}{\mathrm{d}t} &  = -i[\hat{H}_Z,\hat\rho(t)] + \nonumber\\
& \sum_{k=1}^{n} \left[ \hat L_k\hat\rho(t) \hat L^{\dagger}_k - \frac{1}{2} \hat{L}^{\dagger}_k\hat L_k \hat\rho(t) -  \frac{1}{2} \hat\rho(t)\hat{L}^{\dagger}_k\hat L_k \right] \,\, , \label{eq:evo}
\end{align}
where $n$ is the number of jump operators $L_k$. 

We neglect spin-orbit coupling in treatment of electron-phonon interaction, so that we disregard phonon-induced spin relaxation (its significance has been discussed in Sec.~\ref{sec:hotspot}). The $\hat{L}_k$ operators must then be spin-diagonal. Their dependence on spin degree of freedom is not however obvious: for $\delta \omega \! \neq \! 0$ the transitions for the two spin directions correspond to transfer of unequal energy quanta into the environment. 

If time-energy uncertainty allows for resolving of the $\delta \omega$ energy difference, which is the case when $\delta \omega \! \gg \! \gamma$, where $\gamma$ is the $e$-$g$ transition rate, we should use a separate jump operator $\hat L_{k}$ for each spin. For phonon emission, the physical picture is the following: in this situation the phonon wave packets emitted for transitions involving each spin are not overlapping in frequency, so in principle the information on which spin-diagonal transition occurred is imprinted on the environment \cite{Gawelczyk18}, and the act of phonon emission amounts to measurement of the spin projection. It is straightforward to check that if we use $\hat{L}_s = \sqrt{\gamma}_s \ket{g}\bra{e}\otimes \ket{s}\bra{s}$ with $s\! =\! \uparrow$, $\downarrow$, there is no pumping of coherence from $\rho_{e\uparrow,e\downarrow}$ to $\rho_{g\uparrow,g\downarrow}$, and while the relaxation leads to repopulation of $\ket{g}$ state, i.e. $\rho_{gs,gs}$ occupation returns to its pre-excitation value, spin coherence remains suppressed by $1-p_e$ factor. Note that using the terminology of Section \ref{sec:nonadiab_dephasing}, $\delta \omega \! \gg \! \gamma$ is equivalent to $\delta \phi \! \approx \!  \delta\omega/\gamma \! \gg \! 1$, so that we recover the result of $\delta C \! \approx \! p_e$. 

On the other hand, when $\delta \omega \! \ll \! \gamma$, the traces left in the environment by transitions involving each of the spin states are indistinguishable, and we should use a single jump operator that is blind to the spin degree of freedom, $\hat L_0 \! =\! \sqrt{\gamma} \ket{g}\bra{e} \otimes \mathds{1}_s$ where $\mathds{1}_s$ is an unit operator in spin space. With such a jump operator, Eq.~(\ref{eq:evo}) leads to the following time evolution of the ground-state spin coherence
\begin{equation}
\rho_{g\uparrow,g\downarrow}(\Delta t) =    \rho_{g\uparrow,g\downarrow}(t_e) + \rho_{e\uparrow,e\downarrow}(t_e) \frac{\gamma\left( 1-e^{-i\delta\omega \Delta t} e^{-\gamma\Delta t} \right) }{\gamma + i\delta\omega} \,\, , 
\end{equation}
in which $\Delta t \! =\! t-t_e$. For $\gamma \Delta t \! \gg \! 1$ we obtain the asymptotic value of coherence after relaxation:
\begin{equation}
    |\rho_{g\uparrow,g\downarrow}(\Delta t \gg 1/\gamma)| \approx |C(-\infty)| \left |1-p_e + p_e\frac{\gamma}{\gamma +i\delta \omega} \right | \,\, , \label{eq:Cinfty}
\end{equation}
in which we used $\rho_{g\uparrow,g\downarrow}(t_e)\! =\! (1-p_e)C(-\infty)$ and $\rho_{e\uparrow,e\downarrow}(t_e)\! =\! p_e C(-\infty)$. Expanding this result to the lowest order in $p_e$ and $\delta \omega / \gamma \! \ll \! 1$ we obtain $\delta C \approx p_e \delta\omega^2/\gamma^2$, which is in agreement with the previous qualitative estimate $\delta C \! \approx \! p_e \delta \phi^2/2$ valid for $\delta \phi \! \ll \! 1$ once we identify $\delta \omega/\gamma$ (equal to $2(\delta g/g) \mu_B B \tau_e$ using the quantities used in Section \ref{sec:nonadiab_dephasing}) with $\delta \phi/\sqrt{2}$.

\section{Effects of the electrostatic disorder on the electron in the moving dot}
\label{app:excitation}
\begin{figure}[ht!]
	\includegraphics[width=\columnwidth]{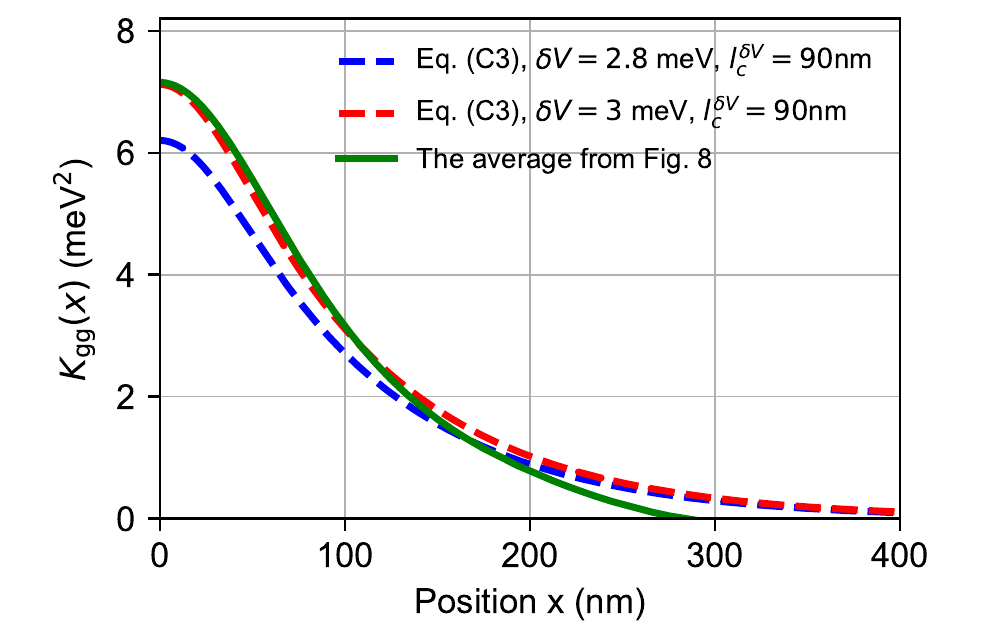}
	\caption{Correlation function of the matrix element for the electron occupying ground orbital state $K_{\text{gg}}(|x|)$. With dashed lines we plot numerically integrated formulas Eq.~\eqref{eq:kgg_dx} assuming exponential form of the correlation function of effective 1D potential $\langle \delta V(x) \delta V(0) \rangle = \delta V^2 e^{-|x|/l_c^{\delta V}}$, with $l_c^{\delta V} = 90\,\mathrm{nm}$ and two values of noise amplitude $\delta V = 2.8\,\mathrm{meV}$ (blue) and $\delta V = 3\,\mathrm{meV}$ (red) . For comparison using green solid line we plot $K_{\text{gg}}(|x|)$, obtained from the numerical averaging of different realisations of electrostatic disorder (See Fig.~\ref{fig:corr_fun}).  }\label{fig:correlations}.
\end{figure}

Let us first reconstruct the correlation function of electrostatic disorder felt by the electron in the ground state of a moving dot. Following definition from Eq.~\eqref{eq:kgg} we write such a correlation function as:
\begin{align}
    K_\text{gg}(\Delta x) = \int \text{d}x_1\text{d}x_2 &     |\psi_g(x_1)|^2 |\psi_g(x_2)|^2  \nonumber \\ &K_{\delta V}(x_1-x_2+\Delta x).
\end{align}
In the above the bare correlation function of electrostatic disorder is assumed to be translationally invariant, and have an exponential form:
\begin{equation}
    K_{\delta V}(x) = \langle \delta V(x)\delta V(0) \rangle = \delta V^2 e^{-|x|/l_c^{\delta V}},
\end{equation}
which is parameterized by the amplitude $\delta V$ and correlation length $l_c^{\delta V}$. Using Fourier transform, the $K_\text{gg}(\Delta x)$ can be written as:
\begin{equation}
\label{eq:kgg_dx}
    K_{\text{gg}}(\Delta x) =\int \frac{\text{d}k}{2\pi} \frac{2\delta V l_c}{1+(kl_c)^2} e^{ik\Delta x} |F_{\text{gg}}(k)|^2,
\end{equation}
in which we introduced the Fourier transform of electron probability density:
\begin{equation}
F_{\text{gg}}(k) = \int \text{d}x |\psi_g(x)|^2 e^{-ikx} = e^{-\frac{1}{2}kL_\text{dot}^2}.
\end{equation}
In Fig.~\ref{fig:correlations} we plot the numerically integrated Eq.~\eqref{eq:kgg_dx}, and compare against the result of Fig.~\ref{fig:corr_fun} in which $K_\text{gg}(x)$ was obtained by averaging over realisations of the electrostatic disorder. In this way prove that the exponential form of $K_{\delta V}(x)$ allows for reconstruction of the numerical result.

We now use the exponential correlation function $K_{\delta V}(x)$ to compute the transition rate from ground to excited orbital state. Following Eq.~\eqref{eq:gamo_meg}, such a rate is related to the temporal Fourier transform of the correlation function of the matrix element between ground and excited orbital states (see Eq.~\eqref{eq:mat_elem}). In analogy to Eq.~\eqref{eq:kgg_dx} such correlation function can be written as:
\begin{equation}
    K_{\text{eg}}(\Delta x)=\int \frac{\text{d}k}{2\pi} \frac{2\delta V l_c}{1+(kl_c)^2} e^{ik\Delta x} |F_{\text{eg}}(k)|^2 \,\, ,
\end{equation}
in which the filtering of spatial disorder depends on the shape of the wavefunctions of ground and excited orbital states. For harmonic potential we have:
\begin{equation}
    F_{\text{eg}}(k) = \int \text{d}x \,\psi_g(x)\psi_e(x) e^{ikx} = ikL_\text{dot} e^{-\frac{1}{2}k^2L_\text{dot}^2}.
\end{equation}
For the moving dot the argument of $K_{\text{eg}}(\Delta x)$ becomes time-dependent $\Delta x = vt$, which allows to compute the transition rate as:
\begin{align}
    \Gamma_{+,o} &= \int \text{d}t K_{\text{eg}}(vt) e^{-iE_\text{orb}t}=  \\ &= \frac{\delta V^2 l_c^{\delta V}}{v} \frac{E_\text{orb}^2\ldot^2}{v^2 + (E_\text{orb} \lcorb)^2} \exp(-\frac{E_\text{orb}^2\ldot^2}{2v^2})\nonumber,
\end{align}
Finally in the experimentally relevant limit of $v \ll E_\text{orb} \,l_c^{\delta V} $, we arrive at Eq.~\eqref{eq:gam_po} from the main text.

\section{Landau-Zener approximation of multiple step passage}
\label{app:LZ_step}
We provide here more detailed description of the electron moving over multiple atomistic steps, the model of which has been used in Sec.~\ref{sec:single_step}. We derive here the probability of coherent transition to excited valley state on a single atomistic step, and then verify result of multiple step passage against numerical simulations. 
\subsection{Single step passage}
\begin{figure}[ht!]
	\includegraphics[width=\columnwidth]{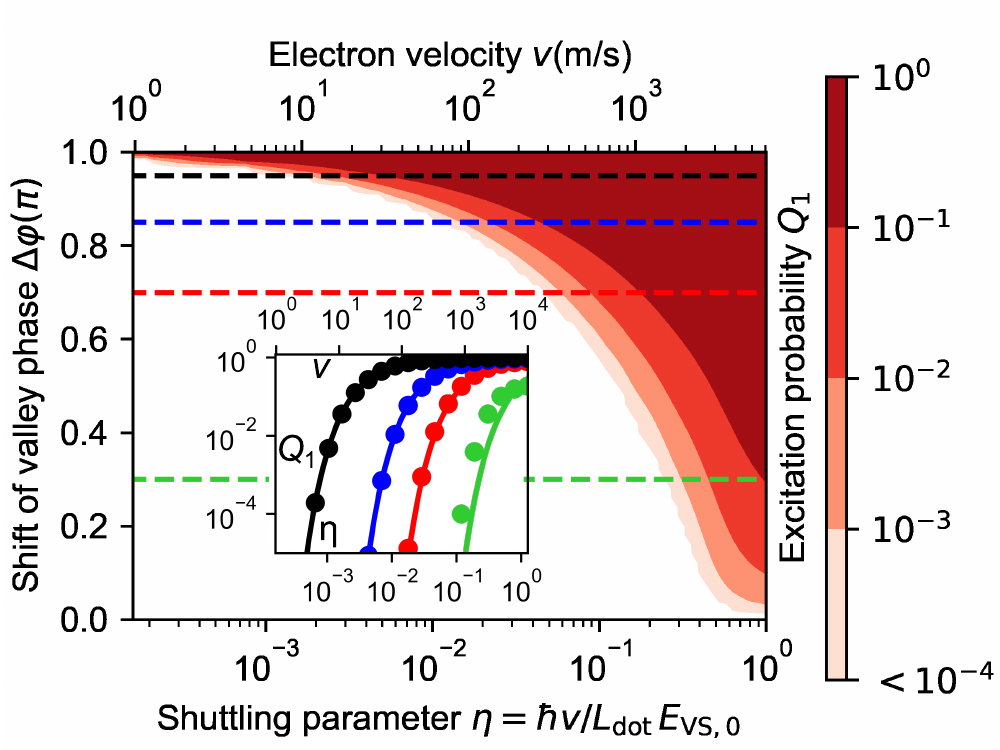}
	\caption{Occupation of higher valley state after single step passage $Q_1$ as a function of shuttling parameter $\eta$ (lower scale) and valley field angle difference $\Delta \varphi$. Inset: Numerical solution to Eq.~\eqref{eq:gradientmodel_Hv} is compared against effective L-Z approximation \eqref{eq:qstep} for selection of angles $\Delta \varphi/\pi = 0.3,0.7,0.85,0.95$ (dashed lines in larger figure). The upper x-axis translates $\eta$ into electron velocity $v= \ldot E_{\text{VS},0} \eta/\hbar$ assuming typical values of $\ldot= 20\,$nm and $E_{\text{VS},0} = 200\,\mu$eV.}
	\label{fig:single_step}
\end{figure}

First we consider passage over single step localized at $x = 0$. In the harmonic approximation the probability of occupying left ($x<0$) and right regions ($x>0$) are given by $p_{\text{L}/\text{R}}(t) = \frac{1}{2}(1\mp\erf[vt/L])$. For a step of single atomistic height, the relative valley phase between those regions is shifted by $\Delta \varphi = 0.85\pi$. For the symmetry reasons we thus assume the valley phases are given by $\varphi_{\text{VS},L} = -\Delta \varphi/2$ in the left and $\varphi_{\text{VS},R} = \Delta \varphi/2$ in the right region. This allows to write Hamiltonian \eqref{eq:Hv_int} in the form:
\begin{equation}
H_{\text{v,s}}(t) = \frac{\vsz}{2}\bigg(\cos(\frac{\Delta \varphi}{2}) \hat \sigma_x + \erf(vt/L)\sin(\tfrac{\Delta \varphi}{2}) \hat \sigma_y\bigg).
\end{equation}
Next we apply a simple basis transformation (rotation around x-axis), that allows us to cast the above Hamiltonian into Landau-Zener form $\hat H_{\text{LZ}} = \frac{1}{2}(\epsilon(t) \hat \sigma_z + \Delta \hat \sigma_x)$, i.e.:
\begin{equation}
H_{\text{v,s}}(t) =  \frac{\vsz}{2}\big(\cos(\tfrac{\Delta \varphi}{2}) \hat \sigma_x + \erf(vt/\ldot)\sin(\tfrac{\Delta \varphi}{2}) \hat \sigma_z\big),
\end{equation}
where the $\erf(x) = (2/\sqrt{\pi})\int_0^x e^{-y^2} \mathrm{d}y$ is the Gaussian error function. We can identify time-independent part as:
\begin{equation}
    \zeta  = \vsz \cos(\tfrac{\Delta \varphi}{2}),
\end{equation}
while the time-dependent term reads:
\begin{equation}
    \epsilon(t) = \vsz \erf(vt/\ldot) \sin(\tfrac{\Delta \varphi}{2}).
\end{equation}
Since L-Z formula of non-adiabatic exciation $Q_{\text{LZ}} = \exp(-\pi \zeta^2/2\hbar a)$ is applicable for linear sweep, $\epsilon(t) = a t$, we linearize the error function around $t=0$, which produces an effective sweep rate:
\begin{equation}
    a \approx \frac{2\vsz v}{\sqrt{\pi}\ldot} \sin(\tfrac{\Delta \varphi}{2}).
\end{equation}
This finally allow us to estimate probability of non-adiabatic excitation on a single step as:
\begin{align}
    Q_1 &= \exp(- \frac{\pi}{2\hbar} \frac{\sqrt{\pi}\ldot}{2\vsz \sin(\tfrac{\Delta \varphi}{2})} \vsz^2\cos^2(\tfrac{\Delta \varphi}{2})) =\nonumber\\&=
    \exp(-\pi^{3/2} \frac{ E_{\text{VS},0}\,\ldot}{4\hbar  v} \frac{\cos^2\tfrac{\Delta \varphi}{2}}{\sin\tfrac{\Delta \varphi}{2}}). 
\end{align}
In Fig.~\ref{fig:single_step} we plot occupation of excited valley state after single step passage $Q_1$ as a function of $\eta = \hbar v/\vsz \ldot$. In the inset we compare numerical solution (dots) against Landau-Zener approximation (lines) for $\Delta \varphi/\pi = 0.3,0.7,0.85,0.95$, which shows that $Q_1$ formula is applicable in the relevant range of $\Delta \varphi \sim 0.85\pi$, i.e. for wide enough sweep of effective adiabatic parameter $\epsilon(t)$. Otherwise, for small angles $\Delta \varphi$ the initial and final adiabatic ground states would be not "orthogonal enough" since for small angles $\epsilon(\infty)/t_c = \tan(\Delta \varphi/2) \approx \Delta \varphi/2$.
\subsection{Numerical test of multiple step passage}
In Sec.~\ref{sec:single_step} we computed occupation of excited valley state by concatenating number of isolated single-step passages. We argued that for small enough probability of non-adiabatic excitation on a single step $Q_1\ll 1$, the probability of coherent return to the ground state scales as $Q_1^2 \ll Q_1$ and hence can be neglected. As a result the probability of occupying higher valley state is the effect of quantum interference between transitions on single atomistic steps, i.e.
\begin{equation}
\label{eq:Q_n2}
    p_{e,v} \approx  Q_1 \left|\sum_{n=1}^{N} e^{i\Phi_{n}} \right|^2,
\end{equation}
which is bounded from above by $p_{e,v}\leq N^2 Q_1$. Additionally we argued that due to intrinsic random fluctuations of electron velocity the possibility of coherent addition of probability amplitudes is negligibly small, and in particular in the limit of large number of uncorrelated phases gives $\big|\sum_{n=1}^{N} e^{i\Phi_{n}} \big|^2 \sim N$. 

\begin{figure}[t!]
    \centering
    \includegraphics[width=\columnwidth]{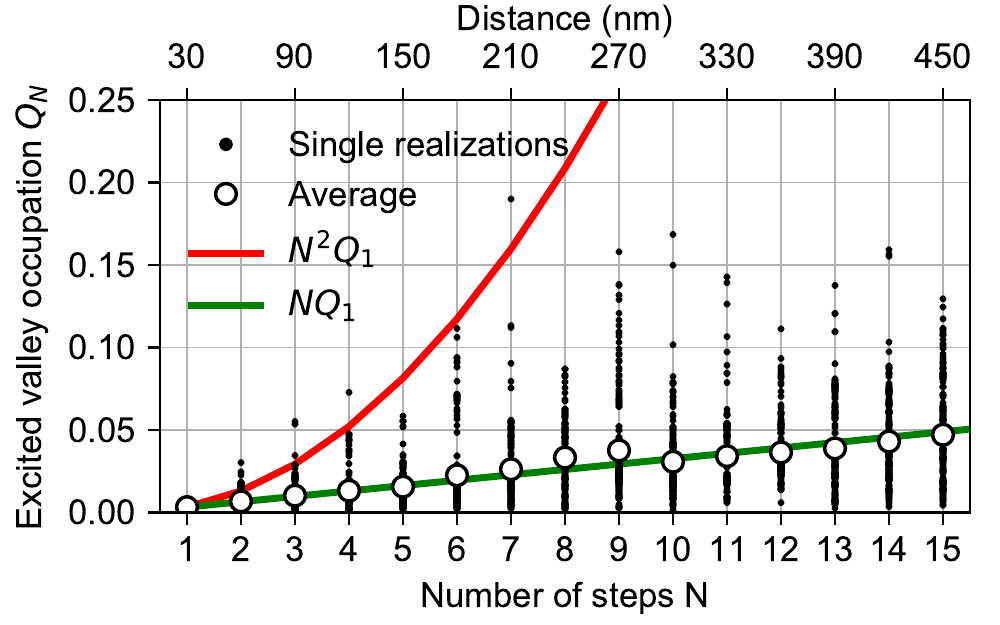}
    \caption{Probability of occupying higher valley state after N-step passage $Q_N$. We use a simple model of disorder in which electron velocity $\vd$ fluctuates between the realizations (filled dots) with rms $\delta v = 0.1v$, and for remaining parameters we take $L=15$\,nm, $E_{\text{VS},0} = 100\,\mu$eV, $\overline v= 30$\,m/s, $\vartheta = 0.25^\circ{}$, which corresponds to $\overline d \sim 30$\,nm. To show statistical behaviour, for each number of steps $N$ we draw a new arrangement of steps with terrace lengths given by the distribution Eq.~\eqref{eq:dist_tarraces} and for that arrangement simulate 200 realisations of velocity.}
    \label{fig:many_steps}
\end{figure}

Here we use Fig.~\ref{fig:many_steps} to test this hypothesis using numerical simulation of multi-step passage, with quasistatic fluctuations in electron velocity. For concreteness we take $v = 30$ m/s and assumed its uncertainty (rms) to be  $\Delta v/v= 10\%$. For each value of velocity we numerically solve the evolution generated by the time-dependent Hamiltonian: 
\begin{equation}
    \hat H_{\text{v,s}}(x_0) = \frac{E_{\text{VS},0}}{2}\sum_{n=0}^N p_{n}(x_0)\Big[\cos(n\Delta \varphi)\hat \tau_x + \sin(n\Delta \varphi)\hat \tau_y\Big],
\end{equation}
where $p_n(x_0) = \int_{x_n}^{x_{n+1}} \rho(x-x_0) \text{d}x$, in which $\rho(x-x_0)$ is the ground state probability density function of the traveling electron centered at $x_0 = vt$ and $x_n$ is the location of nth step. By averaging over many realizations of electron velocity we obtain average occupation of excited valley state:
\begin{equation}
    Q_N = \langle \Tr{\ketbra{e_v} \varrho(\tau)}\rangle_v.
\end{equation}
To reflect realistic conditions for fixed $N$ we use the same arrangement of steps for all realization of electron velocity. However we change the arrangement for each $N$ such that multiple arrangements are investigated.  We follow \cite{Zandvliet00-2, Swartzentruber93} and draw length of consecutive terraces $d_n \equiv x_{n+1} - x_n$ from independent Gaussian distributions:
\begin{equation}
\label{eq:dist_tarraces}
    P(d_n) \sim \mathcal{N}(\overline d,  (\overline d/5)^2),
\end{equation}
where $\overline d$ is the average distance between the steps, which can be related to an average interface miscut $\theta_\text{rad} = h/\overline d$. In Fig.~\ref{fig:many_steps} we plot the results of single realization (small dots) their average (large hollow dots) and compare them against the coherent limit $Q_N^{\text{max}} = N^2 Q_1$ (red solid line) and average limit $Q_N \approx N Q_1$ (green solid line) as a function of number of steps, proportional to shuttling distance $L_s = N \overline d$.

As it can be observed the average occupation of excited valley state after traveling over $N$ steps agrees with linear prediction, even for the parameters which exaggerates typical valley excitation, i.e. $Q_1 \approx 10^{-2}$ . What is more, single realizations of the shuttling are generally contained below $N^2 Q_1$ red line, with an exception of few points for which a single realization of $v$ allowed for more then a single transition between adiabatic levels. Notably most results lies also below the average, the value of which is however increased by the outliers close to $N^2Q_1$ line. We stress out that the predictions are valid for typical for conveyor $N Q_1 \ll 1$, since otherwise one has to take into account more then a single transition between the levels and more complex interference pattern.

\section{Spin-valley hot spot}
\label{app:hot_spot}
\subsection{Estimation of spin-valley coupling}
\addt{We start the analysis by characterizing the spin-valley avoided crossing visible in Fig.~\ref{fig:hotspot}. To do so we estimate typical values of the spin-valley coupling  $\Delta_{\text{s-v}}$ (the gap) and the effective sweep rate $a\equiv\mathrm{d}(\vs(t)-E_z)/\mathrm{d}t$, which together give an estimate of time spent around the avoided crossing $\tau_{\text{s-v}} \approx \Delta_{\text{s-v}}/a$. The spin-valley coupling can be computed in second order perturbation theory as a result of spin-orbit and valley-orbit couplings:
\begin{align}
\Delta_{s-v}^{\text{art}}\approx&2\frac{|\langle g_o\downarrow| H_\mathrm{SO}|e_o\uparrow\rangle\langle e_oe_v| H_\mathrm{VO}|g_0g_v\rangle|}{E_\mathrm{orb}}.
\end{align}
We start with the artificial spin-orbit coupling (also termed synthetic in the literature \cite{Burkard21}), that is caused by the presence of transverse magnetic field gradients.
For the gradients of the order of $(\Delta B/\Delta x) = 0.1\,\mathrm{mT/nm}$, we have $|\bra{g_o\downarrow} H^\mathrm{art}_\mathrm{SO}\ket{e_o\uparrow}|\sim 0.2\,\mu$eV. Note that we can afford to reduce the gradient magnetic field by one order of magnitude compared to e.g. Ref.~\cite{Yoneda18}, since in a quantum computing architecture based on shuttling the QD displacement for electric dipole spin resonance can be increased to $\sim 10$\,nm, such that no compromise in the Rabi frequency is expected. The valley-orbit coupling element can be computed as in Sec.~\ref{sec:valley_relax}, and its order of magnitude can be conservatively estimated to be comparable to the valley splitting $\langle e_o e_v| H_\mathrm{VO}|g_o g_v\rangle\approx 100\,\mathrm{\mu eV}$. 
Together, the gradient dominated SOC produces the spin-valley coupling $\Delta_{\text{s-v}}^\text{art} \leq\, 20\,\mathrm{neV}$. In absence of a strong enough magnetic gradient, the intrinsic spin-orbit interaction gives the same order of magnitude $\Delta_{\text{s-v}}^\text{int} \leq 20$\,neV, since for typical couplings ($\alpha,\beta \approx 50 \,\mathrm{m/s}$ \cite{Tanttu19}), despite larger matrix element $|\langle g_o\downarrow| H^\mathrm{int}_\mathrm{SO}|e_o\uparrow\rangle|\sim 2\,\mu$eV, the additional reduction of $\Delta_{\text{s-v}}^{\text{art}}$ by a factor of $E_\mathrm{Z}/E_\mathrm{orb}$ is caused by the Van-Vleck cancellation \cite{VanVleck40,Abrahams57,Khaetskii01,Hanson07}. Next, we estimate the effective sweep rate from the expression $a\approx \vs v/\ldot$, which conservatively assumes that the valley splitting $\vs$ may change by values comparable to its own magnitude over the range of a QD length scale. Taking $\vs\approx 100\,\mathrm{\mu eV}$, we can estimate $a\approx 50 \,\mathrm{\mu eV}/\mathrm{ns}$ for a shuttling velocity of $v=10\,\mathrm{m/s}$.}

\subsection{Estimation of the spin flip probability due to valley relaxation around the hot-spot (mechanism 1)}
\addt{
We compute now $\delta p_\uparrow$ due to the mechanism of temporal spin-valley mixing (labeled as mechanism 1). It can be done by multiplying spin relaxation rate and the time spent around avoided crossing $\Delta_{s-v}/a$. 
For a conservative estimate, we assume that the spin relaxation rate is upper-bounded by the inverse of the valley lifetime evaluated at the Zeeman splitting $\tau_{r,v}(E_Z)$, and thus reads
\begin{equation}
    \delta p_\uparrow^{(1)}\sim \Gamma_{s}(t) \frac{\Delta_{\text{s-v}}}{a} \geqslant \frac{1}{\tau_{r,v}(E_Z)} \frac{\Delta_{\text{s-v}}}{a} \to \frac{0.1}{\tau_{r,v}(E_z)[\mathrm{ns}] v[\mathrm{\tfrac{m}{s}}]} \,\, ,
\end{equation}
where we have used the values computed above $\Delta_{s-v}^{\text{art}} \sim 200$\,neV and $\vs \sim 100\,\mu$eV. From Fig.~\ref{fig:valleyT1_step_grad}, we conclude that phonon-dominated relaxation yields $\tau_{r,v}(E_z)\geq10^4\,\mathrm{ns}$\, for $E_z \leq 100\,\mathrm{\mu eV}$. Thus, $\delta p_\uparrow^{(1)} \leq  10^{-5}/v[m/s]$ which is at least order of magnitude smaller then a spin-flip caused by an adiabatic transition (See Eq.~\ref{eq:hot_spot_2}). We point out that other mechanisms of relaxation can dominate over phonon emission at smaller $E_Z$. However, those smaller values of $E_Z\sim 10$\,$\mu$eV would most likely eliminate the occurrence of hotspots in the first place. Finally, let us note that the above value most probably overestimates the relaxation experienced by an electron going through the hotspot anticrossing in mostly diabatic way. We have taken the maximum relaxation rate from a spin-valley hybridized lower-energy state at the anticrossing, and while the electron spends the time $\Delta_{s-v}/a$ in state of this character during the mostly adiabatic evolution, its time-averaged state is probably closer to being non-hybridized for the mostly for the mostly diabatic evolution.
}
\end{appendix}


%

\end{document}